\documentclass[aps,reprint,groupedaddress]{revtex4-2}
\usepackage{graphicx}
\usepackage{amssymb,amsmath}
\usepackage{epstopdf}
\usepackage{subcaption}
\usepackage{xcolor}
\usepackage{hyperref}
\usepackage[normalem]{ulem}

\begin{document}


\title{A Model for the Influence of Media on the Ideology of Content in Online Social Networks}


\author{Heather Z. Brooks}
\affiliation{Department of Mathematics, University of California Los Angeles, Los Angeles, California 90095, USA}

\author{Mason A. Porter}
\affiliation{Department of Mathematics, University of California Los Angeles, Los Angeles, California 90095, USA}


\date{\today}

\begin{abstract}

Many people rely on online social networks as sources of news and information, and the spread of media content with ideologies across the political spectrum influences online discussions and impacts actions offline. To examine the impact of media in online social networks, we generalize bounded-confidence models of opinion dynamics by incorporating media accounts as influencers in a network. We quantify partisanship of content with a continuous parameter on an interval, and we formulate higher-dimensional generalizations to incorporate content quality and increasingly nuanced political positions. We simulate our model with one and two ideological dimensions, and we use the results of our simulations to quantify the ``entrainment'' of content from non-media accounts to the ideologies of media accounts in a network. We maximize media impact in a social network by tuning the number of media accounts that promote the content and the number of followers of the accounts. Using numerical computations, we find that the entrainment of the ideology of content spread by non-media accounts to media ideology depends on a network's structural features, including its size, the mean number of followers of its nodes, and the receptiveness of its nodes to different opinions. We then introduce content quality --- a key novel contribution of our work --- into our model. We incorporate multiple media sources with ideological biases and quality-level estimates that we draw from real media sources and demonstrate that our model can produce distinct communities (``echo chambers") that are polarized in both ideology and quality. Our model provides a step toward understanding content quality and ideology in spreading dynamics, with ramifications for how to mitigate the spread of undesired content and promote the spread of desired content.

\end{abstract}


\maketitle


\section{Introduction}

Online social media (such as Twitter, Facebook, Instagram, and others) have become
extremely influential sources of news in daily life. For example, over two thirds of American adults who participated in a recent survey responded that social media, and their associated networks, is their primary source for obtaining news \cite{pew2018}. Given this large audience and the ease of sharing content online, the content that spreads on online social networks can have important consequences on public opinion, policy, and voting \cite{bessi2016social, bond201261}.

The spread of content in a social network depends on the response and biases of the individuals in that network. Individual user preferences play a strong role in the choice to consume news that conforms to (or even enhances) previously held views \cite{bakshy2015exposure}. One reason that propaganda, misinformation, and disinformation has become so widespread on social media networks is that users are more likely to share a false or misleading story if it seems to confirm or support their biases \cite{bessi2014social,cinelli2019selective}. Another challenge is that content is also spread and amplified through bot, cyborg, and sockpuppet accounts \cite{ferrara2016rise}. 

The political biases of content and the accounts that spread it goes beyond a naive separation of `liberal' versus `conservative', and the political spectrum of ideology can include axes for social views (`progressive' versus `conservative'), economic views (`socialist' versus `capitalist'), views on government involvement (`libertarian' versus `authoritarian'), and others. Opinions can also be issue-dependent. In a large body of research, Poole and Rosenthal \cite{poole2005spatial, poole2011ideology} developed methods to classify ideological positions in roll-call voting in the United States Congress in ideological space, and they found that two dimensions is typically sufficient to capture an overwhelming majority of the variance of ideologies. Their work laid the foundation
of the Voteview project \cite{voteview}, which characterizes the ideologies of legislators in the United States Congress based on their voting records. Multidimensional political ideologies have also been explored in the context of the United States Supreme Court. For example, by examining singular value decompositions, Sirovich \cite{sirovich2003pattern} demonstrated that decisions by the judges in the second Rehnquist Court are well-described by a two-dimensional subspace. This has typically also been true of voting on legislation in the United States Congress \cite{porter2005}.
 
 Political biases are not the only way in which people
 judge the news that they consume and share on social media. As the prevalence of misinformation, disinformation, and ``fake news" becomes increasingly prominent in everyday life and the global conversations of talking heads, social-media users are also considering the quality of news content. For example, in the aforementioned 
 Pew Research Study \cite{pew2018}, the majority of respondents acknowledged that the news that they see, consume, and spread on social-media platforms may be inaccurate. There is evidence that poor-quality content 
  travels deeper, more broadly, and faster than high-quality content (such as fact-based reporting) \cite{vosoughi2018spread}. The same study concluded that these differences in spreading patterns are primarily the result of human judgement, rather than arising from accounts such as bots.

A large body of work has studied the spread of content on social media \cite{guille2013information,vosoughi2018spread}. One approach is to mathematically model the spread of content on social media using ordinary differential equations, where the equations can capture changes in time of the proportion of a population that is ``susceptible to", ``exposed to", ``infected with", or ``immune to" the propagation of such content (e.g., a rumor) \cite{jin2013epidemiological,liu2017analysis,wang2014siraru,zhao2013sir}. Such compartmental models have the advantage of being analytically tractable, but they do not capture the effects of either network structure or heterogeneity in account characteristics. Other studies have focused instead on data-driven or computational approaches \cite{bakshy2015exposure,batrinca2015social,bessi2014social,vosoughi2018spread}. One important research direction in this vein is quantifying the existence of distinct communities and ``echo chambers" \cite{sunstein2009,flaxman2016,del2016echo,cota2019quantifying} on social media, wherein individuals interact primarily with other like-minded individuals. Some recent work on modeling the spread of content has tried to bridge the gap between these two approaches by introducing mathematical models that capture 
network features of social media \cite{yy-book,obrien2018spreading}, including some very recent mechanistic models of radicalization dynamics \cite{baumann2019modeling} and filter bubbles \cite{chitra2019understanding}.
Our approach in the present paper has some similarities with those in Prasetya and Murata \cite{prasetya2018modeling}, who generalized an independent-cascade model to explore the effects of selective exposure and connection strength on news propagation, and Martins \cite{martins2008mobility}, who developed a continuous-opinion, discrete-action model to explore the emergence of extremism. Extremism in models of opinion dynamics was also studied recently in the context of radicalization and terrorism \cite{chuang2019mathematical}.

In the present paper, we formulate and study a model for the influence of media accounts on the ideology and quality of content that is shared in a social network. First, we develop a general content-spreading model with an $n$-dimensional continuous ideology space and spreading choices based on a bounded-confidence mechanism \cite{opinion-review,deffuant2000mixing,hegselmann2002opinion}. We also model how to include media accounts as influencer nodes in a network. This has been considered in other recent work, such as in the context of voters models with discrete-valued opinions \cite{bhat2019a,bhat2019b}. In our work, we 
conduct extensive numerical simulations to examine the impact of media accounts on content in examples with one and two ideological dimensions. We then introduce content quality into our model; this is a key novel feature of our work. Using the resulting enhanced model, we employ numerical simulations to examine the effect of multiple media sources on the ``entrainment'' (with one possible uncharitable interpretation as ``brainwashing'') of non-media nodes in the network by media nodes. As a parallel to our consideration of media nodes, radical or charismatic leaders have also been introduced into voter models \cite{mobilia2007role,acemouglu2013opinion}, DeGroot models \cite{yi2019disagreement}, and bounded-confidence models \cite{hk2015}.

Our work advances the study of online content spreading in several ways. In our model, we propose a measure of media impact; this allows us to quantify the influence that a set of media accounts has on the ideology of content in a social network 
at consensus. Our model has the advantageous feature of supporting a multidimensional continuous ideology space for content, although we are not the first to employ a multidimensional opinion space. See \cite{lorenz2008fostering} for a discussion of consensus in bounded-confidence models in $\mathbb{R}^d$ and \cite{stamoulas2018convergence} for some results on the convergence and stability of such models. For wider-ranging discussions of consensus dynamics and opinion models, we see \cite{opinion-review,bullo2019}.
A key novelty of our work is the introduction of content quality into the spreading dynamics; to our knowledge, our model is the first to explore the effects of content quality on spreading. 
This is an important advancement for two reasons: (1) the quality of news content has a major effect on its spreading dynamics, and it is therefore important to study it from a dynamical perspective; and (2) the spread of poor-quality news content is a major social problem. Investigations into such problems --- especially ones that produce potential mechanistic insights into them --- have the potential to help motivate strategies for mitigating their effects.

Our paper proceeds as follows. In Sec.~\ref{sec:overview}, which gives a concise overview of our work, we give a non-technical sketch of our model and briefly outline our key results. In Sec.~\ref{sec:model}, we develop our bounded-confidence model with content spreading. We describe a network structure that incorporates both non-media and media accounts in Sec.~\ref{sec:networkstructure}, and we introduce our dynamic content updating rule in Sec.~\ref{sec:contentupdating}. In Sec.~\ref{sec:1d}, we examine the dynamics of our model with one ideological dimension for content. We describe how we simulate the model in Sec.~\ref{sec:1dsim}, and we quantify media impact for different parameter values and network architectures in Sec.~\ref{sec:1dimpact}. We present a generalization of our model to two ideological dimensions for content in Sec.~\ref{sec:2d}. In Sec.~\ref{sec:quality}, we develop our model further to incorporate a notion of media quality into the content updating rule. In Sec.~\ref{sec:multi}, we show simulations and results of media impact with the incorporation of content quality, including an example in which we draw media biases and qualities from a public hand-curated distribution. We conclude and discuss several possible extensions of our work in Sec.~\ref{sec:discussion}.


\section{Overview of our Model and Results} \label{sec:overview}

The aim of our paper is to advance understanding of the spread of misinformation and ``fake news" on social media through mathematical modeling. This contrasts with the more common data-driven, computational approach to these problems that tend to use existing models (such as compartmental models). An important facet of our approach is that the underlying spreading mechanisms are based on continuous, multidimensional quantities --- specifically, ideology and quality of content --- that result in discrete actions (spreading a message). By incorporating these continuous, multidimensional variables into the spreading dynamics, our model (1) provides a plausible framework 
 to study the spread of extremism on networks and (2) and lays the foundation for systematic comparison with empirical data of the spread of ideologies online.

Our model takes the form of a dynamical process on a network. It includes (1) a time-independent network structure of accounts and relationships between them in an online social network (see Sec.~\ref{sec:networkstructure}) and (2) a mechanism for the temporal evolution of the ideology of the content that is produced by media accounts in that network (see Sec.~\ref{sec:contentupdating}). To represent the network structure, we construct a graph in which nodes represent accounts and edges represent followership relationships between these accounts. One can either construct these networks from empirical data (such as Facebook or Twitter networks) or using synthetic networks (e.g., from generative models of random graphs). In a network, we designate each account as either a media account or a non-media account. {\em Media accounts} are accounts that do not follow any other accounts but are followed by at least one other account. In our model, we construe media accounts as ``influencers", ``opinion sources", or ``external opinion forcing" in a network. It is for simplicity that we assume that their content is not influenced by any other accounts. All other accounts in a network are {\em non-media accounts}, which each follow at least one other account in the network.

We represent content that spreads in an online social network based on its ideology (see Sec.~\ref{sec:multi}) and eventually also on its quality (see Sec.~\ref{sec:quality}). We suppose that accounts in a network spread content during each discrete time step. We determine the ideology of the content that is spread by an account at each time step with a simple update rule. A helpful way to conceptualize the update rule is as follows. At each time step, each account ``reads" the content from all of the accounts that it follows, and it is influenced only by accounts whose content is sufficiently similar to the ideology of its own content. In the next time step, the account spreads new content in which the new ideology is a mean of its old ideology and the aforementioned sufficiently similar ideologies of the accounts that it reads. As a real-life example, a user may read several articles on a topic and then create a message (e.g., a tweet) based on these articles that has their own ``spin" on the material. We give a precise definition of the update rule in Sec.~\ref{sec:contentupdating}. 

An important question that we are able to study with our model is how much the ideology of the media account(s) in a network entrains the content that is spread by non-media accounts in that network. That is, how strong does the external forcing by media need to be to have a concrete impact on the outcome of the ideological dynamics? To quantify the amount of media influence, we calculate a summary statistic that we call {\em media impact}. Using numerical simulations of our model, we illustrate how the number of media accounts and the number of followers per media account affect the media impact. Surprisingly, we find that the most successful media entrainment occurs when there are a moderate number of media accounts, each of which has a moderate number of followers. We study how network parameters (such as the number of non-media accounts) affect this phenomenon in Sec.~\ref{sec:1d} and Sec.~\ref{sec:2d}.

An exciting outcome of our modeling and analysis is that ``echo chambers" arise naturally from our content-spreading mechanism. Following common usage in sociological research (see, e.g., \cite{sunstein2009} and \cite{flaxman2016}), we use the phrase ``echo chambers" to refer to groups of non-media nodes who primarily or solely influence only each other's ideologies and are not influenced much by those accounts that are outside the group. 
These ideological echo chambers emerge even when there are many followership connections between accounts in the different ideological echo chambers. (In our model, the follower relationships do not change, although it would be fascinating to study a generalization of our model in the form of a coevolving network.)
 In Sec.~\ref{sec:multi}, we use a hand-curated distribution of ideologies and qualities of real-world media outlets as an input to our model. We observe the emergence of two polarized communities: one with high-quality, ideologically moderate content and one with conservative, low-quality content. Our model is very flexible, as it allows many generalizations and modifications to study features of interest in online social networks. For example, one can study the influence of multiple social-media platforms by generalizing our networks to multilayer networks, and one can also draw media content and ideology from different probability distributions. It is also possible to adapt our model to study other applications in which one combines continuous parameters (for ideology, quality, self-confidence, or something else) with discrete actions based on those parameters. A few examples include models for gambling in sports, choices among competing products, and choices of students of majors and courses to take at a university.


\section{A Bounded-Confidence Model with Content Spreading} \label{sec:model}

\subsection{Social network structure}\label{sec:networkstructure}

Consider a social network with $N$ non-media accounts. We represent this social network with a graph $G(\mathcal{V}_N,\mathcal{E}),$ where each nodes in the set $\mathcal{V}_N$ represent an account and each edge in the set $\mathcal{E}$ represents a follower relationship from one account to another. Specifically, we say that account $i$ is a {\em follower} of account $j$ if there is a directed edge from $j$ to $i$. With this structure, we can represent this social network as an adjacency matrix $A$, where $A_{ij}=1$ if account $i$ follows account $j$ and $A_{ij}=0$ otherwise. We do not assume reciprocal follower relationships, so $A$ is not symmetric in general. We also assume that the network structure is fixed, so follower relationships do not change in time. 

We now introduce media accounts into the above social network. Suppose that media accounts do not change their ideology, as encapsulated in the content that they produce, so they are not influenced by other accounts. Therefore, media accounts only have edges that are directed outward (i.e., their in-degree is always $0$). Of course, this is a simplification of real-world media outlets, which may (in some cases) be swayed or affected by individuals or by public opinion. We make this simplifying assumption for two reasons: (1) we suppose that the media ideology represents the ideology of a particular topic or news story (and such a view of a given media outlet is unlikely to change much on the time scale of online content spread of a story); and (2) it allows us to examine the effect of an ``external forcing" of ideological content. 
Our assumption is analogous to the inclusion of zealots, which have been studied in various opinion models, including
voter models \cite{mobilia2007role} and the Deffuant model \cite{sobkowicz2015extremism}. 
We let $M$ denote the number of media accounts and $n_M$ denote the number of followers (i.e., out-edges) per media account. We assume that each media account has the same number of followers, with one exception in Sec.~\ref{sec:multi} (where we impose a media follower-number distribution based on data from real-world media outlets). Once we add media accounts, the total number of nodes in a network is $N+M$.

We represent the ideology of each account in a network at discrete time $t$ by ${\bf x}_i^t$ (where $i \in \{1,\dots,N+M\}$), and we take the ideology space to be continuous and bounded, such that ${\bf x} \in  [-1,1]^d$. We allow the ideology space to be $d$-dimensional, so we can be more nuanced than the typical choice in prior research of using $d = 1$ \cite{meng2018opinion}. At time $t$, each account spreads content (perhaps with their own spin, as has been studied for memes using Facebook data \cite{fb-meme}) that reflects its current ideology ${\bf x}_i^t$.  


\subsection{Content updating rule}\label{sec:contentupdating}

We update content synchronously at each discrete time step in the following way. Accounts update their content by averaging the ideologies of observed content 
that is sufficiently close to their own opinion; we say that an account is {\em receptive} to such content. Accounts include their own ideology at the previous time step when determining their location in ideological space in the next step. During that step, they share content with that ideology. Accounts see content only from accounts that they follow; therefore, media accounts are not receptive to any other content. We define a receptiveness parameter $c$, which gives the distance in ideology within which an account is receptive. With this notation, we define the set of accounts to which account $i$ is receptive to be $I_i = \{ j \in \{1, \dots, N+M\} | A_{ij}=1; \mathrm{dist}({\bf x}_j,{\bf x}_i )<c\}$. We also have to decide how to measure distance $\mathrm{dist}(y)$ in ideological space. One can consider any metric, and we choose to use the $\ell_p$ norm.

\begin{figure}[h!]
\centering
\includegraphics[width=0.5\textwidth]{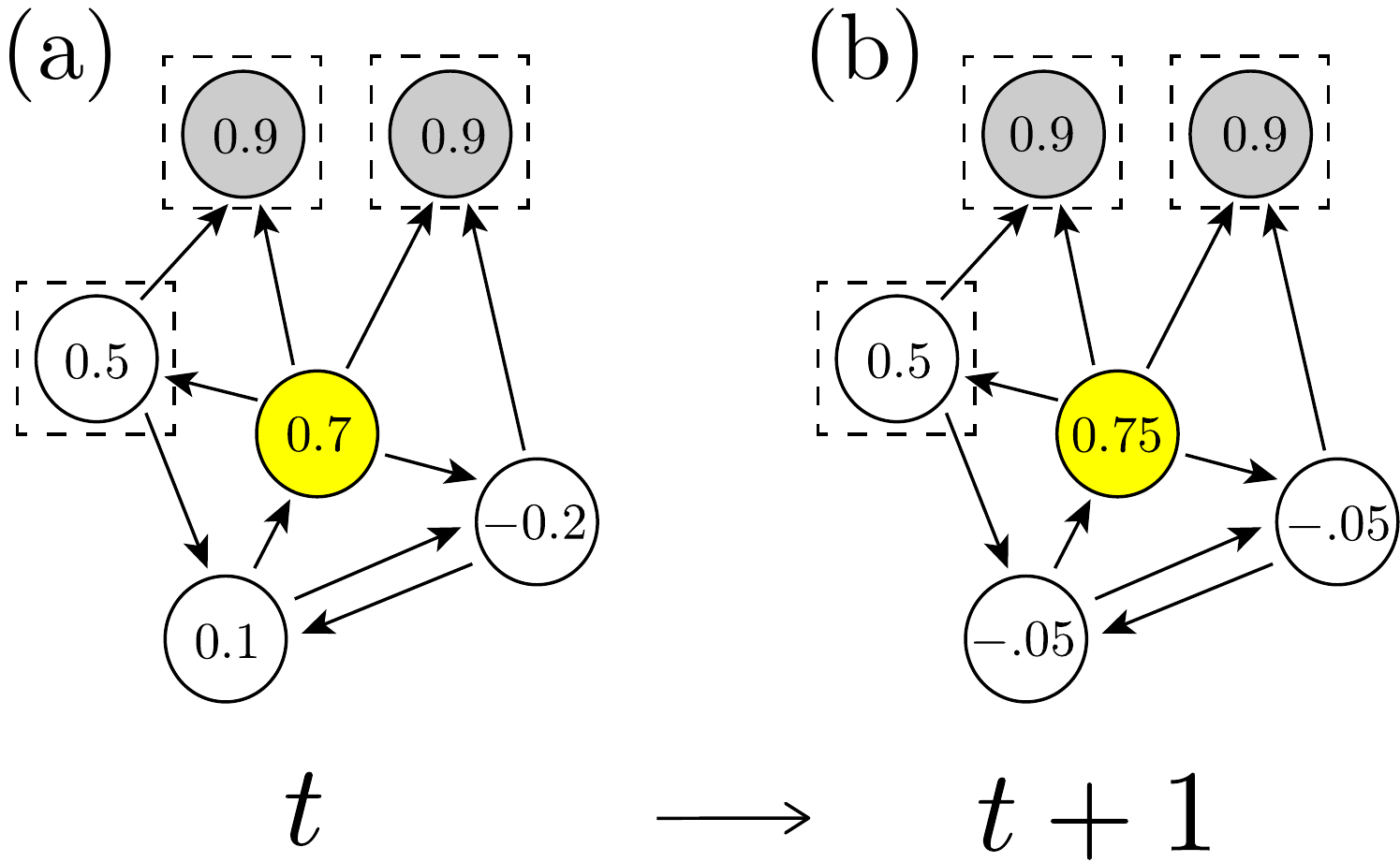}
\caption{A schematic of the ideology updating rule in our model. In panel (a), we show the ideology of nodes in a social network at time $t$. In panel (b), we show their ideologies in the next time step (after we have applied the updating rule). In this example, there are $M=2$ media nodes (gray circles) and $N=4$ non-media nodes. To make this example more concrete, let's focus on the yellow highlighted node. The yellow node follows four other accounts, two of which are media accounts and two of which are non-media accounts. If we take the receptiveness parameter to be $c=0.5$, the yellow node is receptive only to content from three of these accounts (specifically, the ones
that are surrounded by dashed boxes). After averaging the content from the accounts to which it is receptive, the yellow account updates its ideology for the next time step to be
 $x^{t+1}= \frac{1}{4}\left( 0.7+0.5+0.9+0.9\right) = 0.75$. During time step $t + 1$, this account shares content with that ideology.
 }
\label{fig:schematic}
\end{figure}

The updating mechanism for the ideology of an account is
\begin{align}
	{\bf x}_i^{t+1} = \frac{1}{\vert I_i \vert+1}\left( {\bf x}_i^t + \sum_{j\in I_i} {\bf x}_j^t\right)\,,
\end{align}
which we can also write as
\begin{equation}
	{\bf x}_i^{t+1} = \frac{1}{\vert I_i \vert+1}\left( {\bf x}_i^t + \sum_{j=1}^{N+M} A_{ij}{\bf x}_j^tf({\bf x}_j^t,{\bf x}_i^t)\right)\,,
\label{eqn:generalmodel}
\end{equation}
where $f({\bf x})=1$ if  $\mathrm{dist}({\bf x}_j,{\bf x}_i )<c$ and $f({\bf x})=0$ otherwise. In Fig.~\ref{fig:schematic}, we show a schematic of this updating rule with a small, concrete example. Our model builds on previous work in opinion dynamics. Our consensus-forming mechanism from averaging ideology is reminiscent of the influential work of DeGroot \cite{degroot1974reaching}. Although this is a standard choice, a recent paper by Mei et al.~\cite{mei2019occam} suggested that weighted-median influence is also a reasonable choice for consensus formation. In the present paper, we draw our bounded-confidence mechanism from the Hegselmann--Krause (HK) \cite{hegselmann2002opinion} and Deffuant \cite{deffuant2000mixing,meng2018opinion} models of continuous opinion dynamics. Bounded-confidence updates are also related to the DeGroot--Friedkin (DF) model of social power \cite{jia2015opinion}.

\section{Dynamics with a one-dimensional ideology} 
\label{sec:1d}

As an example, we examine content spreading when our model has one ideological dimension, where we interpret the value $x_i \in [-1,1]$ as the political perspective of the content of account $i$ on a liberal--conservative axis. We take $x=-1$ to be very liberal, $x=1$ to be very conservative, and $x=0$ to be moderate. We use the distance $\mathrm{dist}({\bf x}_j,{\bf x}_i ) = \vert {\bf x}_j-{\bf x}_i \vert$. We then write the content updating rule for these evolving pieces of one-dimensional political content (see Eq.~\ref{eqn:generalmodel}) by writing
\begin{align}
	f(x_j,x_i) = \begin{cases} 1\,, & \text{ if } \vert{\bf x}_j-{\bf x}_i \vert<c \\
		0\,, & \text{ otherwise}
		\end{cases}\,.
\end{align}
In this example, we suppose that there are $M$ media accounts that all have the same political opinion $x_M \in [-1,1].$ There are 
multiple possible interpretations of this assumption.
 For example, one can interpret it as $M$ different media outlets that share the same message, which has a given political bias (as represented by a particular location in ideological space). Another way to interpret it is as one media account with $M-1$ affiliated ``sockpuppet" accounts, which it is using to help spread its message with the specified political bias.


\subsection{Simulations}
\label{sec:1dsim}

In each network, we suppose that there are $N$ non-media accounts, and we vary the number $M$ of media accounts and the number $n_M$ of followers of each media account. We assume that all media accounts in a given simulation have an equal number of followers, although it would be interesting to extend our model by 
drawing $n_M$ from a distribution of numbers of followers. Unless we note otherwise, we use $200$ trials for each numerical experiment, and we interpret a trial to have ``converged" if $\vert x_i^{t}-x_i^{t-1}\vert <10^{-4}$ for all $i$ for ten consecutive steps. For simplicity, we set the receptiveness parameter to take the value $c=0.5$, except for Fig.~\ref{fig:ERvaryc}, where we examine the effects of varying this parameter. For a detailed discussion on the effects of the receptiveness parameter in bounded-confidence models, see \cite{meng2018opinion}. 

We initialize the non-media accounts to have evenly-distributed initial ideologies, so
\begin{equation*}
	{\bf x}_0 = \left[-1, -1+\frac{2}{N-1}, \dots, 1\right]\,.
\end{equation*}	 
We uniformly randomly permute the starting values of non-media nodes for each trial, so they are not spatially ordered in the network. In each trial, each media account has $n_M$ distinct followers, which we choose uniformly at random from the non-media accounts. We choose each media account's followers independently, so it is possible for a non-media account to follow multiple media accounts.

\begin{figure}[h!]
	\begin{subfigure}[t]{0.5\textwidth}
		\includegraphics[height=1.75in]{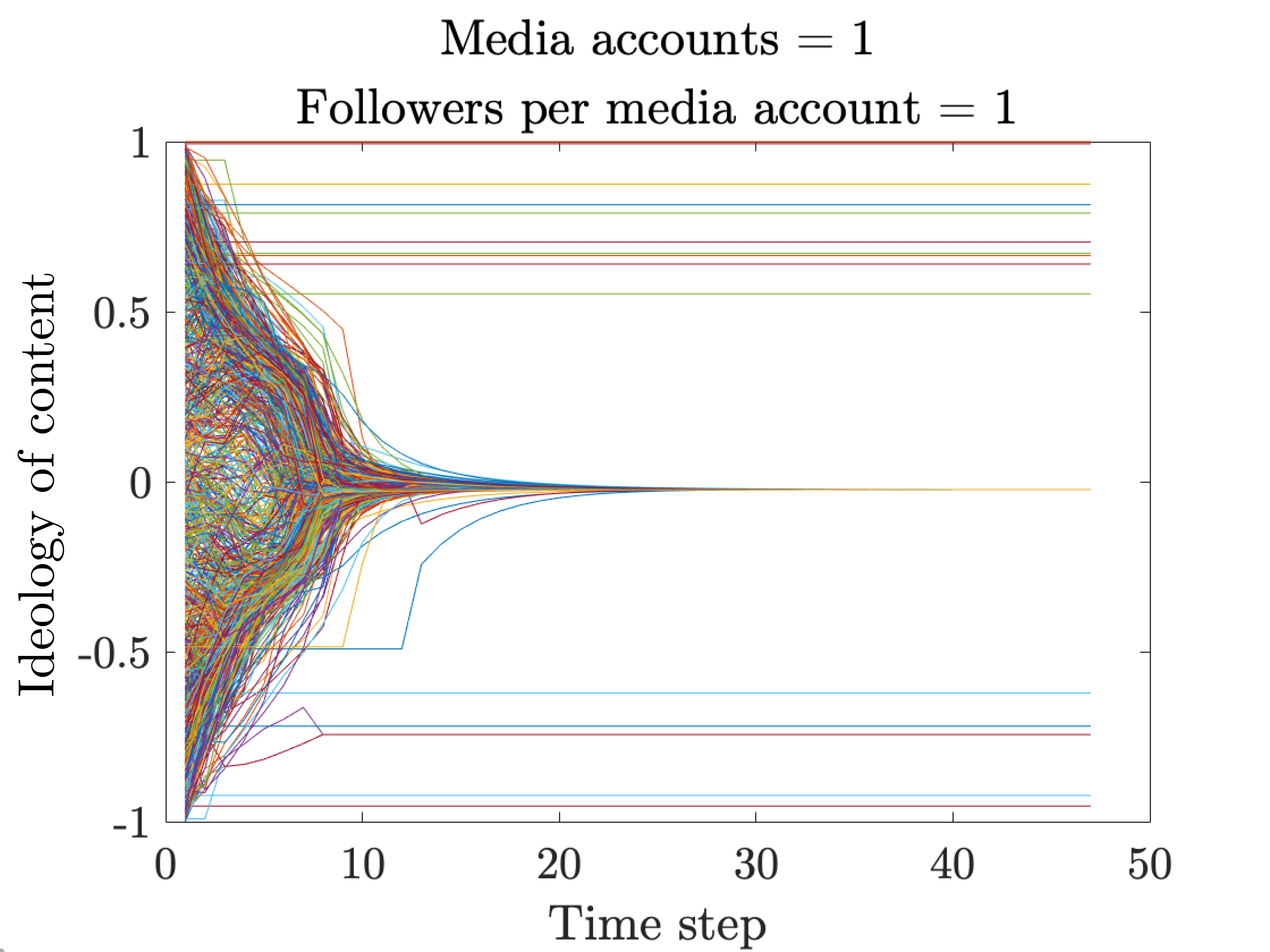}
	\caption{}
	\label{fig:ReedSmallM}
	\end{subfigure}
	\begin{subfigure}[t]{0.5\textwidth}
		\includegraphics[height=1.75in]{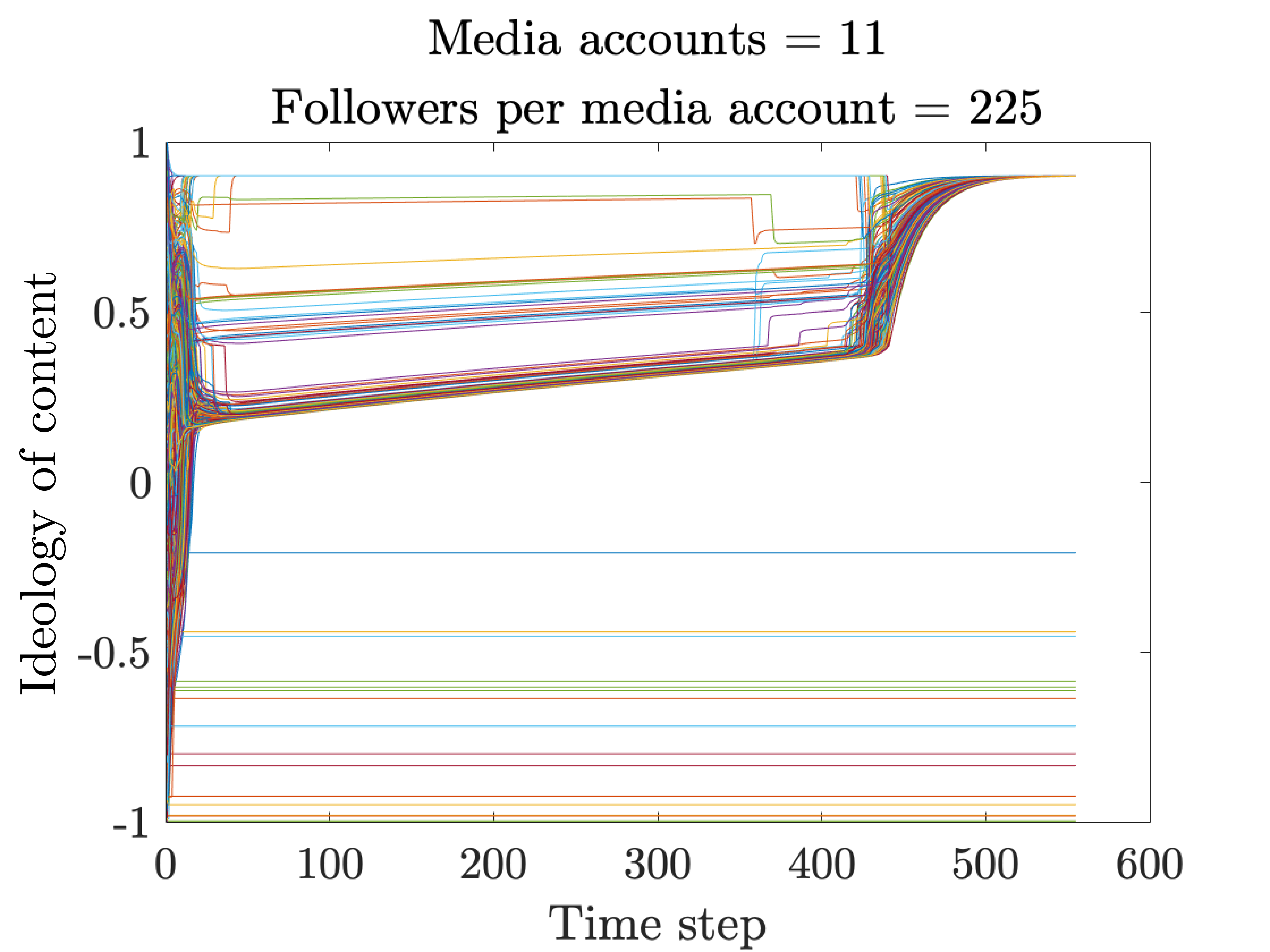}
	\caption{}
	\label{fig:ReedMedM}
	\end{subfigure}
	\begin{subfigure}[t]{0.5\textwidth}
		\includegraphics[height=1.75in]{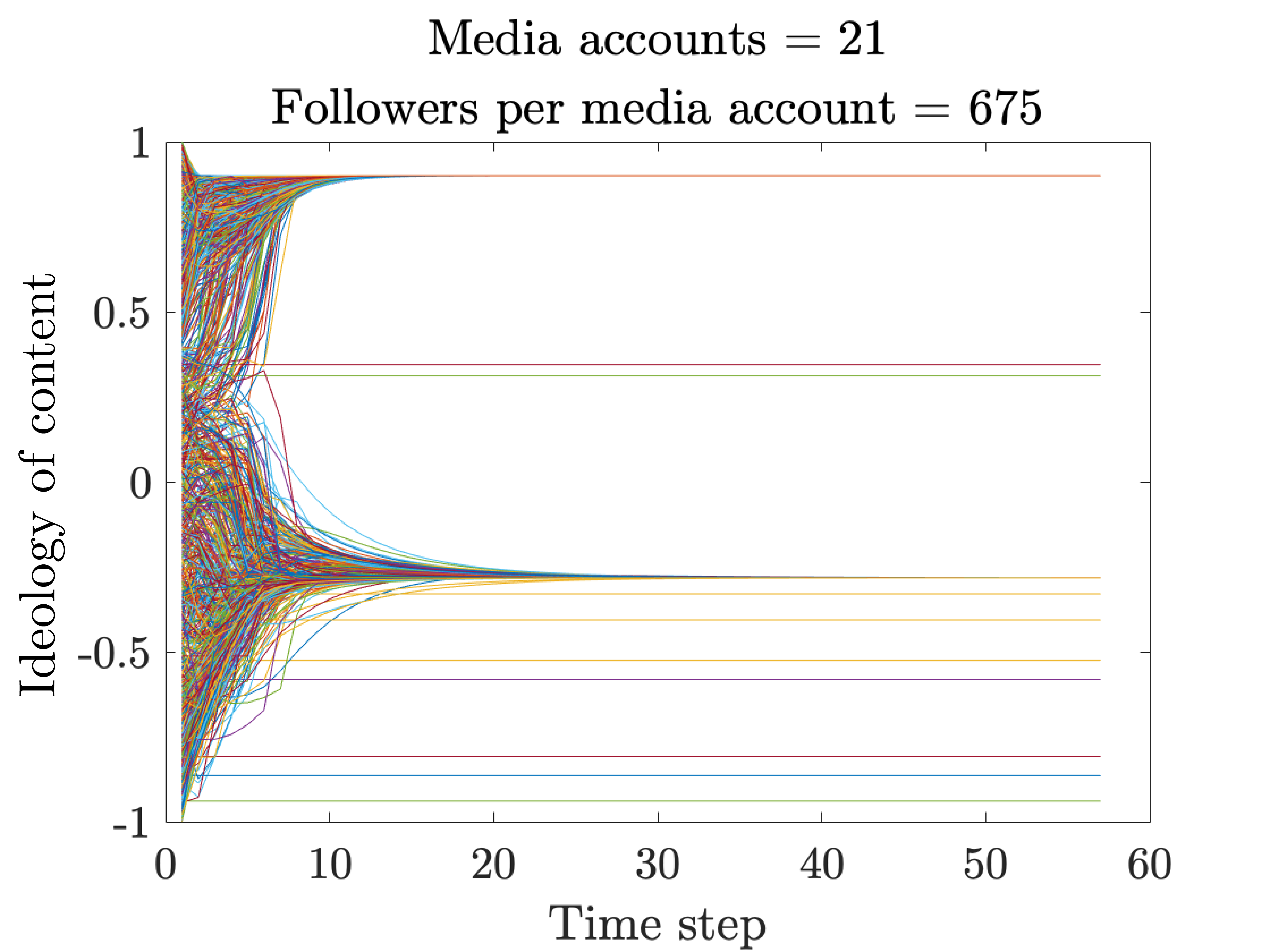}
	\caption{}
	\label{fig:ReedBigM}
	\end{subfigure}
\caption{Individual trials of the Reed College Facebook network (with $N=962$ non-media nodes) with different levels of media influence. We influence the convergence time and content-spreading dynamics by varying the media-account parameters $M$ and $n_M$. (a) With low media influence, most
accounts converge to the location $x=0$ in ideological space. (b) For larger values of $M$ and $n_M$, a large fraction of the non-media accounts converge to the ideological position of the media nodes. (c) For still larger values of these parameters, splitting in ideology occurs very rapidly, resulting in a smaller 
fraction of non-media accounts that converge to the media's ideology. In the simulations of all three panels, we evenly distribute the initial ideologies of non-media accounts, and we set the ideological position of all media accounts to $x_M=0.9$.}
\label{fig:Reedexamples}
\end{figure}

In Fig.~\ref{fig:Reedexamples}, we show examples of individual trials with different numbers of media accounts and numbers of followers per media account. We run our model on the Reed College Facebook network from the {\sc Facebook100} data set \cite{traud2012social}. These networks are Facebook friendship networks on university campuses from one day in fall 2005. In our simulations, we use only the largest connected component of each university Facebook network to represent non-media accounts, and we note that the friendship connections between non-media accounts in these networks are bidirectional. In Fig.~\ref{fig:Reedexamples}, we show the ideology of each of the $N = 962$ non-media accounts on the vertical axis and the simulation time $t$ on the horizontal axis. We take the media opinion to be $x_M=0.9$. These trials suggest that media nodes influence both convergence time and content-spreading dynamics. We explore these ideas further in Sec.~\ref{sec:1dimpact}.


\subsection{Media influence with a one-dimensional opinion space}
\label{sec:1dimpact}

Our simulations of our content-spreading model suggest that the number of non-media accounts that converge to the ideological position of the media nodes --- as we just illustrated for an example (see Fig.~\ref{fig:Reedexamples}) in which all media nodes have the same ideology --- depends nontrivially both on the number of media accounts and on the number of followers per media account. In this subsection, we present simulations that illuminate the level of media influence on a variety of real and synthetic networks. We continue to assume that all media nodes have the same ideological position.

We introduce an order parameter that we use as a diagnostic to measure the impact of media nodes on the ideological positions at convergence. Let $x^b_i$ denote the ideological position of account $i$ at convergence in the absence of media influence. We establish a mean baseline ideology $R_0$ by computing the mean distance between the ideological positions ($x^b_i$) of non-media accounts at convergence and the media ideology $x_M$. In mathematical terms, this baseline is
\begin{equation}
\label{eqn:baselinestat}
	R_0 = \frac{1}{N} \sum_{i=1}^N \|x_i^b-x_M\|_2 \,.
\end{equation}
The function $R_0$ characterizes the mean effect on non-media nodes, for one trial, of the dynamics for a given network structure in the absence of media influence. Because the outcome depends on the initial ideological positions of the non-media accounts, we average $R_0$ over many trials.

Once we have calculated the baseline ideology $R_0$, we construct a similar order parameter $R_M$ that characterizes the mean outcome of the dynamics for the same network when we introduce media nodes. Let $x^*_i$ denote the ideological position of account $i$ at convergence in a network with media influence. The associated
media-influenced ideology diagnostic is
\begin{equation}
\label{eqn:influencestat}
	R_{M} = \frac{1}{N} \sum_{i=1}^N \|x_i^*-x_M\|_2 \,.
\end{equation}

The order parameters $R_0$ and $R_M$ allow us to quantify the impact of media nodes on 
content-spreading dynamics, where we note that one can also define time-dependent analogs of \eqref{eqn:baselinestat} and \eqref{eqn:influencestat}. We define the media impact for one trial to be the ratio of the mean baseline ideological distance to the media-influenced ideological distance in that trial:
\begin{equation}
	R = \frac{\overline{R_0}}{R_{M}}\,.
\end{equation}
We can also obtain an overall media impact 
\begin{equation}
\label{eqn:impactstat}
	\overline{R} = \frac{\overline{R_0}}{\overline{R_{M}}}
\end{equation}
by averaging the media-influenced opinion function $R_{M}$ over some number (e.g., $200$) of trials. We interpret the media impact in the following way. If $\overline{R}=1$, the media has not had an influence on the ``average'' (specifically, the mean) ideological position of the non-media accounts (and the content that they spread) at convergence. If $\overline{R}\in[0,1)$, the media nodes have driven the mean ideological position of the accounts to be farther away from the media's ideological position than is the case without the media accounts. Finally, if $\overline{R}>1$, the mean ideological position of the non-media accounts (and hence of the content that they spread) is closer to the media ideology than it would be without the introduction of the media accounts, with larger values of $\overline{R}$ indicating a stronger impact.

\begin{figure}[h!]
\centering
\includegraphics[width=0.5\textwidth]{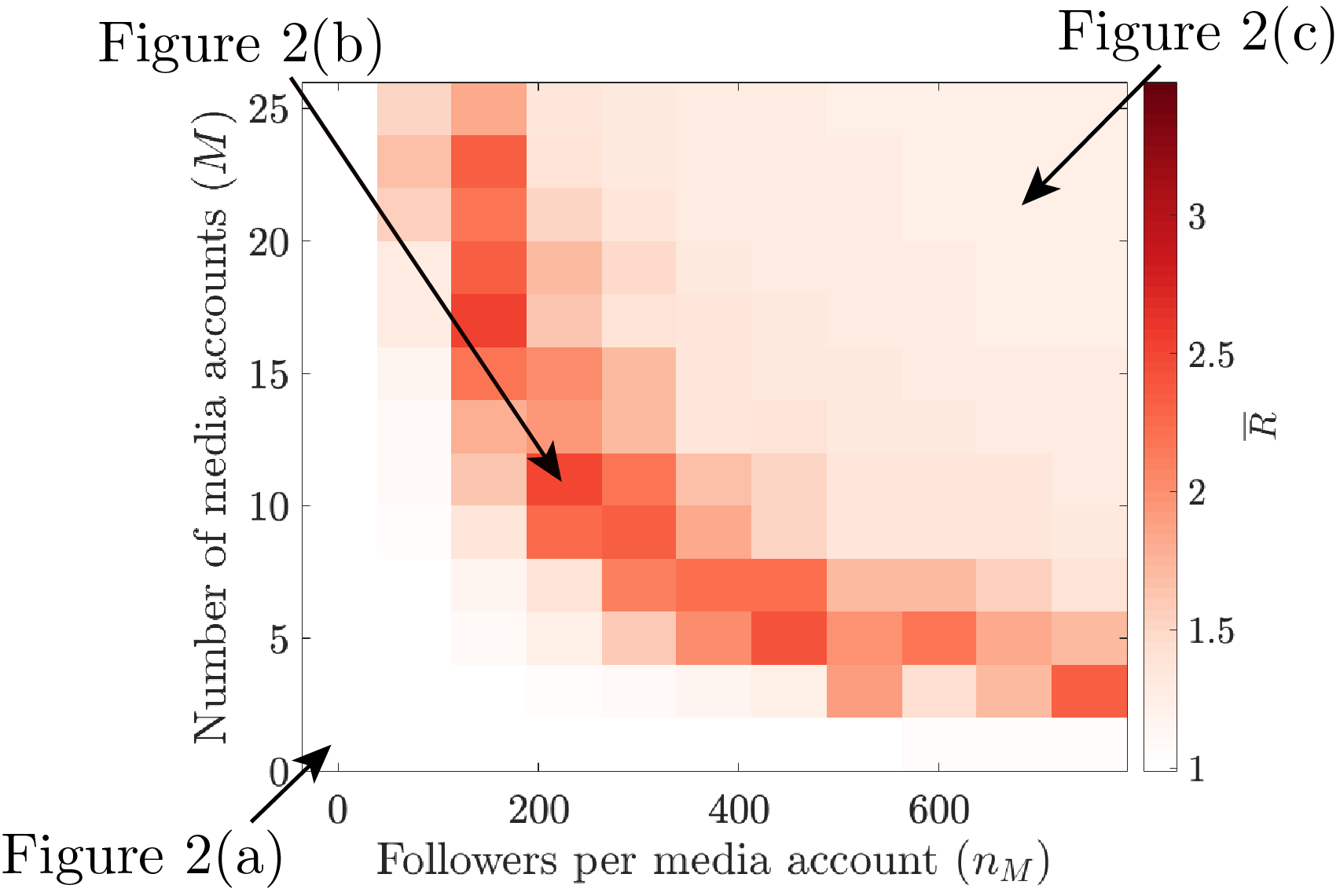}
\caption{Media impact for the Reed College network ($N=962$) with different numbers of media accounts ($M$) on the vertical axis and followers per media account ($n_M$) on the horizontal axis. All media accounts have the same ideological position of $x_M=0.9$. The colors indicate the values of $\overline{R}$, the mean of the impact summary diagnostic [see Eq.~\eqref{eqn:impactstat}] for distance from media ideology over $50$ trials. 
 Dark red indicates the most media impact (i.e., the largest values of $\overline{R}$), and white indicates the least.
The media nodes' ideological position $x_M$ and the initial ideological positions of the non-media accounts are the same as in Fig.~\ref{fig:Reedexamples}. The arrows designate the regions of the plot that correspond to the mean media impact for each of the three panels in Fig.~\ref{fig:Reedexamples}.
}
\label{fig:ReedArrows}
\end{figure}

Equation \eqref{eqn:impactstat} gives one of multiple possible summary diagnostics to measure media influence. Another option is to measure the mean of the distance without rescaling. (That is, one can use only the expected value of $R$ from Eq.~\eqref{eqn:influencestat}.) This alternative entails an impact level of $\overline{R}_M\in[0,2]$, where small values of $\overline{R}_M$ represent considerable
``entrainment" to the media ideology. That is, the media nodes have successfully influenced (e.g., ``brainwashed'', to describe it in an uncharitable way) many non-media accounts in the network. Large values of $\overline{R}_M$ represent low levels of media influence. Using this alternative impact function for the experiments in this section gives the same qualitative results to what we report using \eqref{eqn:impactstat}.

In Fig.~\ref{fig:ReedArrows}, we illustrate media impact values for different numbers of media accounts ($M$) and numbers of media followers ($n_M$). In color, we show the mean value of the summary diagnostic $\overline{R}$ over $50$ trials. These simulations suggest a surprising result: the highest levels of adoption of media ideology do not occur for the largest values in $(n_M,M)$ parameter space. Instead, the most successful scenario for promoting widespread adoption of the media ideology is to spread the content through a moderate number of media accounts, each of which has a moderate number of followers. This observation is consistent with 
previous empirical observations that accounts with a small number of followers can significantly promote the spreading of content on Twitter \cite{sune2013}. For a very small number of media nodes (or if the media nodes have very few followers), as in Fig.~\ref{fig:ReedSmallM}, the media ideology has very low impact. A large number of media nodes with many followers per account (as in Fig.~\ref{fig:ReedBigM}) produces some impact, yielding values of $\overline{R}$ in the interval $(1,2)$. However, we observe a larger impact when there are a moderate number of media accounts that each have a moderate number of followers (see Fig.~\ref{fig:ReedMedM}). In this situation, we often obtain $\overline{R}\geq 2$.

\begin{figure*}
	\begin{subfigure}[t]{0.24\textwidth}
		\includegraphics[height=1.3in]{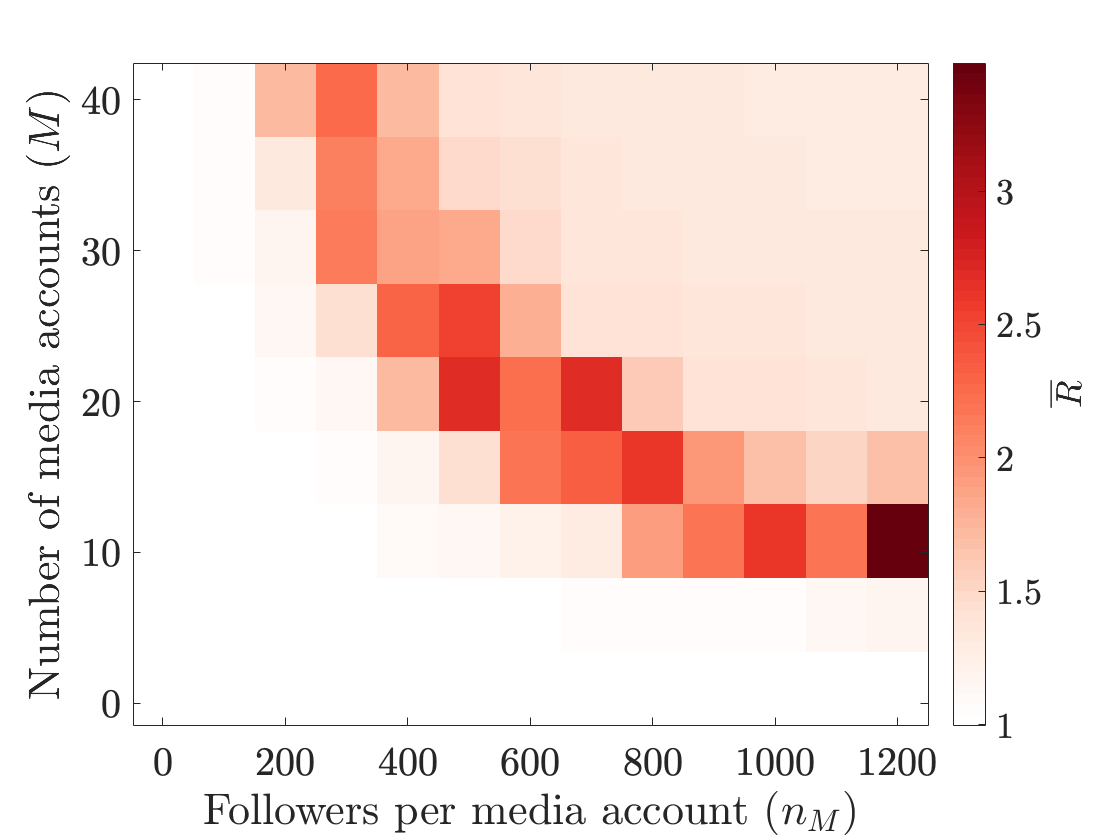}
	\caption{Amherst}
	\end{subfigure}
	\begin{subfigure}[t]{0.24\textwidth}
		\includegraphics[height=1.3in]{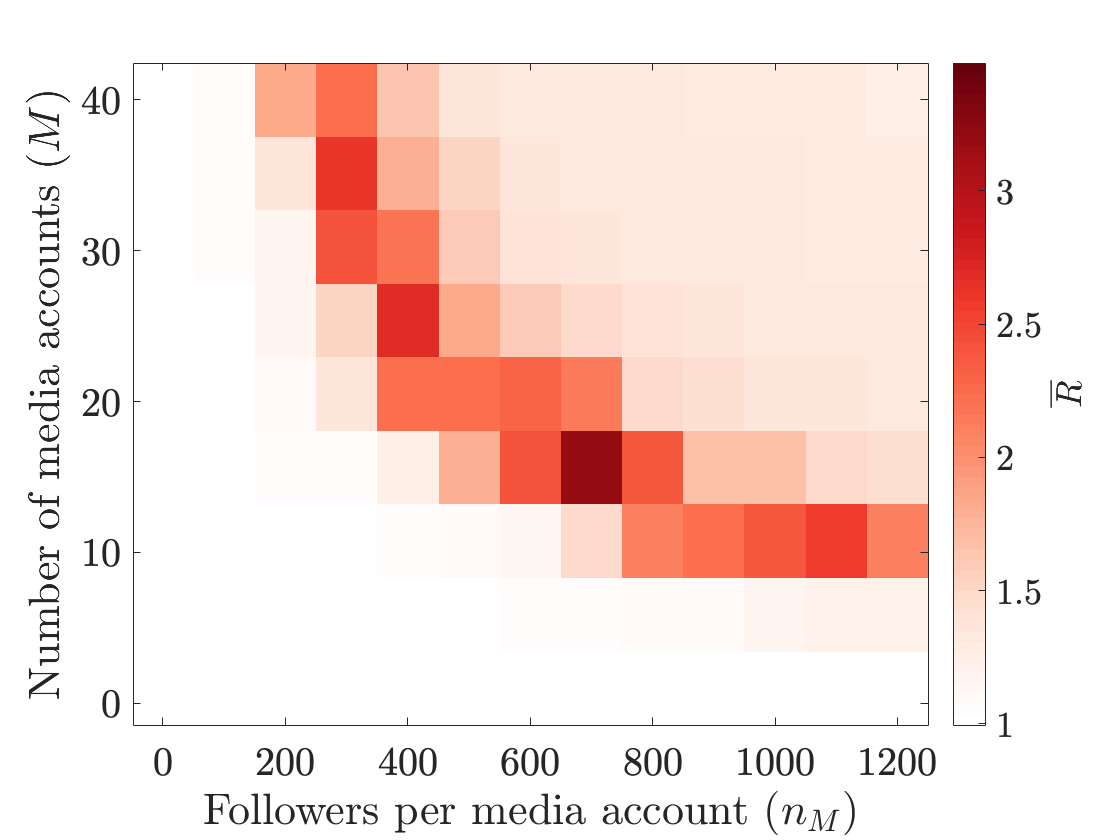}
	\caption{Bowdoin}
	\end{subfigure}
	\begin{subfigure}[t]{0.24\textwidth}
		\includegraphics[height=1.3in]{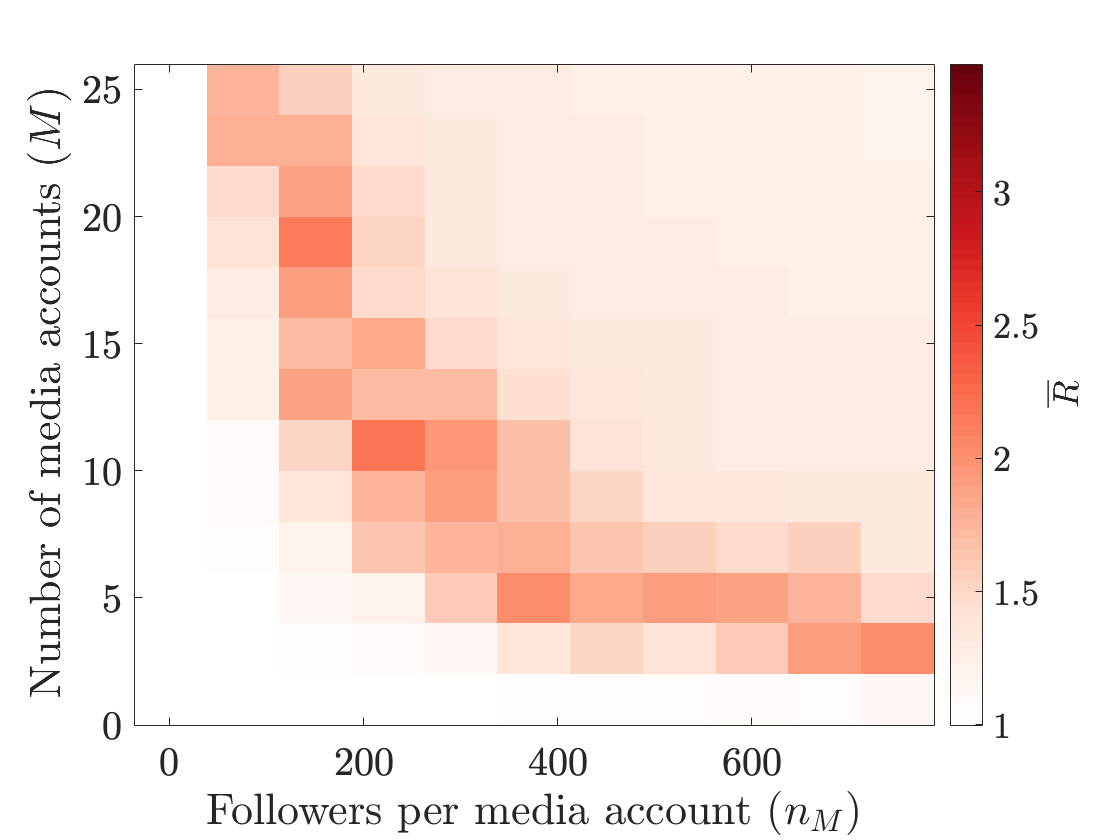}
	\caption{Caltech}
	\end{subfigure}
	\begin{subfigure}[t]{0.24\textwidth}
		\includegraphics[height=1.3in]{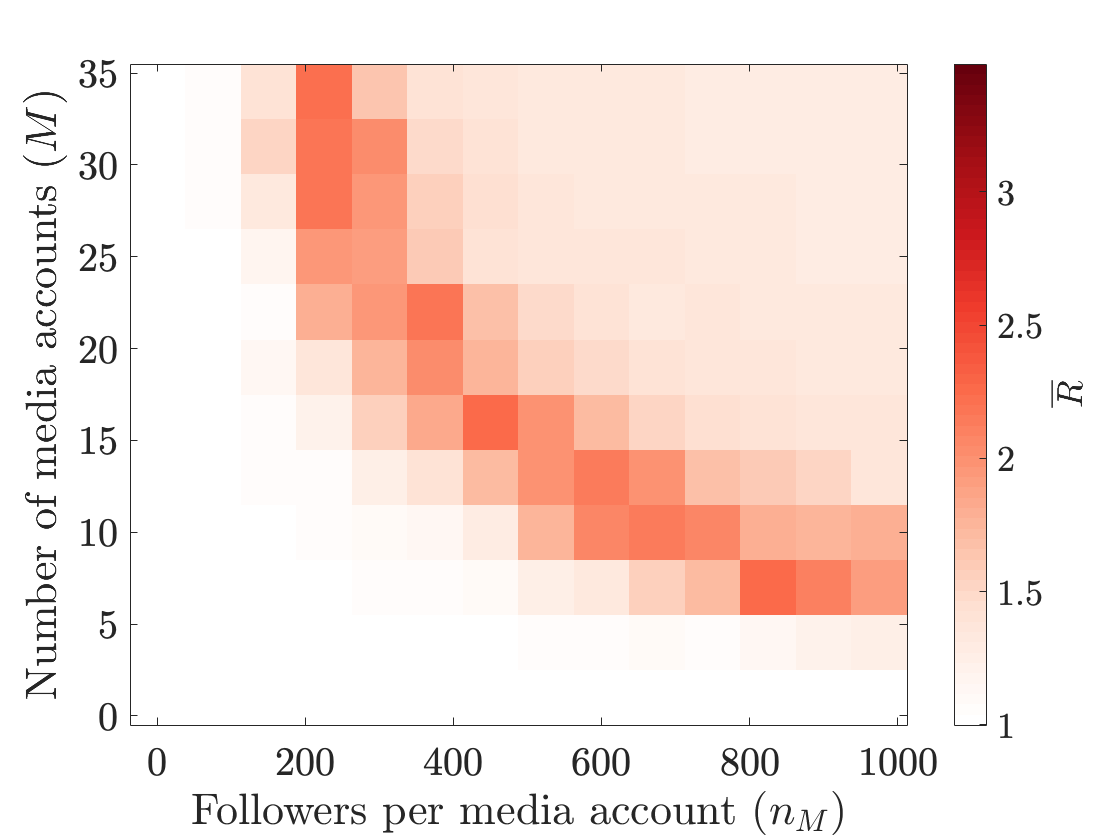}
	\caption{Haverford}
	\end{subfigure}
	\begin{subfigure}[t]{0.24\textwidth}
		\includegraphics[height=1.3in]{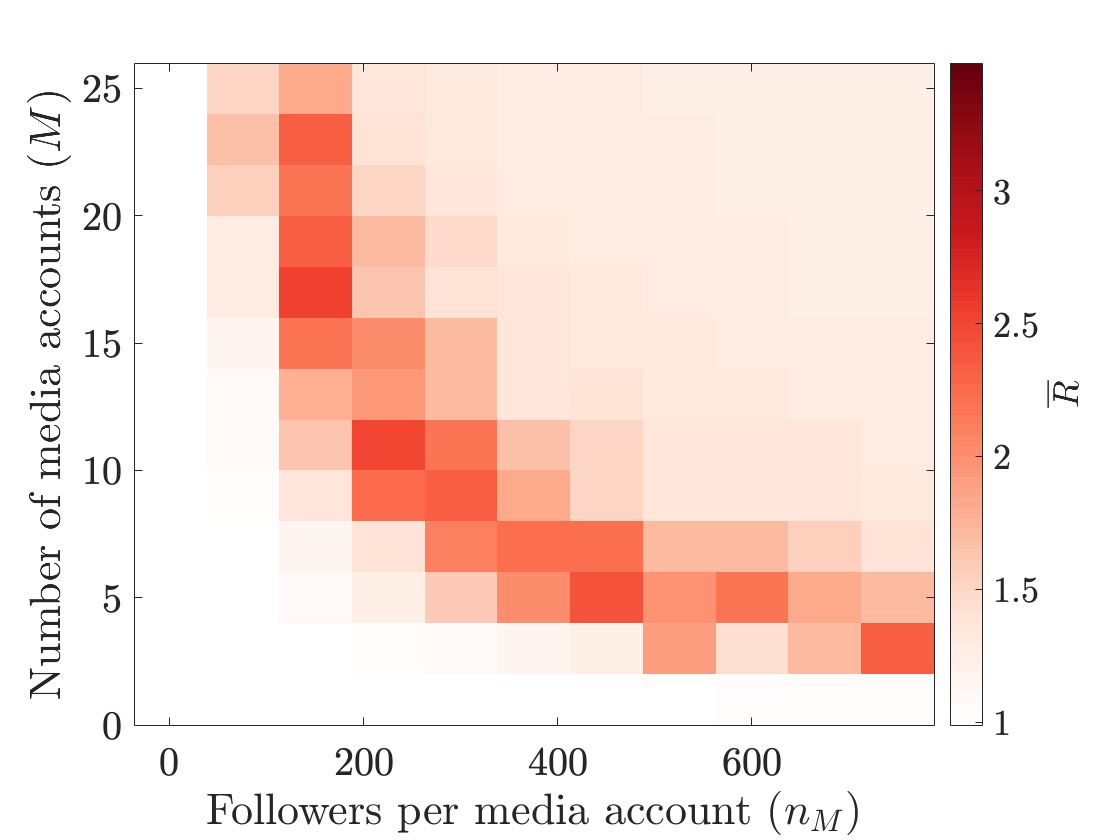}
	\caption{Reed}
	\end{subfigure}
	\begin{subfigure}[t]{0.24\textwidth}
		\includegraphics[height=1.3in]{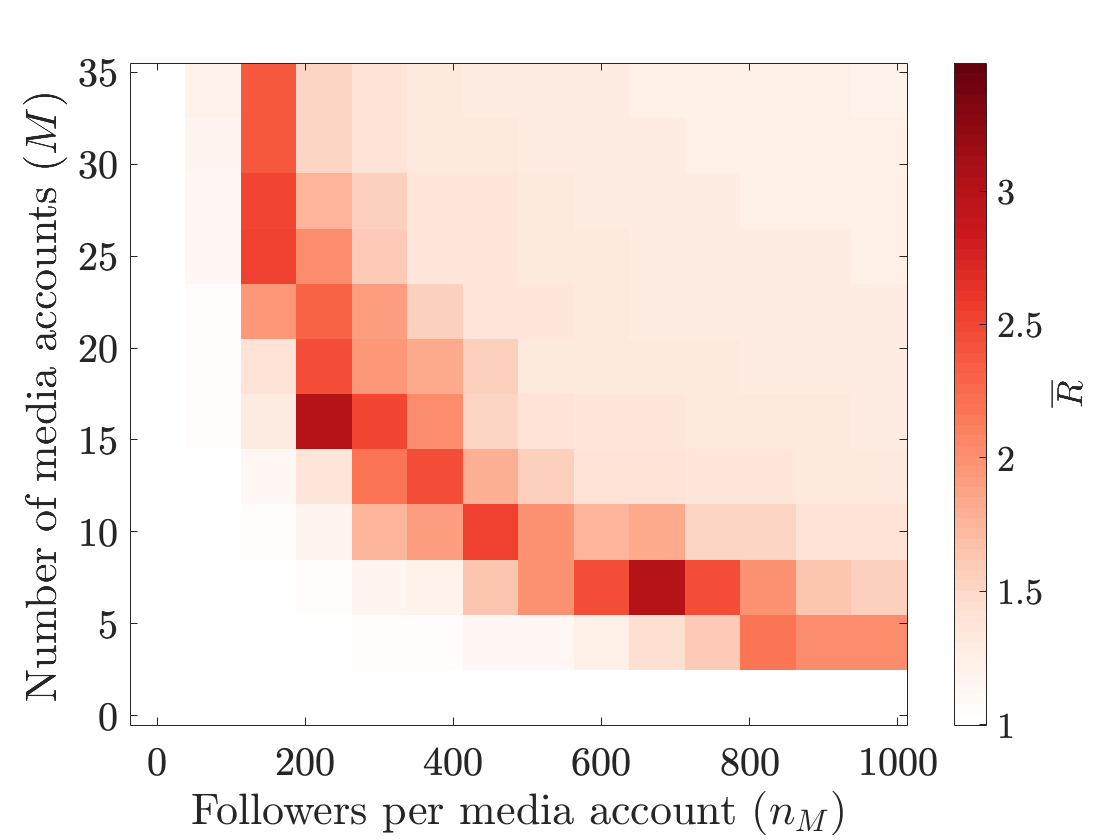}
	\caption{Simmons}
	\end{subfigure}
	\begin{subfigure}[t]{0.22\textwidth}
		\includegraphics[height=1.3in]{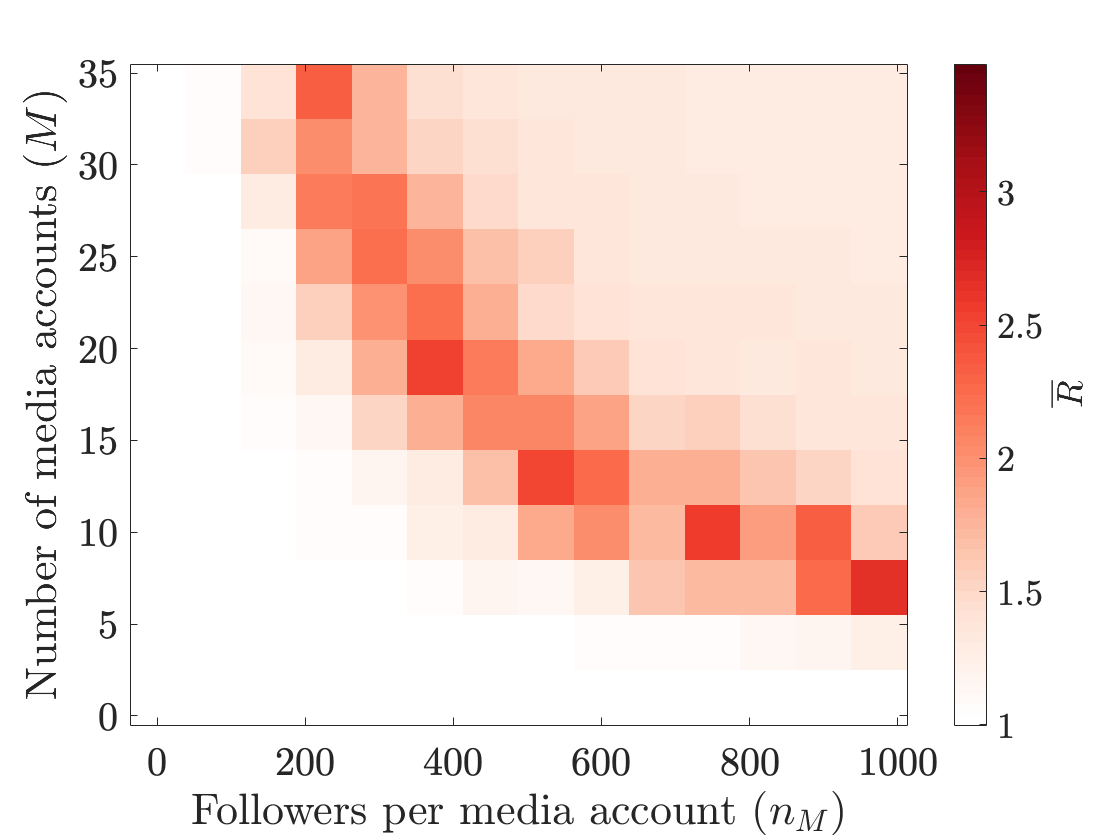}
	\caption{Swarthmore}
	\end{subfigure}
\caption{Heat maps of the summary diagnostic $\overline{R}$ for media impact in the Facebook networks of various universities from the {\sc Facebook100} data set \cite{traud2012social} illustrate that the largest
media impact occurs for moderate numbers of media accounts with moderate numbers of followers.
In each panel, the vertical axis gives the number of media accounts ($M$), the horizontal axis gives the number of followers per media account ($n_M$), and the colors indicate the mean impact summary diagnostic ($\overline{R}$) over $50$ trials.
 Dark red indicates the most media impact (i.e., the largest values of $\overline{R}$), and white indicates the least.
 }
\label{fig:FB100entrain}
\end{figure*}

\begin{figure*}
	\begin{subfigure}[t]{0.3\textwidth}
		\includegraphics[height=1.5in]{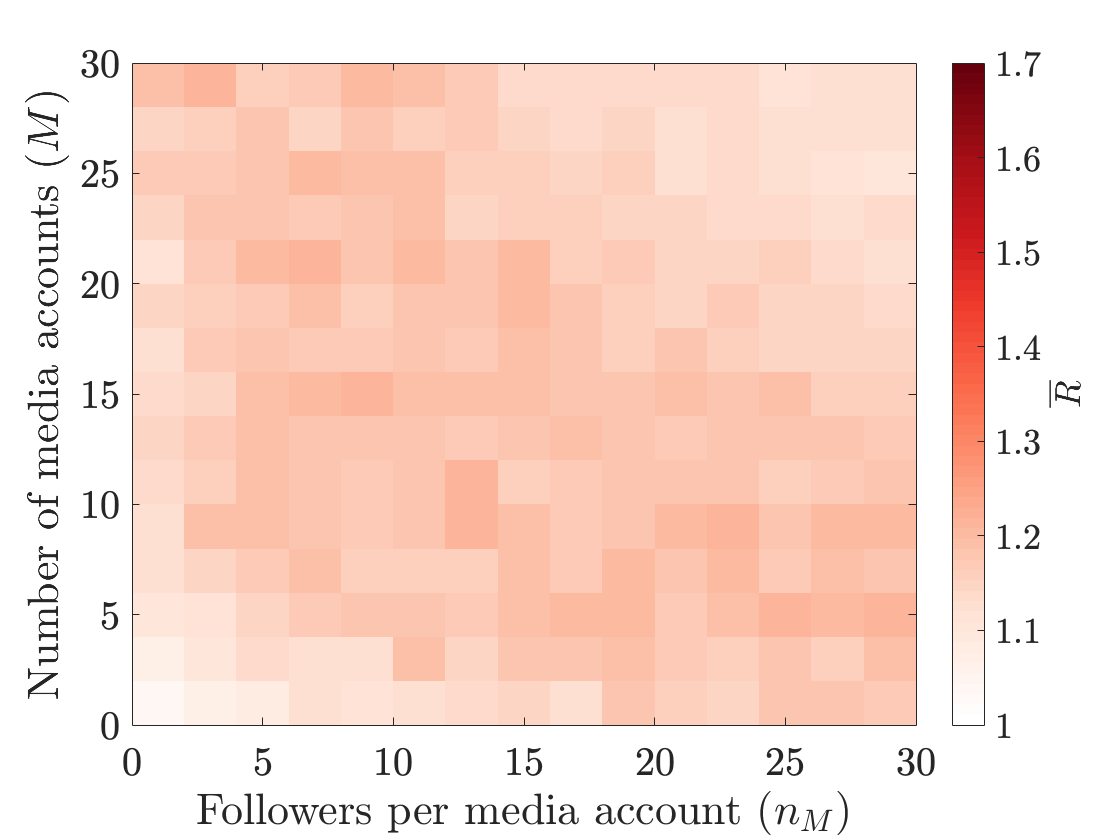}
	\caption{Star network.}
	\end{subfigure}
	\begin{subfigure}[t]{0.3\textwidth}
		\includegraphics[height=1.5in]{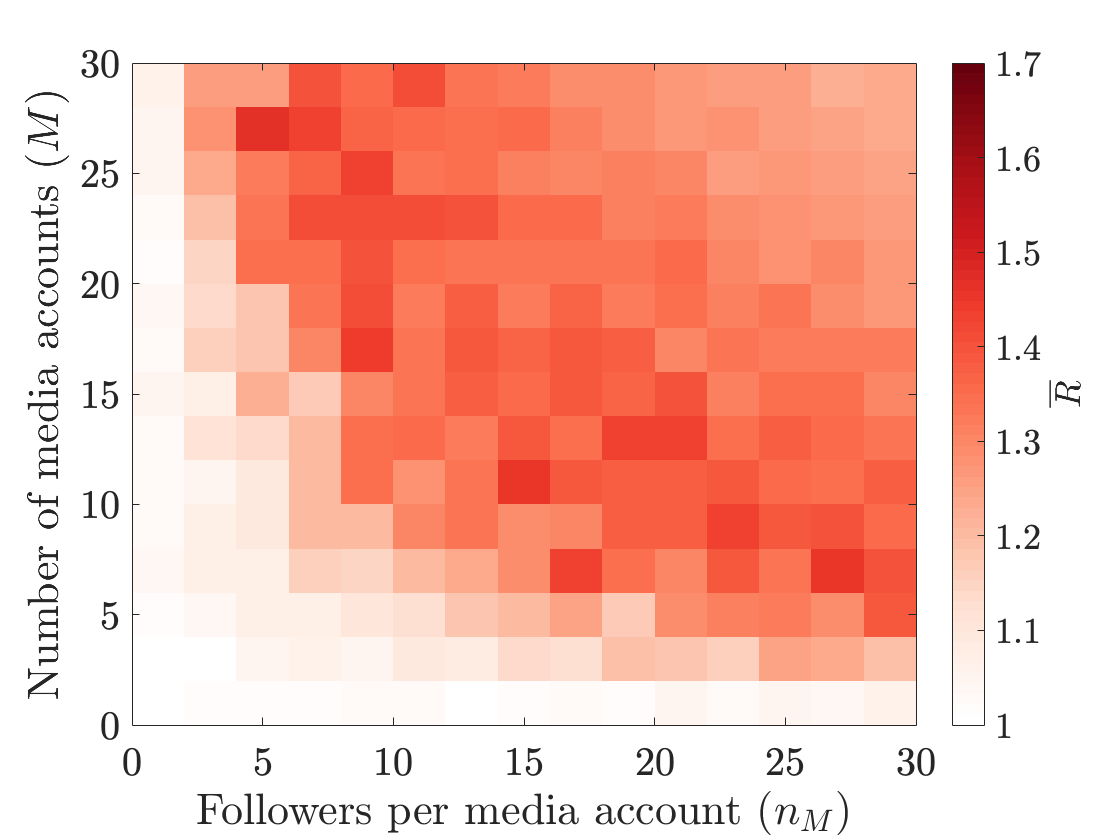}
	\caption{Directed ring lattice 
	with $k=25$.}
	\end{subfigure}
	\begin{subfigure}[t]{0.3\textwidth}
		\includegraphics[height=1.5in]{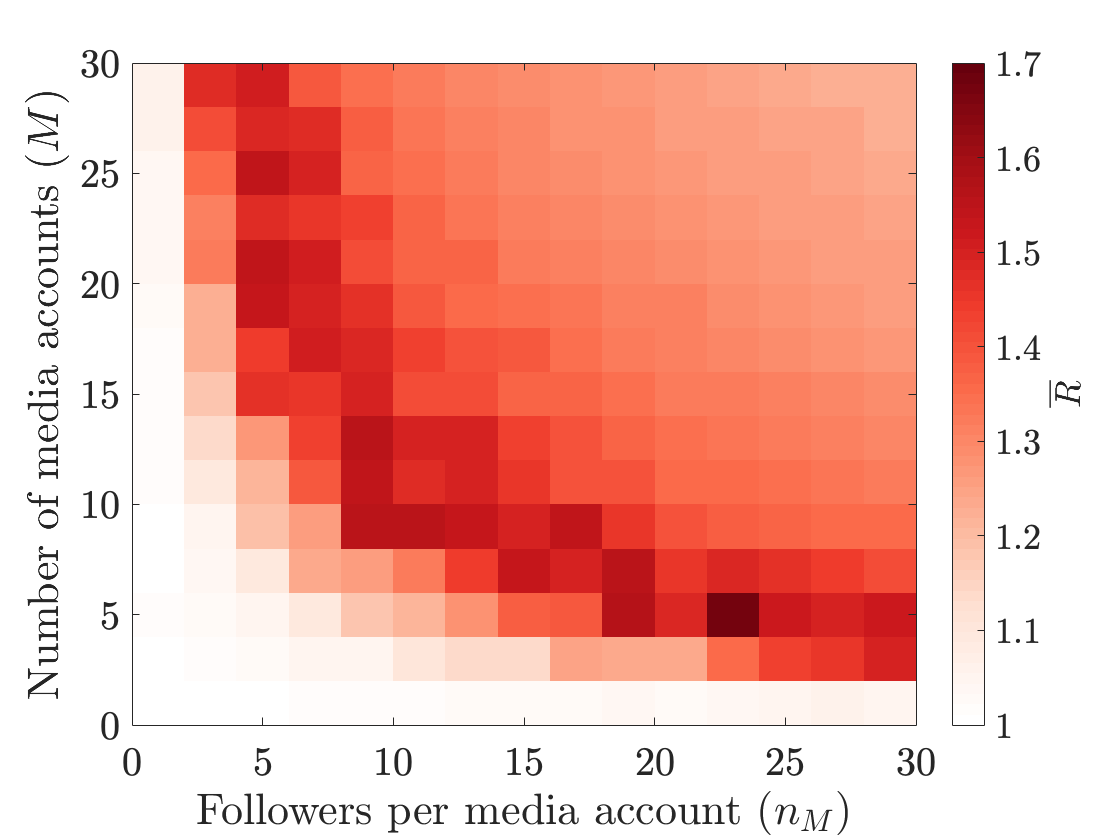}
	\caption{Watts--Strogatz network with $k=25$ and $\beta=0.5$.}
	\end{subfigure}
	\begin{subfigure}[t]{0.3\textwidth}
		\includegraphics[height=1.5in]{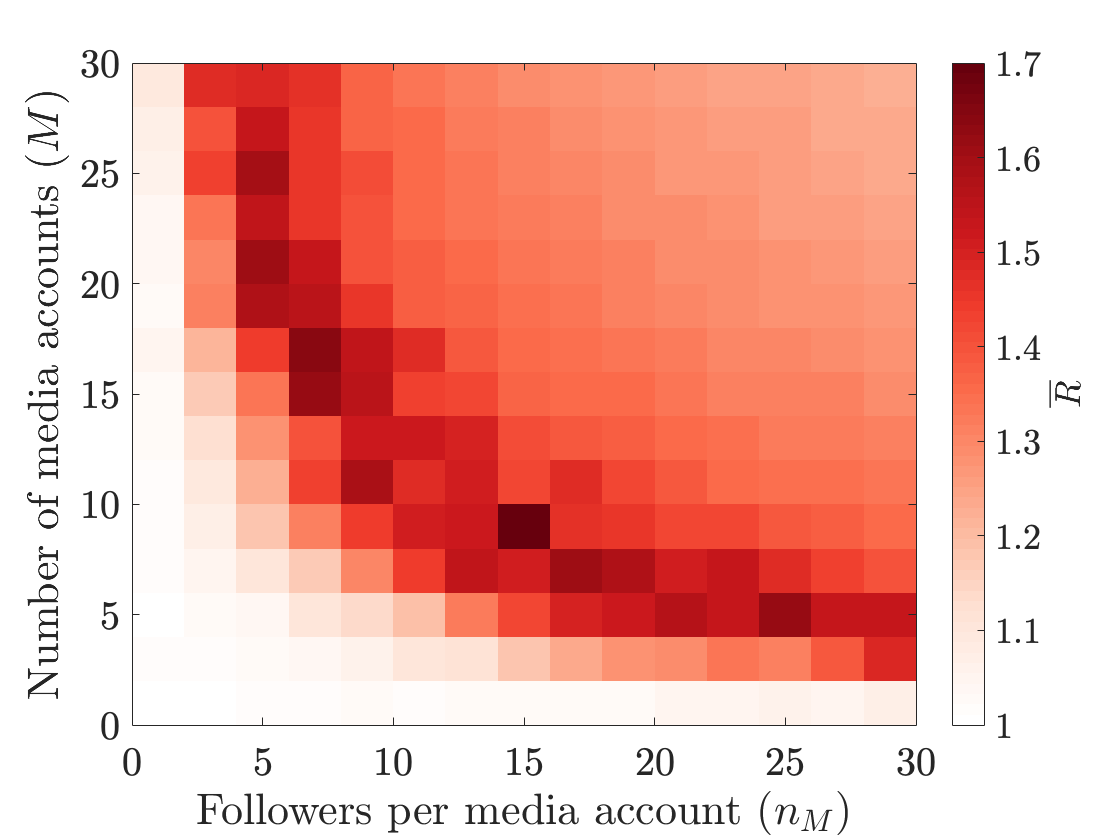}
	\caption{Directed Erd\H{o}s--R\'{e}nyi network with $k=25$.}
	\end{subfigure}
	\begin{subfigure}[t]{0.3\textwidth}
		\includegraphics[height=1.5in]{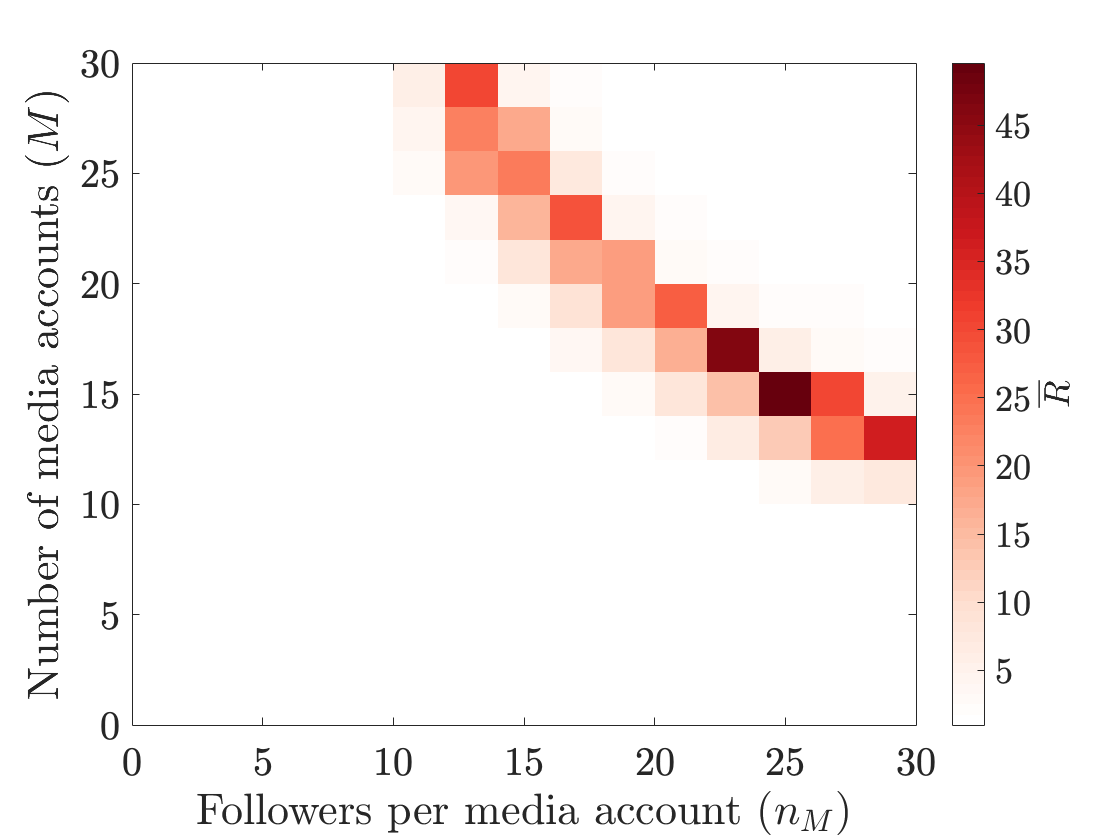}
	\caption{Complete network.}
	\end{subfigure}
\caption{Heat maps of the summary diagnostic $\overline{R}$ for media impact of various synthetic networks illustrate that the largest media impact occurs for moderate numbers of media accounts with moderate numbers of followers for a variety of network architectures. In the directed ring lattice, each non-media account follows $k=25$ neighboring non-media accounts. We construct a directed variant of a Watts--Strogatz network from the directed ring lattice with $k=25$. In it, we rewire follower connections with a probability of $\beta=0.5$. In the directed variant of an Erd\H{o}s--R\'{e}nyi network, we take the expected mean out-degree (i.e., the mean number of non-media accounts that are followed) to be $k=25$. In the star network, all non-media accounts follow one central non-media account, which also follows the peripheral non-media accounts. In each panel, the vertical axis gives the number of media accounts ($M$), the horizontal axis gives the number of followers per media account ($n_M$), and the colors indicate the media impact diagnostic ($\overline{R}$), which we compute as a mean over $200$ trials. 
 Dark red indicates the most media impact (i.e., the largest values of $\overline{R}$), and white indicates the least. 
For the directed ER and WS networks, we construct a new network for each of the $200$ trials.
}
\label{fig:Constructedentrain}
\end{figure*}

In Fig.~\ref{fig:FB100entrain}, we show heat maps of the media entrainment values (using the summary diagnostic $\overline{R}$) for our model for social networks from the {\sc Facebook100} data set \cite{traud2012social}. These networks consists of people at a particular university who are connected to each other via Facebook friendships (which yield bidirectional edges) from one day in fall 2005. We add media accounts to each network in the way that we described previously for the Reed network in Sec.~\ref{sec:1dsim}. For each of the {\sc Facebook100} networks that we examine, our simulations produce a distinct region 
with large media impact (specifically, with $\overline{R}\geq 2$). In each case, the largest amount of media impact does not occur for the largest values of the two quantities in $(n_M,M)$ parameter space.

In Fig.~\ref{fig:Constructedentrain}, we show heat maps of the media entrainment values (using the summary diagnostic $\overline{R}$) for our model 
for a variety of synthetic network architectures, each of which has $100$ non-media accounts. For clarity, we first briefly describe how we construct each of these networks. These networks are a directed ring lattice, a star, a directed variant of Erd\H{o}s--R\'{e}nyi (ER) networks, a directed variant of Watts--Strogatz (WS) networks, and complete networks. In the star network, all but one of the non-media accounts follow one central non-media account. These edges are bidirectional, so the central non-media account 
follows all of the other non-media accounts. We construct a directed ring lattice as follows. Each non-media account follows $k$ other non-media accounts, where the $i$th node follows nodes $\{i+1, i+2, \dots, i+k\} \ (\textrm{mod } N)$. In this case, the edges are not bidirectional (so, for example, account $1$ follows account $2,$ but account $2$ does not follow $1$). We construct a directed variant of a WS network by starting with this directed ring lattice and rewiring each edge with probability $\beta=0.5$ \cite{watts1998collective}. 
We rewire a directed edge from node $i$ to node $j$ by selecting a new node $j'$ uniformly at random from the nodes that account $i$ does not currently follow; we remove the edge from $i$ to $j$ and add an edge from $i$ to $j'$.
We construct directed ER networks, which have an expected mean out-degree of $k=25$, as a variant of the $\mathcal{G}(N,p)$ model. (In other words, the expected mean number of non-media accounts that a node follows is $k=25$.) For the directed ER and WS networks, we construct a new network for each trial, so one should interpret our simulation results as a sample mean over many networks. Although all networks have the same number of non-media accounts, the region of parameter space in which the media has the most impact differs across different networks. This suggests that network architecture plays an important role in determining the level of media involvement that is necessary for media accounts to exert ``global" influence in a network. 

\begin{figure*}
	\begin{subfigure}[t]{0.4\textwidth}
		\includegraphics[height=2in]{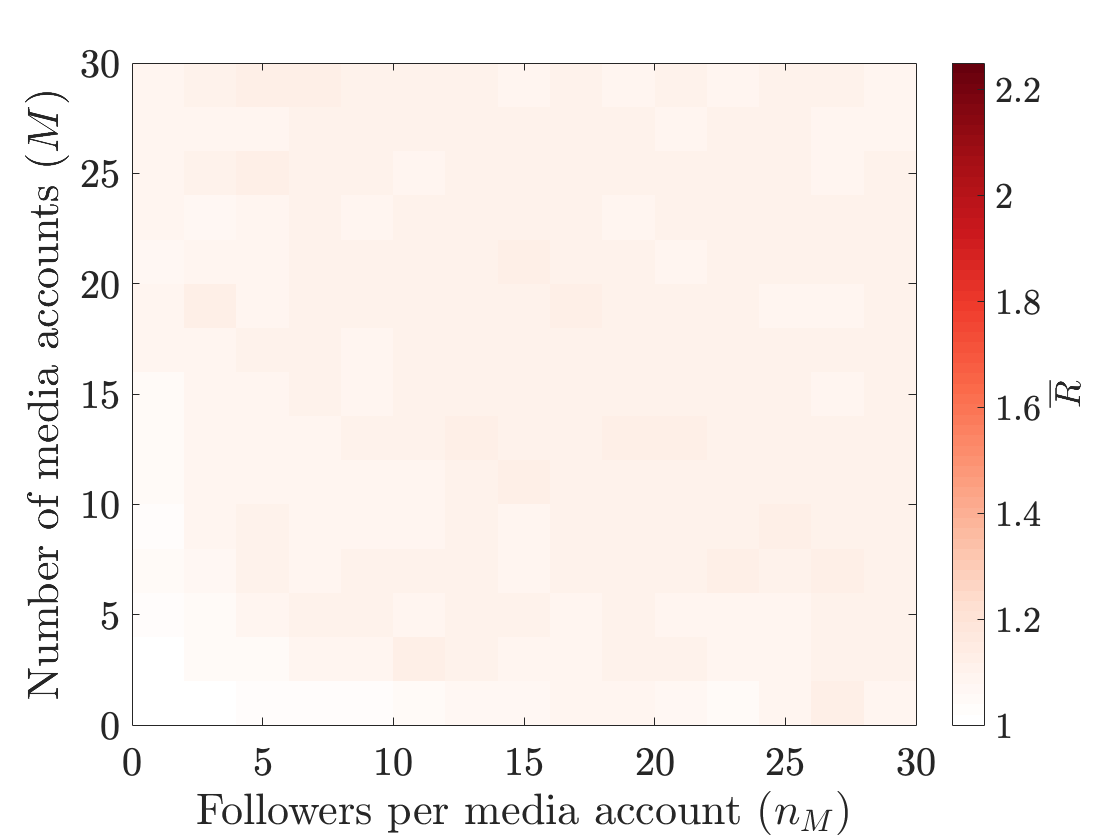}
	\caption{$k=2$}
	\end{subfigure}
	\begin{subfigure}[t]{0.4\textwidth}
		\includegraphics[height=2in]{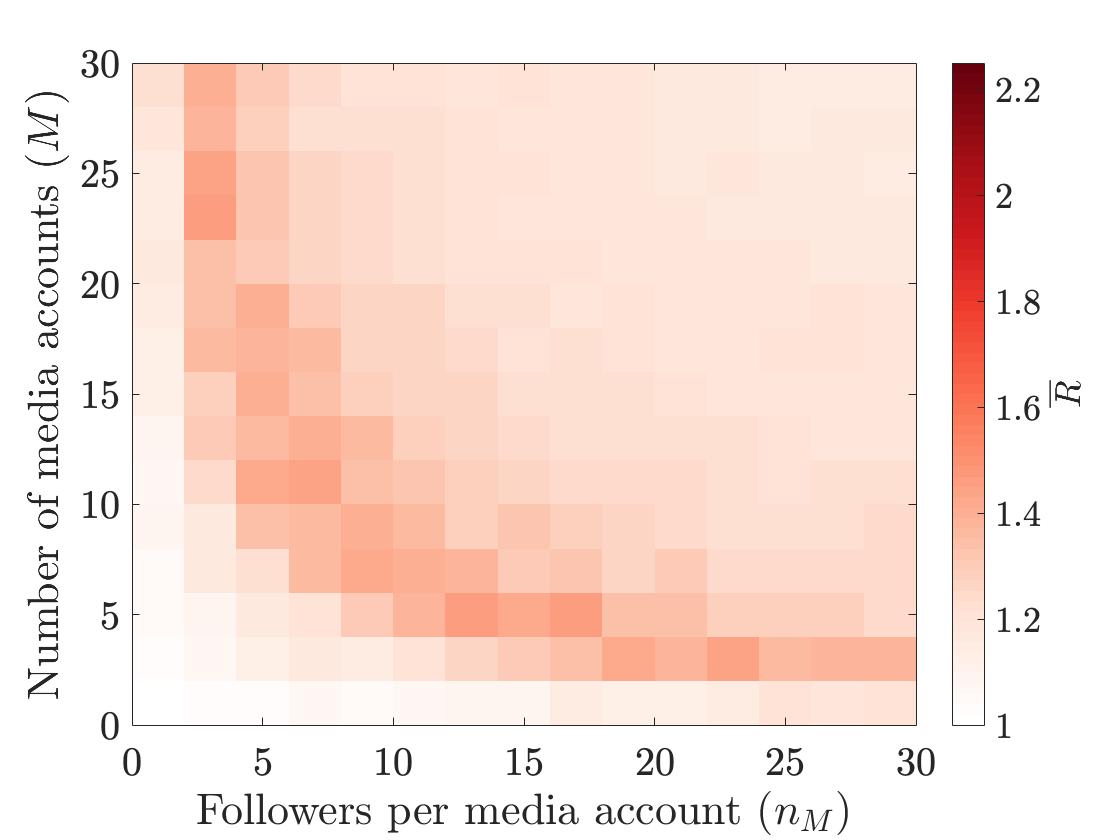}
	\caption{$k=10$}
	\end{subfigure}\\
	\begin{subfigure}[t]{0.4\textwidth}
		\includegraphics[height=2in]{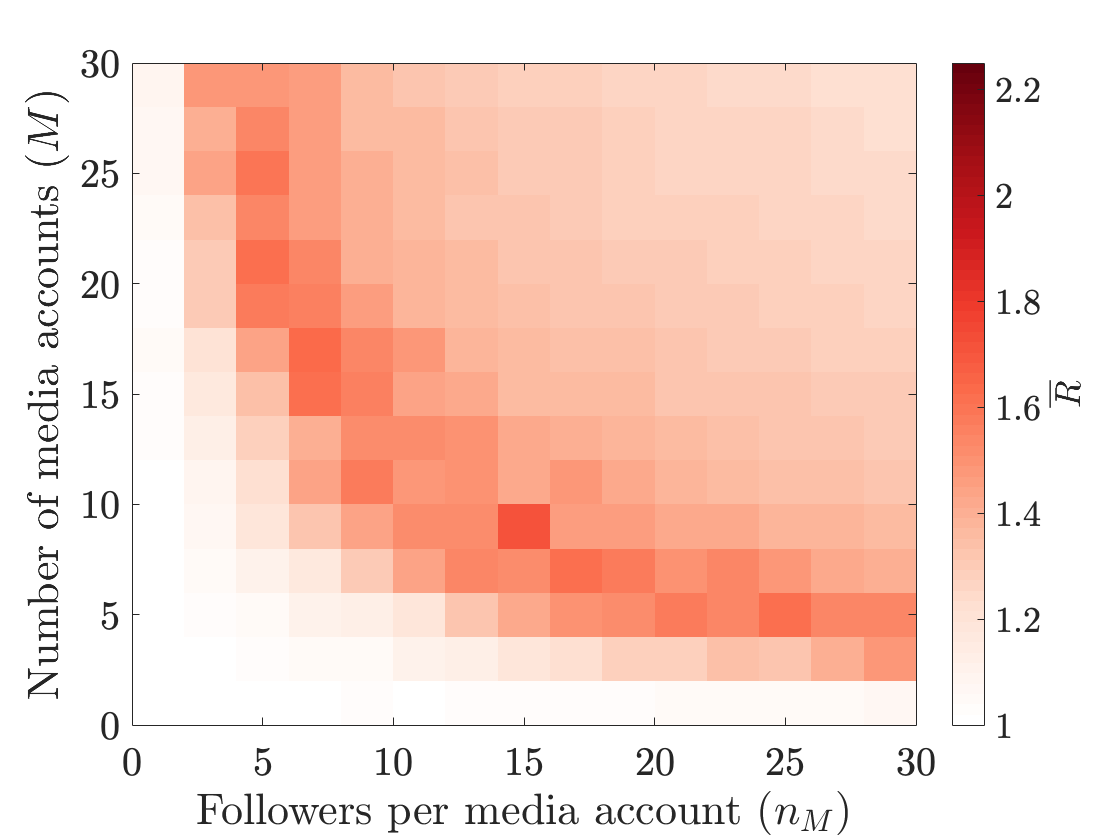}
	\caption{$k=25$}
	\end{subfigure}
	\begin{subfigure}[t]{0.4\textwidth}
		\includegraphics[height=2in]{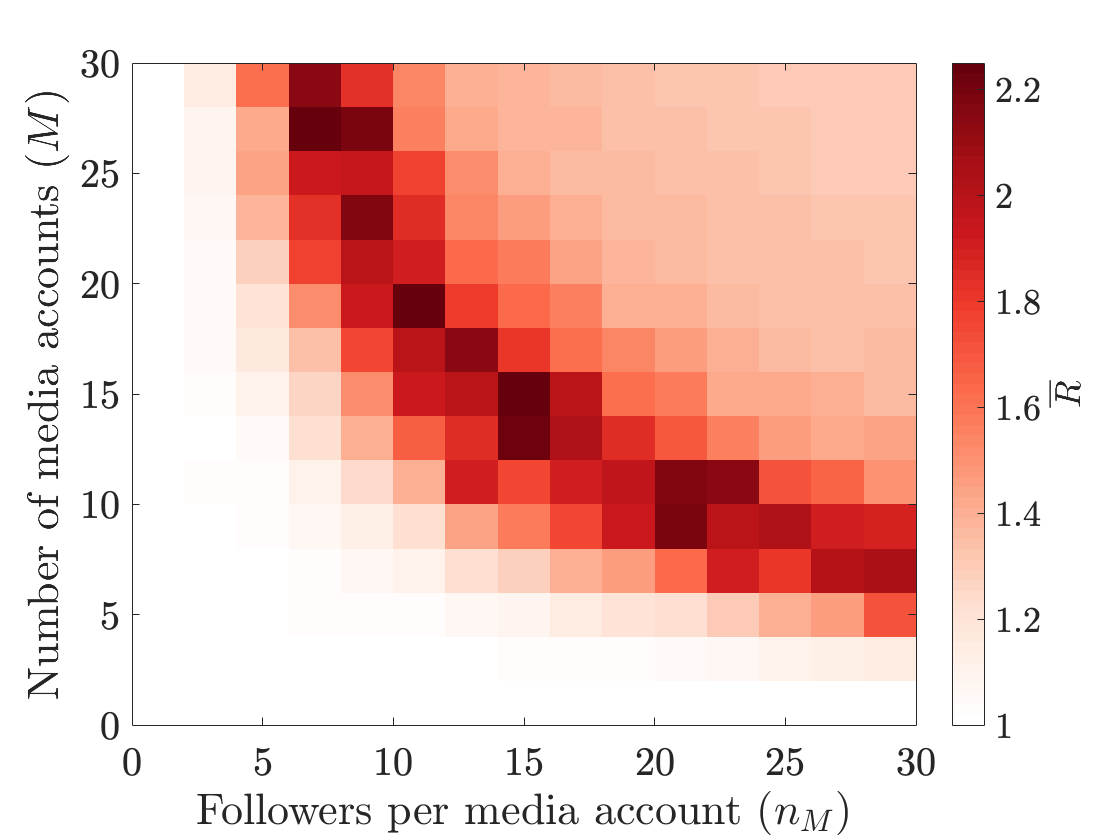}
	\caption{$k=50$}
	\end{subfigure}
\caption{Heat maps of the media impact diagnostic $\overline{R}$ for directed ER networks with different values of the expected mean number $k$ of non-media accounts that non-media accounts follow. Our simulations suggest that media impact is larger for progressively larger $k$. The vertical axis is the number of media accounts ($M$), the horizontal axis is the number of followers per media account ($n_M$), and the colors represent the media impact diagnostic ($\overline{R}$), which we average over $200$ trials. 
 Dark red indicates the most media impact (i.e., the largest values of $\overline{R}$), and white indicates the least.
}
\label{fig:ERvaryk}
\end{figure*}

We now examine the media impact as we vary $k$, the mean number of non-media accounts that a non-media account follows, on similar network architectures. In Fig.~\ref{fig:ERvaryk}, we show heat maps of the media impact summary diagnostic $\overline{R}$ for directed ER networks with different expected mean out-degrees. As we increase the expected mean out-degree $k$, we observe that media impact also increases.

\begin{figure*}
	\begin{subfigure}[t]{0.4\textwidth}
		\includegraphics[height=2in]{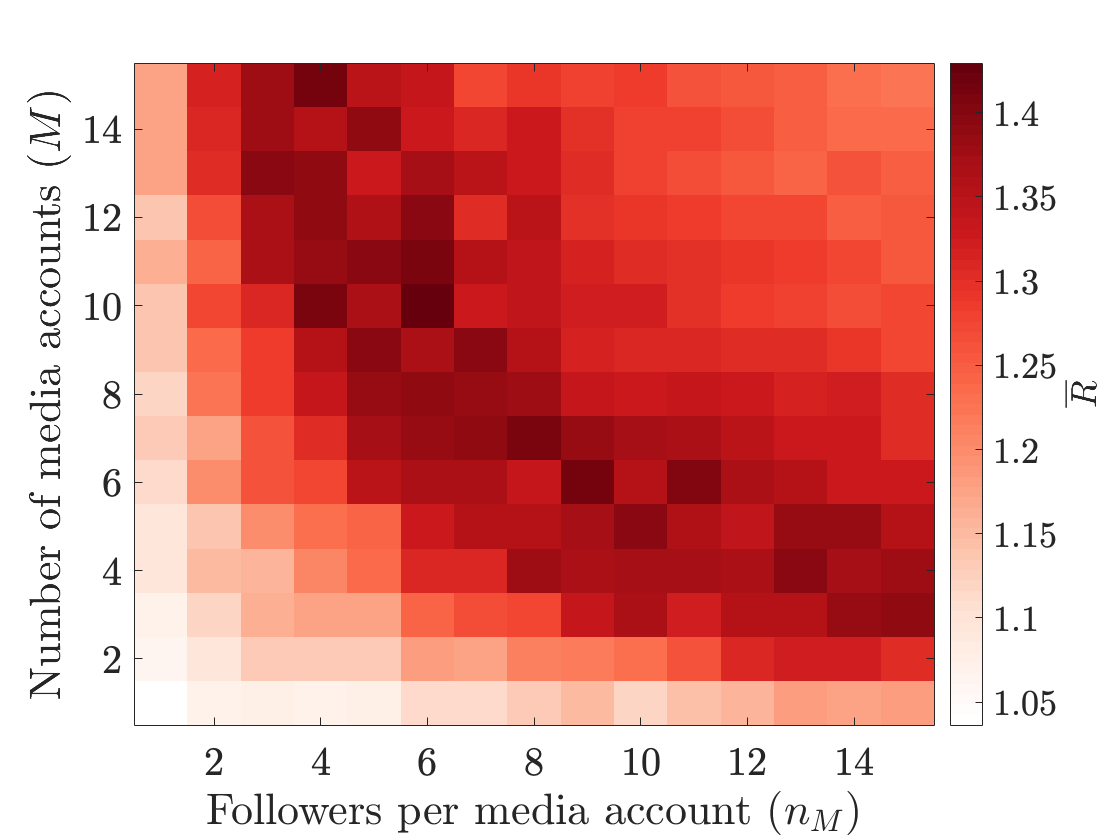}
	\caption{$N=50$ and $k=13$}
	\end{subfigure}
	\begin{subfigure}[t]{0.4\textwidth}
		\includegraphics[height=2in]{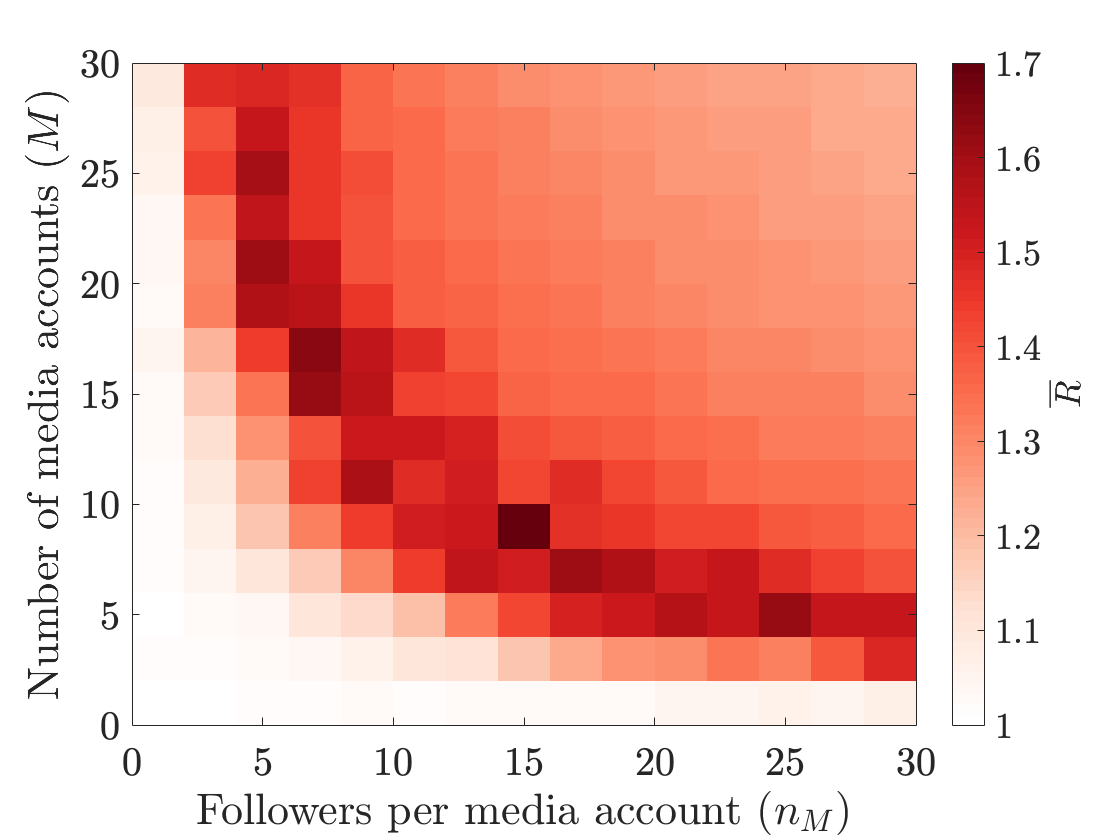}
	\caption{$N=100$ and $k=25$}
	\end{subfigure}\\
	\begin{subfigure}[t]{0.4\textwidth}
		\includegraphics[height=2in]{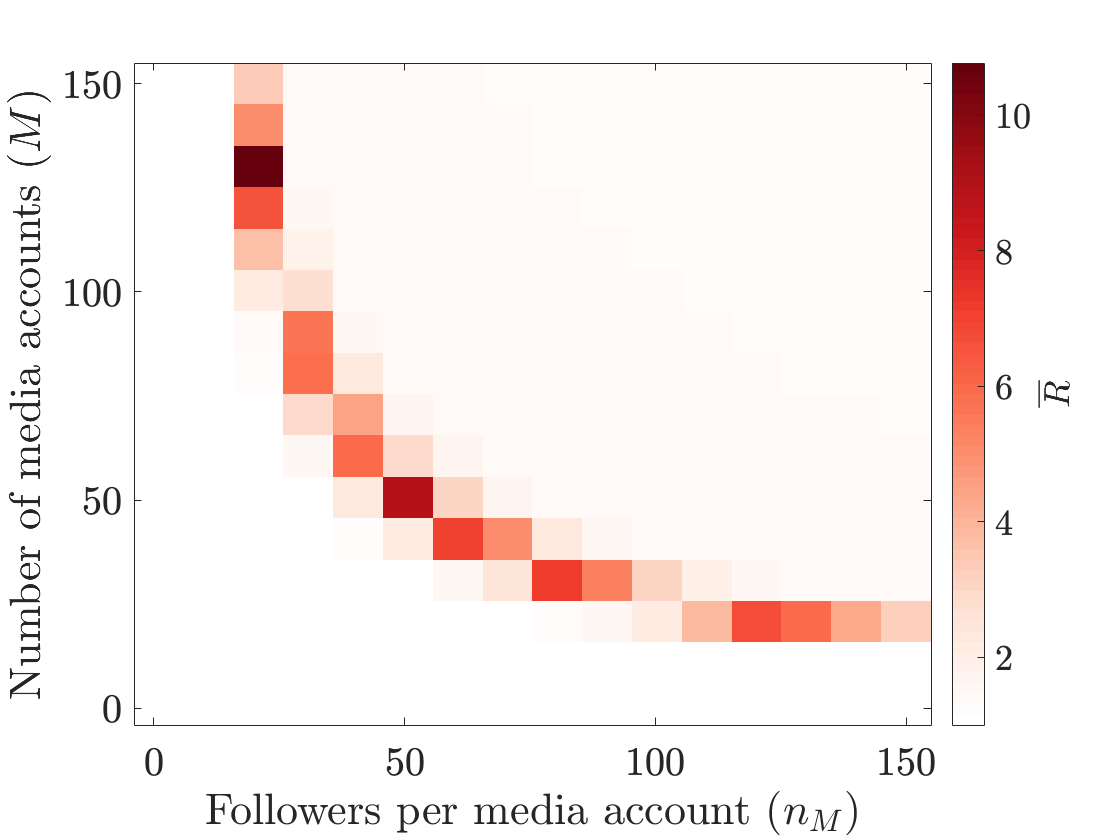}
	\caption{$N=500$ and $k=125$}
	\end{subfigure}
	\begin{subfigure}[t]{0.4\textwidth}
		\includegraphics[height=2in]{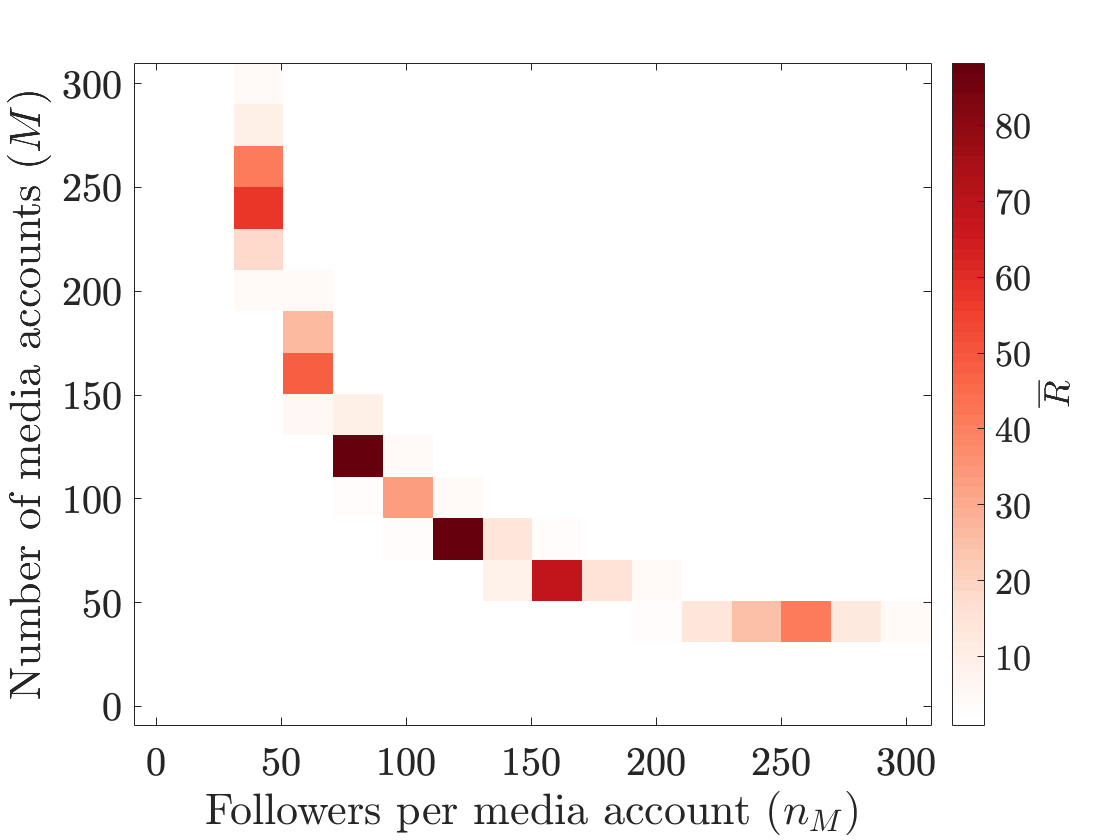}
	\caption{$N=1000$ and $k=250$}
	\end{subfigure}
\caption{Heat maps of the media impact diagnostic $\overline{R}$ for directed ER networks with different numbers of non-media accounts ($N$) for a fixed ratio ($k/N$) of the expected mean number of non-media accounts that non-media accounts follow to the total number of non-media accounts. Our simulations suggest that both media impact $\overline{R}$ and the spread of the media impact (which we define to be $\left(\max\{\overline{R}\}-\min\{\overline{R}\}\right)$ over all $(n_M,M)$ pairs) increase with $N$. The vertical axis is the number of media accounts ($M$), the horizontal axis is the number of followers per media account ($n_M$), and the colors represent the media impact diagnostic ($\overline{R}$), which we average over $200$ trials.  
 Dark red indicates the most media impact (i.e., the largest values of $\overline{R}$), and white indicates the least.
Unlike in our other figures, we have not scaled the color range to be the same across these four figures; this aids in visualizing the difference in the spread of impact values for networks of different sizes.
}
\label{fig:ERvaryN}
\end{figure*}

We now perform numerical experiments in which we vary the size of the directed ER networks by varying the number of non-media accounts. To try to isolate the effect of network size, we fix the ratio of the expected mean number ($k$) of non-media accounts that are followed to the total number of non-media accounts ($N$) for each simulation. As in our earlier observations, we see in Fig.~\ref{fig:ERvaryN} that the media impact increases as we increase the number $N$ of non-media nodes. However, in this case, we observe a progressively larger spread in the media entrainment diagnostic $\overline{R}$ for progressively larger $N$. For example, for $N=1000$ and $k=250$, the spread of the media impact (which we define to be $\left(\max\{\overline{R}\}-\min\{\overline{R}\}\right)$ over all $(n_M,M)$ pairs) is approximately $88.2048 - 0.9980 = 87.2068$. However, for $N=50$ and $k=13$, the spread is approximately $1.4295 - 1.0364 = 0.3931$ in our simulations.

\begin{figure*}
\begin{subfigure}[t]{0.3\textwidth}
\includegraphics[height=1.5in]{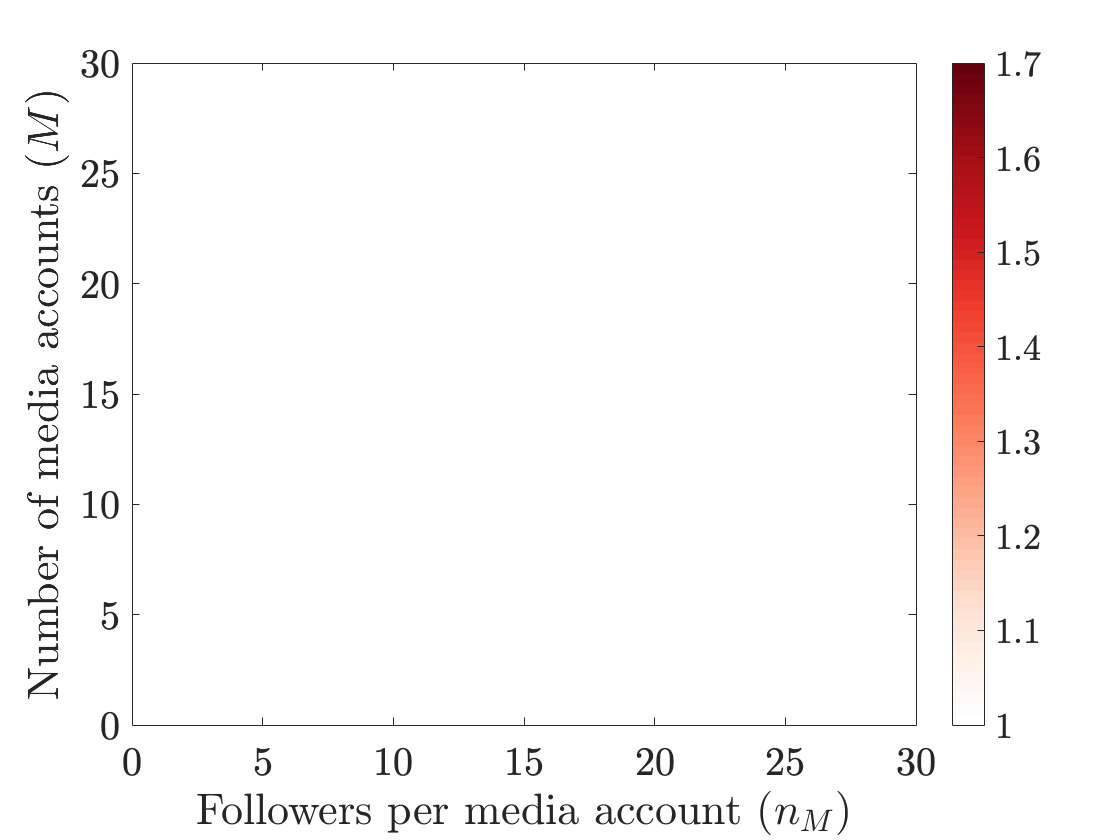}
\caption{$c=0.1$}
\end{subfigure}
\begin{subfigure}[t]{0.3\textwidth}
\includegraphics[height=1.5in]{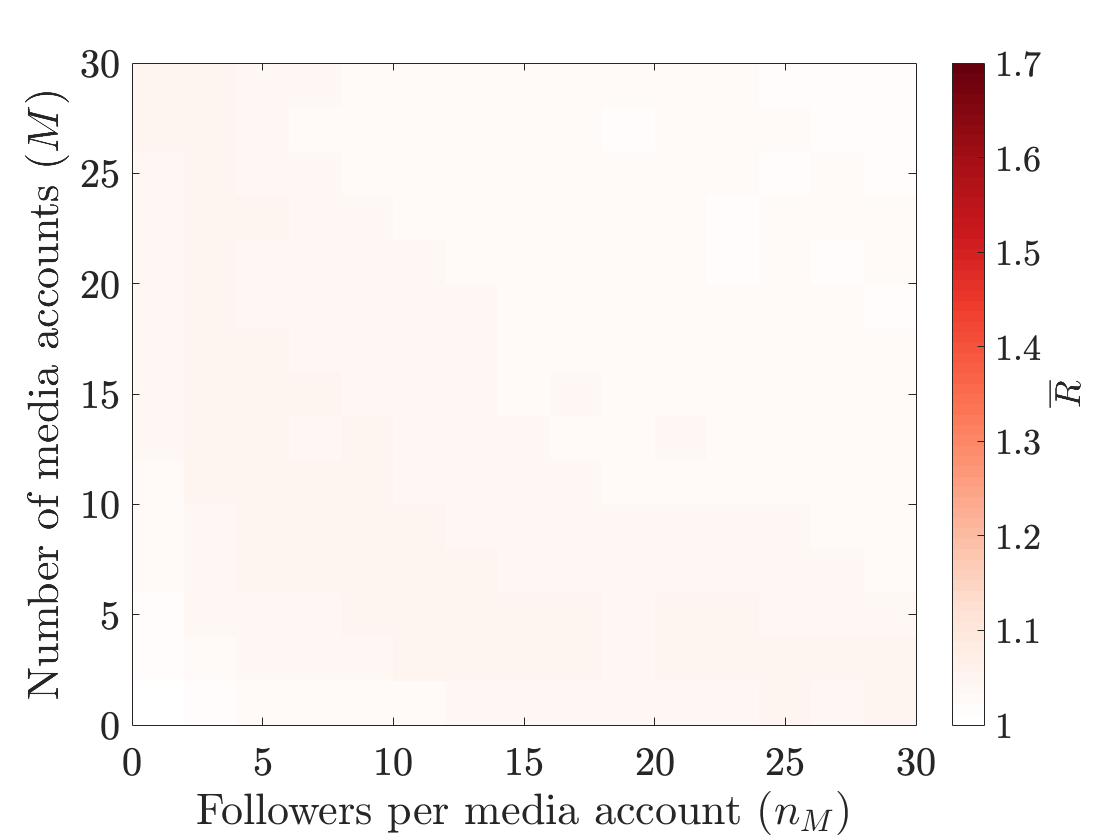}
\caption{$c=0.2$}
\end{subfigure}
\begin{subfigure}[t]{0.3\textwidth}
\includegraphics[height=1.5in]{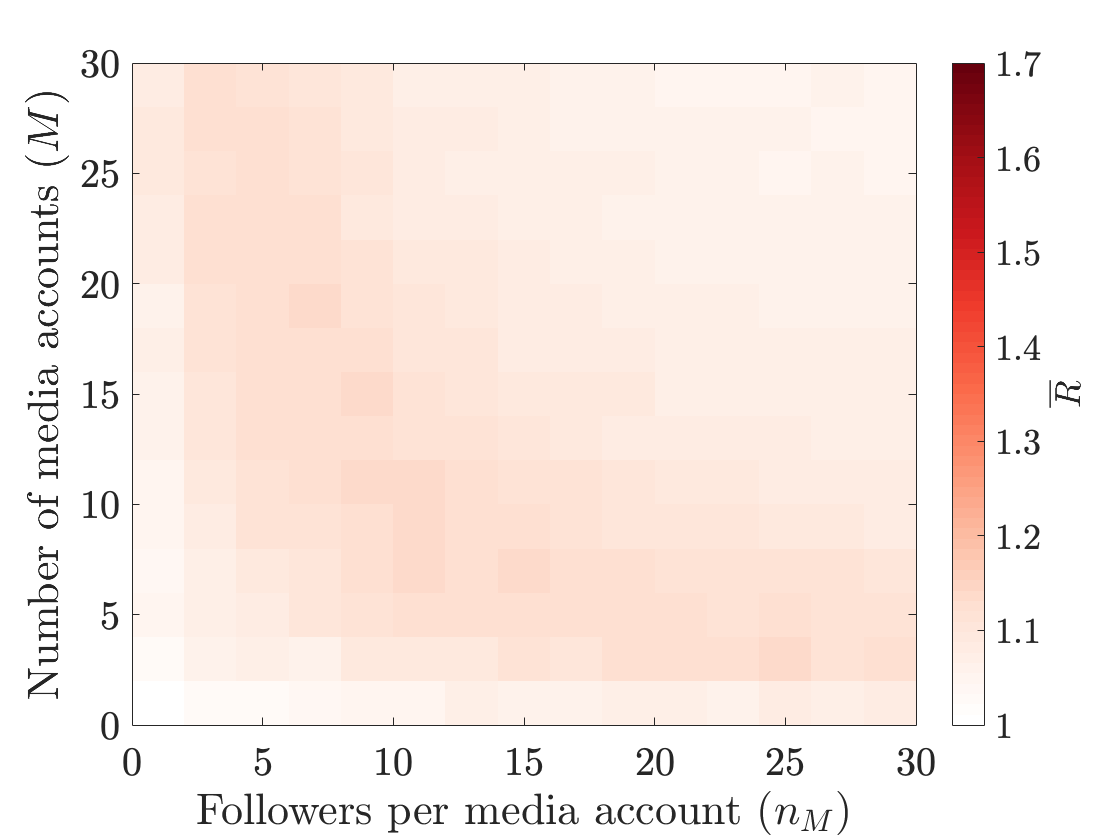}
\caption{$c=0.3$}
\end{subfigure}
\begin{subfigure}[t]{0.3\textwidth}
\includegraphics[height=1.5in]{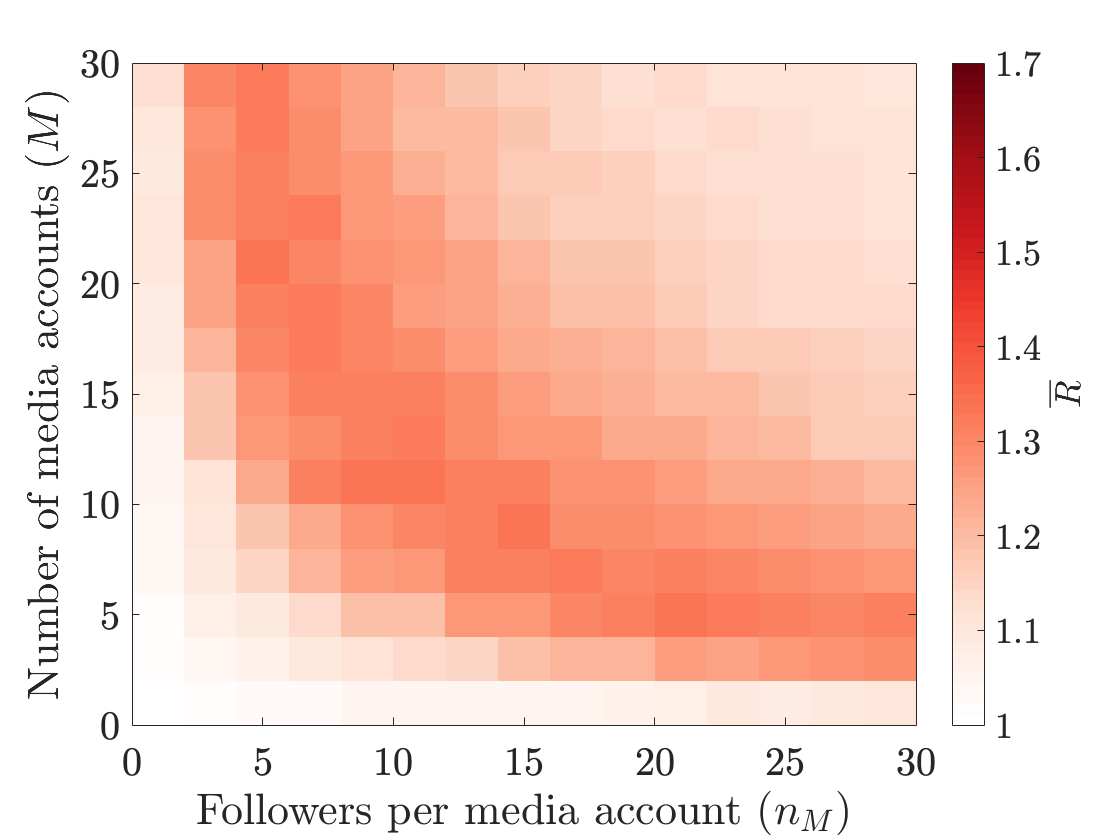}
\caption{$c=0.4$}
\end{subfigure}
\begin{subfigure}[t]{0.3\textwidth}
\includegraphics[height=1.5in]{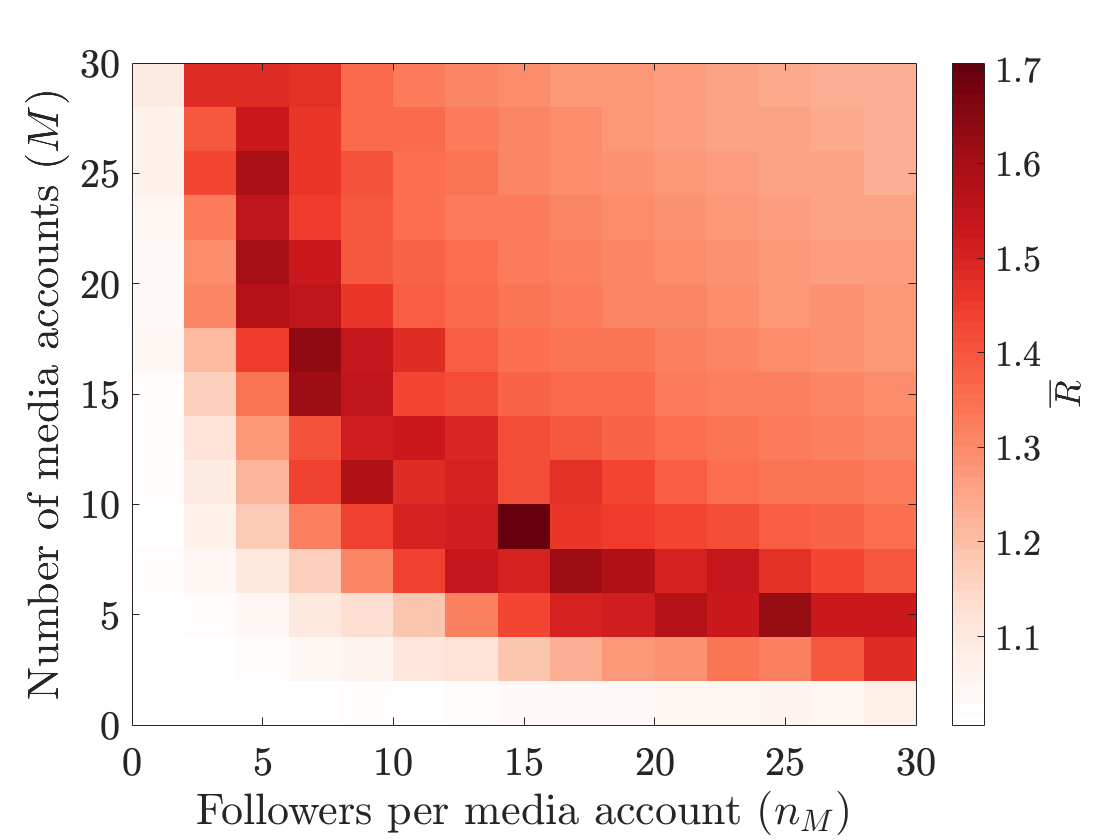}
\caption{$c=0.5$}
\end{subfigure}
\caption{Heat maps of the media impact diagnostic $\overline{R}$ for directed ER networks for different values of the receptiveness parameter $c$. For small values of $c$, the receptiveness of accounts is too small for the media accounts to impact the mean ideological positions of the non-media accounts. For progressively larger values of $c$, we observe progressively larger media impact. The vertical axis is the number of media accounts ($M$), the horizontal axis is the number of followers per media account ($n_M$), and the colors represent the media impact diagnostic ($\overline{R}$), which we average over 200 trials. 
 Dark red indicates the most media impact (i.e., the largest values of $\overline{R}$), and white indicates the least.
}
\label{fig:ERvaryc}
\end{figure*}

In Fig.~\ref{fig:ERvaryc}, we illustrate the effect of varying the receptiveness parameter $c$. If the receptiveness is very small, non-media accounts in a network adjust their ideological positions only if the content that they see is very close to their current ideology. Consequently, we observe little or no media impact on the ideological positions of the non-media nodes in the network. For sufficiently large receptiveness, however, the media do impact
the mean ideological position of non-media nodes, with a larger impact for larger values of $c$.

\begin{figure*}
\begin{subfigure}[t]{0.3\textwidth}
\includegraphics[height=1.5in]{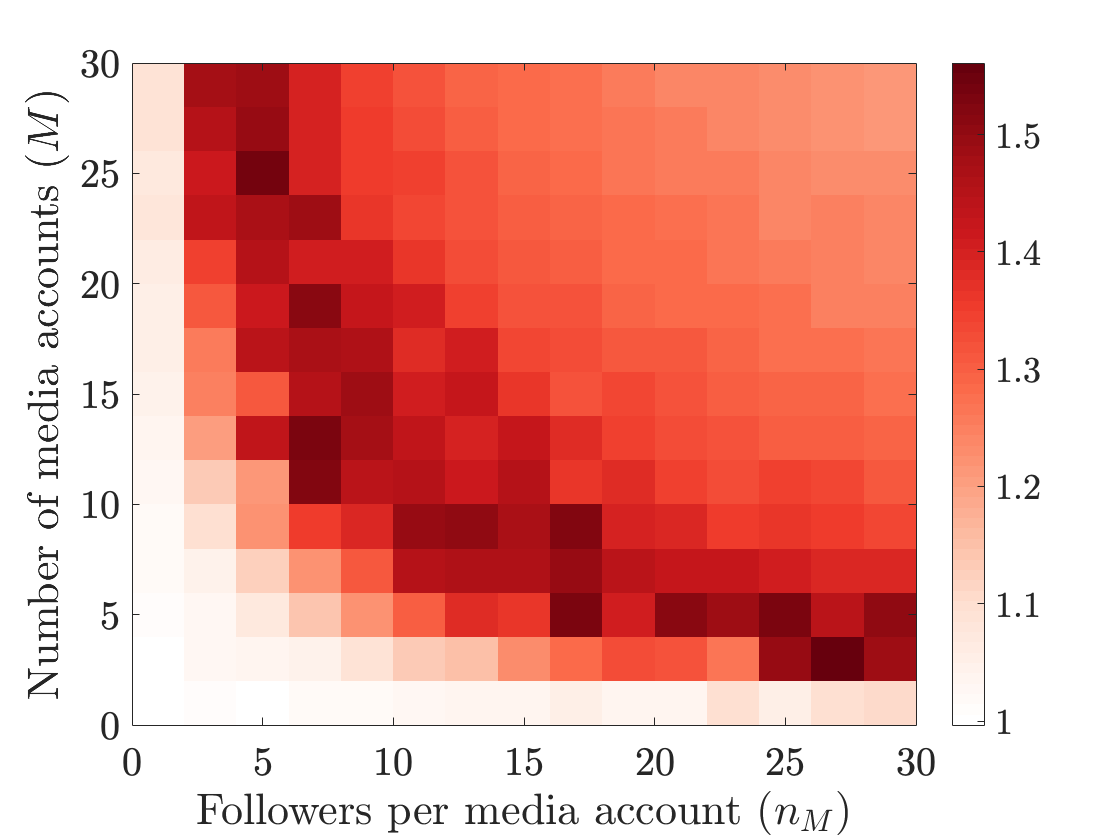}
\caption{$w=3$}
\end{subfigure}
\begin{subfigure}[t]{0.3\textwidth}
\includegraphics[height=1.5in]{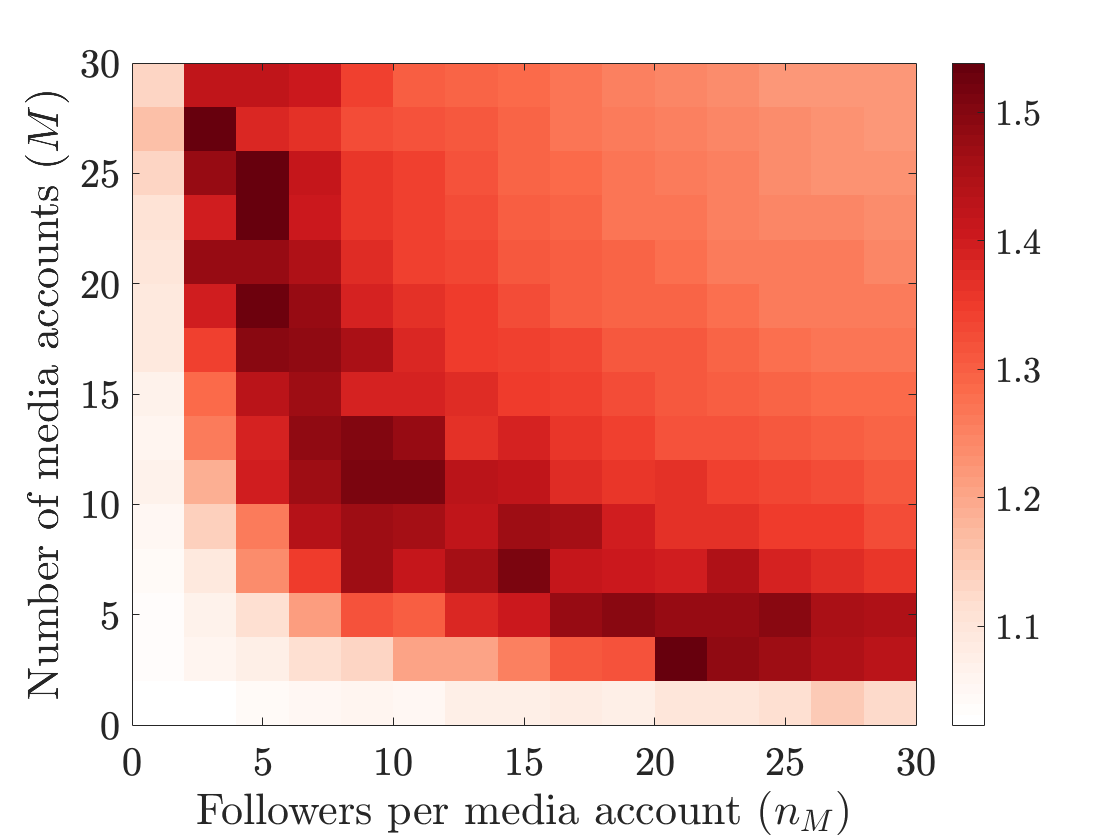}
\caption{$w=5$}
\end{subfigure}
\begin{subfigure}[t]{0.3\textwidth}
\includegraphics[height=1.5in]{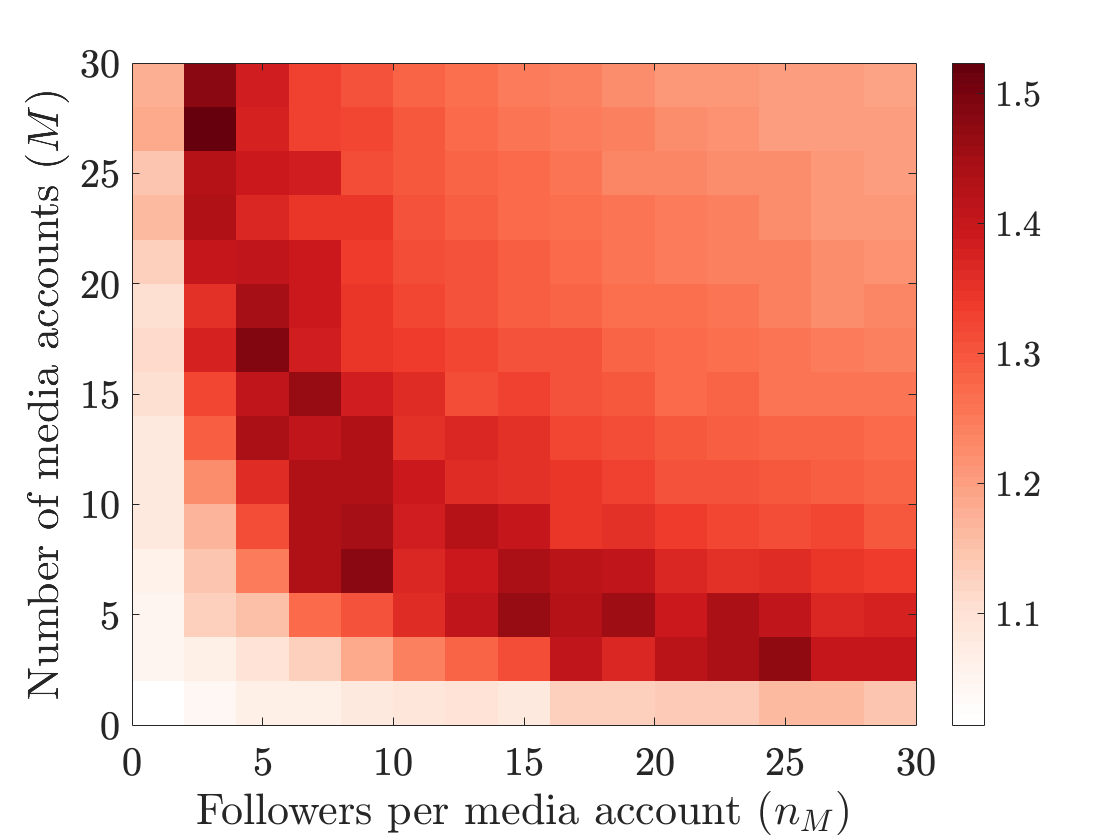}
\caption{$w=10$}
\end{subfigure}
\caption{Heat maps of the media impact diagnostic $\overline{R}$ for directed ER networks for different values of the self-weight $w$. In our numerical experiments, even when accounts weight their own ideology more heavily than those of the accounts that they follow, we observe similar qualitative dynamics as when we do not incorporate the parameter $w$ into our model.
The vertical axis is the number of media accounts ($M$), the horizontal axis is the number of followers per media account ($n_M$), and the colors represent the media impact diagnostic ($\overline{R}$), which we average over 200 trials. 
 Dark red indicates the most media impact (i.e., the largest values of $\overline{R}$), and white indicates the least.
}
\label{fig:ERvaryself}
\end{figure*}

One natural generalization of our model is to consider the effect of varying the weighting of an individual's current ideology in its content updating rule. (This is reminiscent of self-appraisal
in DeGroot models \cite{opinion-review,jia2015opinion}.)
 One way to incorporate such a weighting, such as an account valuing its current ideology more than those of the accounts that it follows,
 is to include a parameter $w \in \mathbb{N}$ into the update rule in Eq.~\ref{eqn:generalmodel}). 
 The new content updating rule with self-weight $w$ is
\begin{equation}
	{\bf x}_i^{t+1} = \frac{1}{\vert I_i \vert+w}\left( w{\bf x}_i^t + \sum_{j=1}^{N+M} A_{ij}{\bf x}_j^tf({\bf x}_j^t,{\bf x}_i^t)\right)\,.
\label{eqn:selfweight_model}
\end{equation}
We perform numerical experiments on directed ER networks in which we increase the self-weight $w$ such that non-media nodes weight their current ideology $3$, $5$, and $10$ times more than the ideologies of the accounts that they follow. We show the results of these simulations in Fig.~\ref{fig:ERvaryself}. In these examples, we observe media impact that is qualitatively similar to our observations without the parameter $w$.

\begin{figure}
	\begin{subfigure}[t]{0.5\textwidth}
		\includegraphics[height=2.5in]{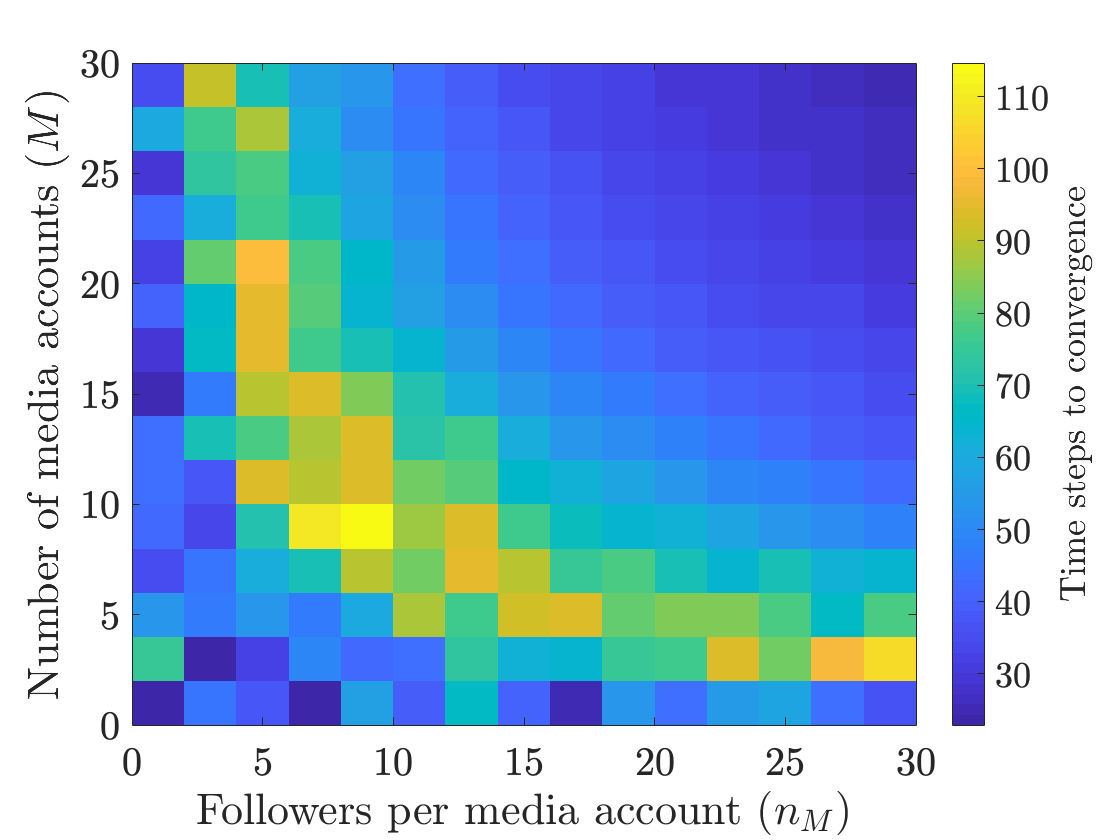}
	\caption{Mean convergence time (in color) for different values of $n_M$ and $M$.}
	\label{fig:ERconvergencea}
	\end{subfigure}
	\begin{subfigure}[t]{0.5\textwidth}
		\includegraphics[height=2.5in]{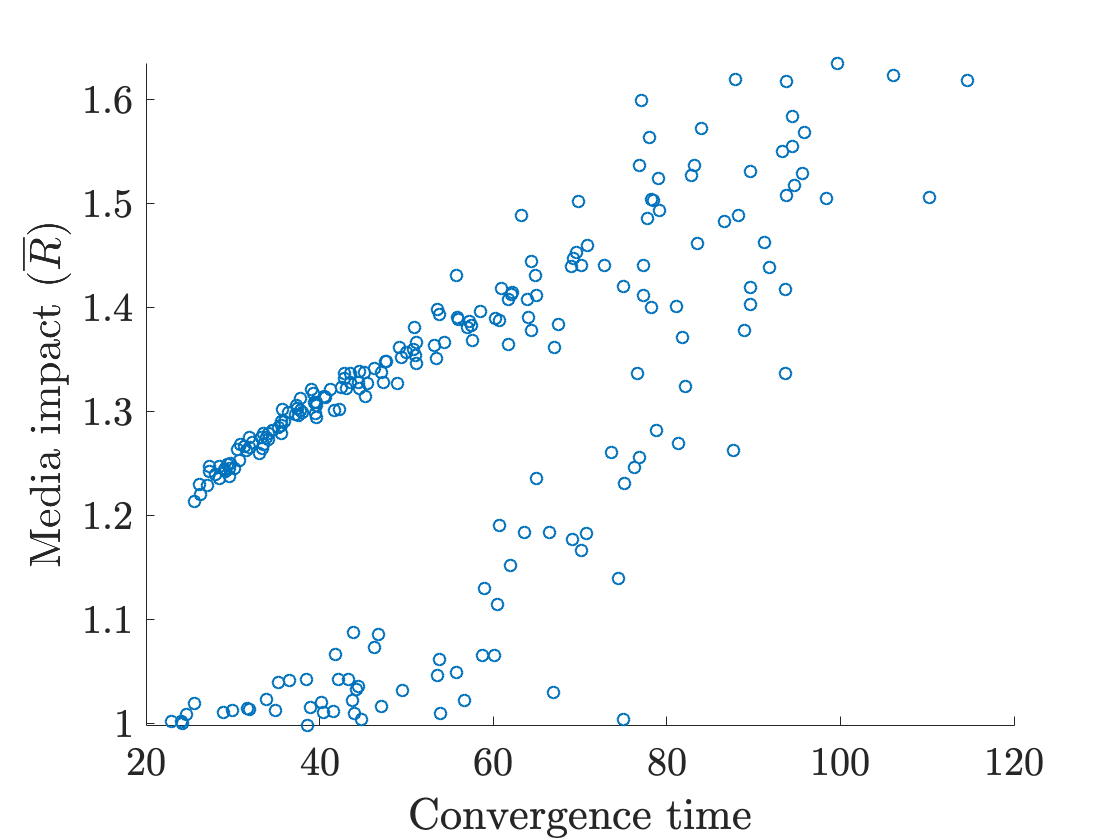}
	\caption{Media impact $\overline{R}$ versus convergence time.}
	\label{fig:ERconvergenceb}
	\end{subfigure}
\caption{In our simulations, we observe an interesting relationship between media impact and convergence time. (a) A heat map of the mean convergence time (in color, where yellow is the longest and blue is the shortest) versus the number of media accounts ($M$) on the vertical axis and the number of followers per media account ($n_M$) on the horizontal axis. (b) Mean media impact $\overline{R}$ versus the mean convergence time. Each circle represents one value in the $(n_M,M)$ parameter plane, so each circle corresponds to one of the square regions from panel (a). A larger media impact is positively correlated with a longer time to convergence. In both panels, we run our simulations on a directed ER network with $N=100$ non-media accounts that follow an expected mean of $k=25$ non-media accounts.}
\label{fig:ERconvergence}
\end{figure}

\begin{figure}[h!]
\includegraphics[width=0.5\textwidth]{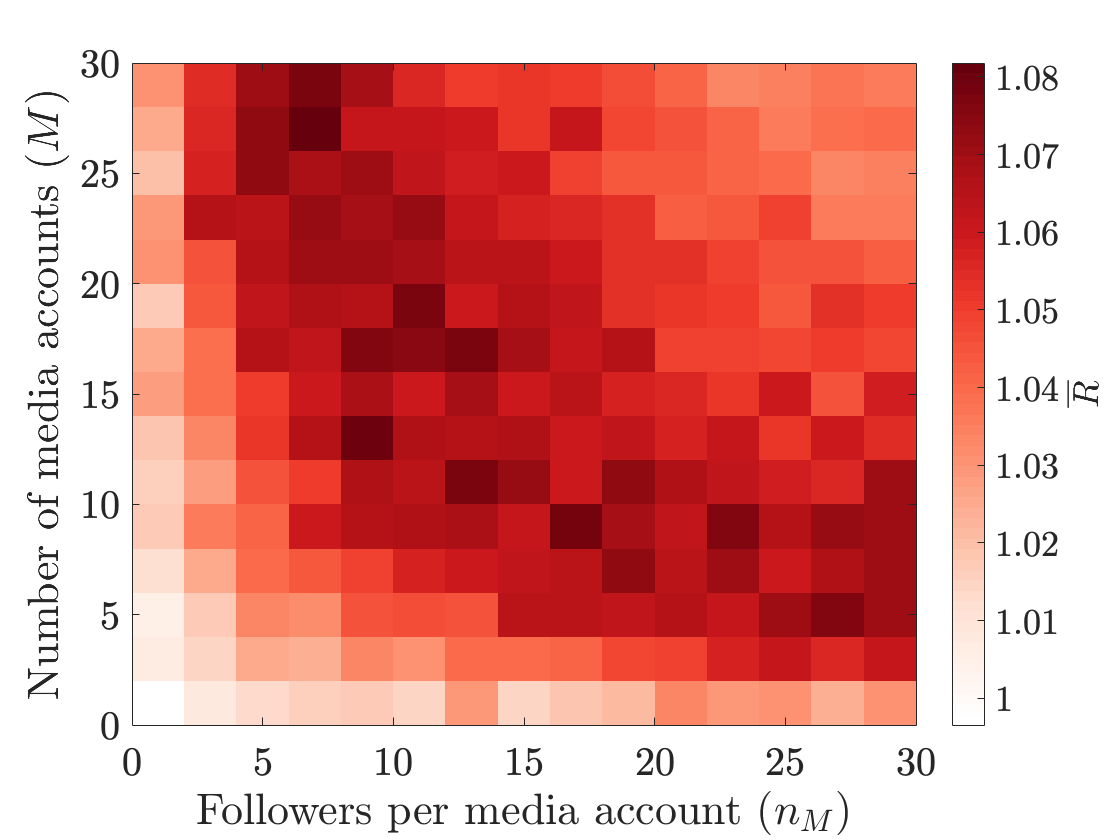}
\caption{An example of a heat map of the media impact diagnostic $\overline{R}$ for our model with political positions with two ideological dimensions. For this example, we simulate our model on directed ER networks with $N=100$ non-media accounts that follow an expected mean of $k=25$ non-media accounts. We obtain qualitatively similar results as in our computations with one ideological dimension, but the mean media impact is less pronounced in this case. Specifically, the mean impact $\overline{R}$ is 
smaller when using two ideological dimensions than what we observed previously when using one ideological dimension for directed ER networks with the same parameter values. The vertical axis is the number of media accounts ($M$), the horizontal axis is the number of followers per media account ($n_M$), and the colors represent the media impact diagnostic ($\overline{R}$), which we average over $200$ trials. 
 Dark red indicates the most media impact (i.e., the largest values of $\overline{R}$), and white indicates the least.
}
\label{fig:2dimpact}
\end{figure}

\begin{figure*}
	\begin{subfigure}[t]{0.24\textwidth}
		\includegraphics[height=1.3in]{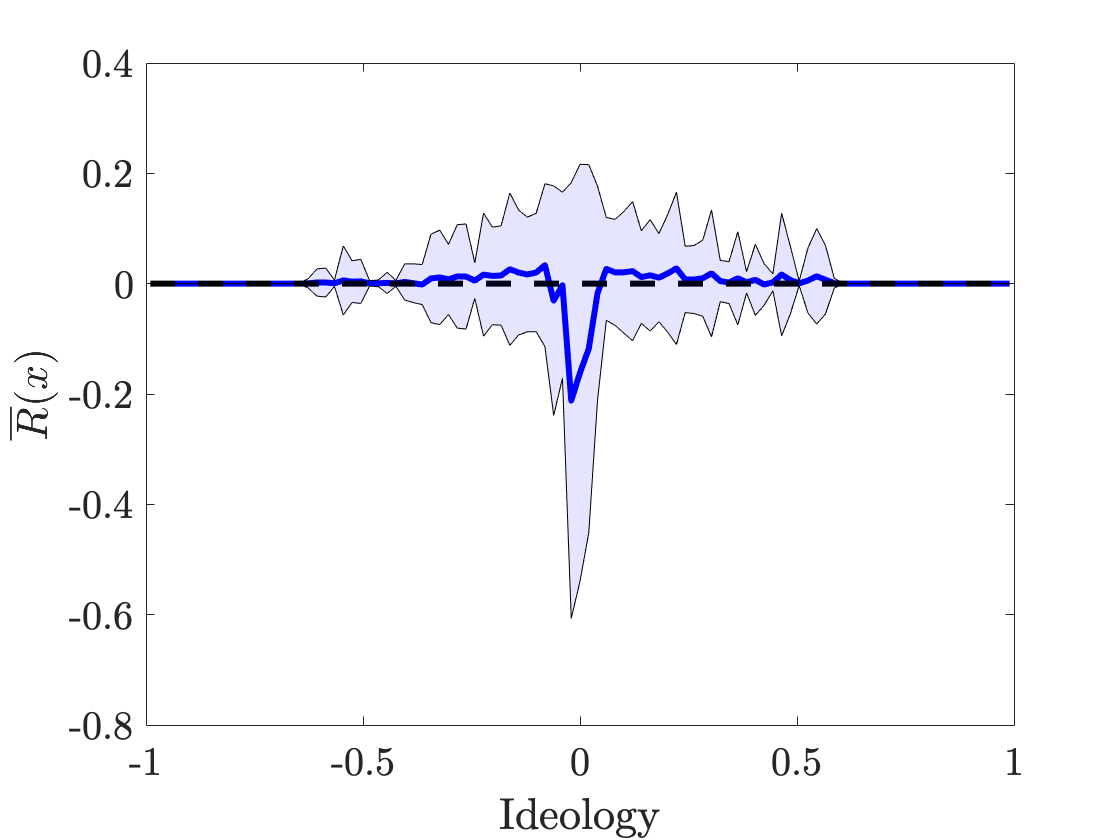}
	\caption{$n_M = 1$}
	\end{subfigure}
	\begin{subfigure}[t]{0.24\textwidth}
		\includegraphics[height=1.3in]{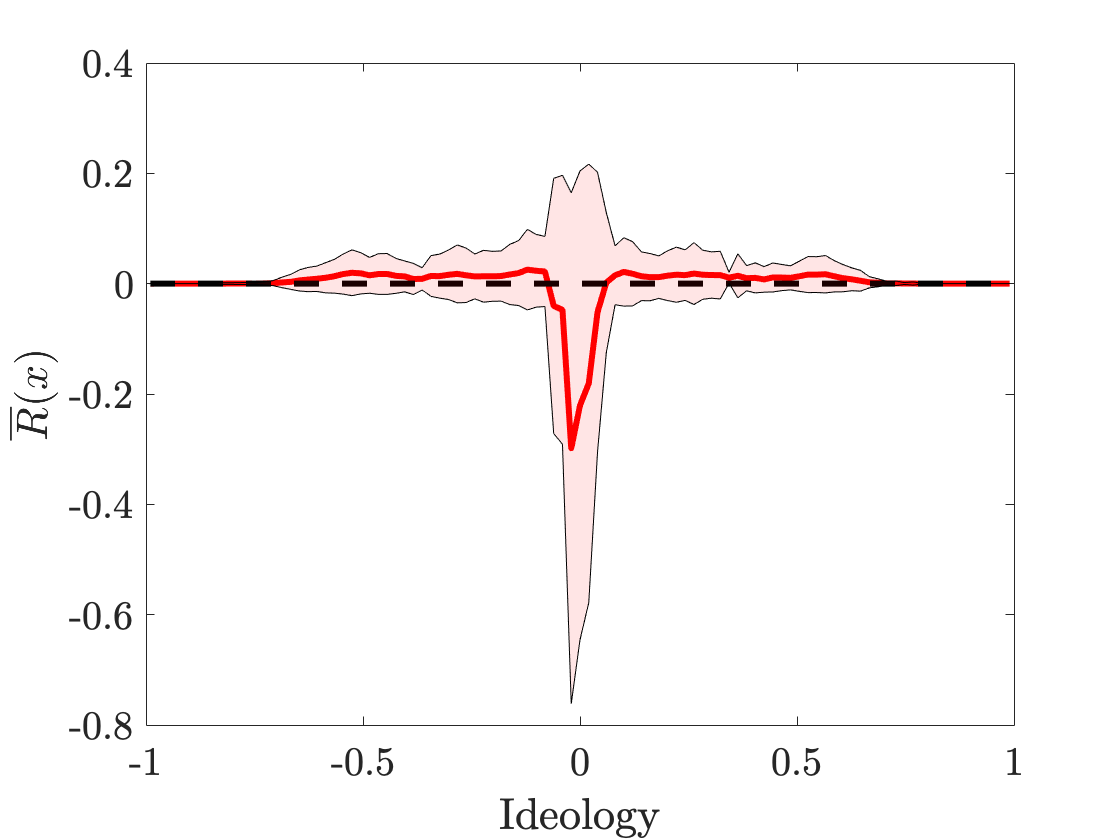}
	\caption{$n_M = 25$}
	\end{subfigure}
	\begin{subfigure}[t]{0.24\textwidth}
		\includegraphics[height=1.3in]{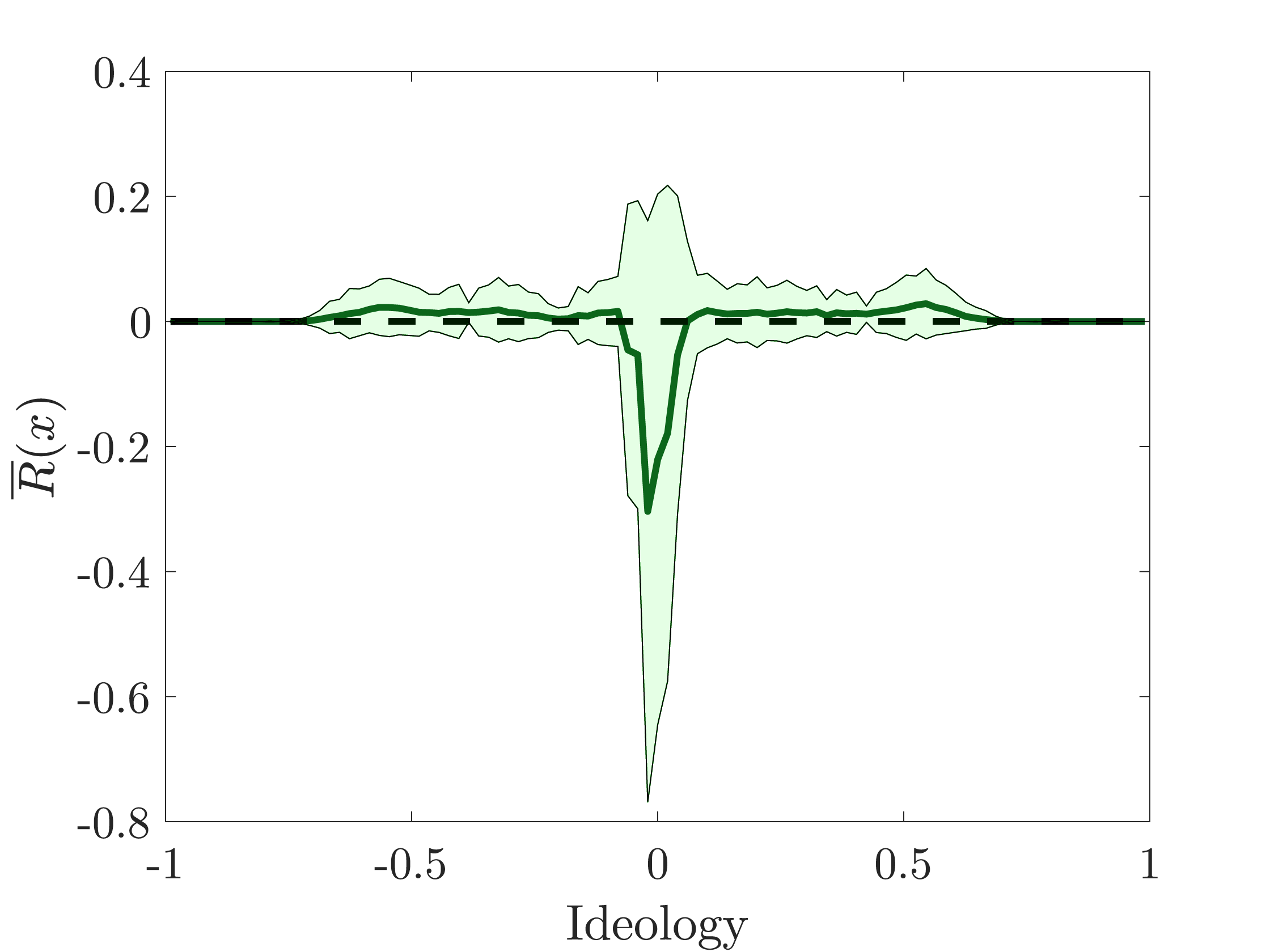}
	\caption{$n_M = 50$}
	\end{subfigure}
	\begin{subfigure}[t]{0.24\textwidth}
		\includegraphics[height=1.3in]{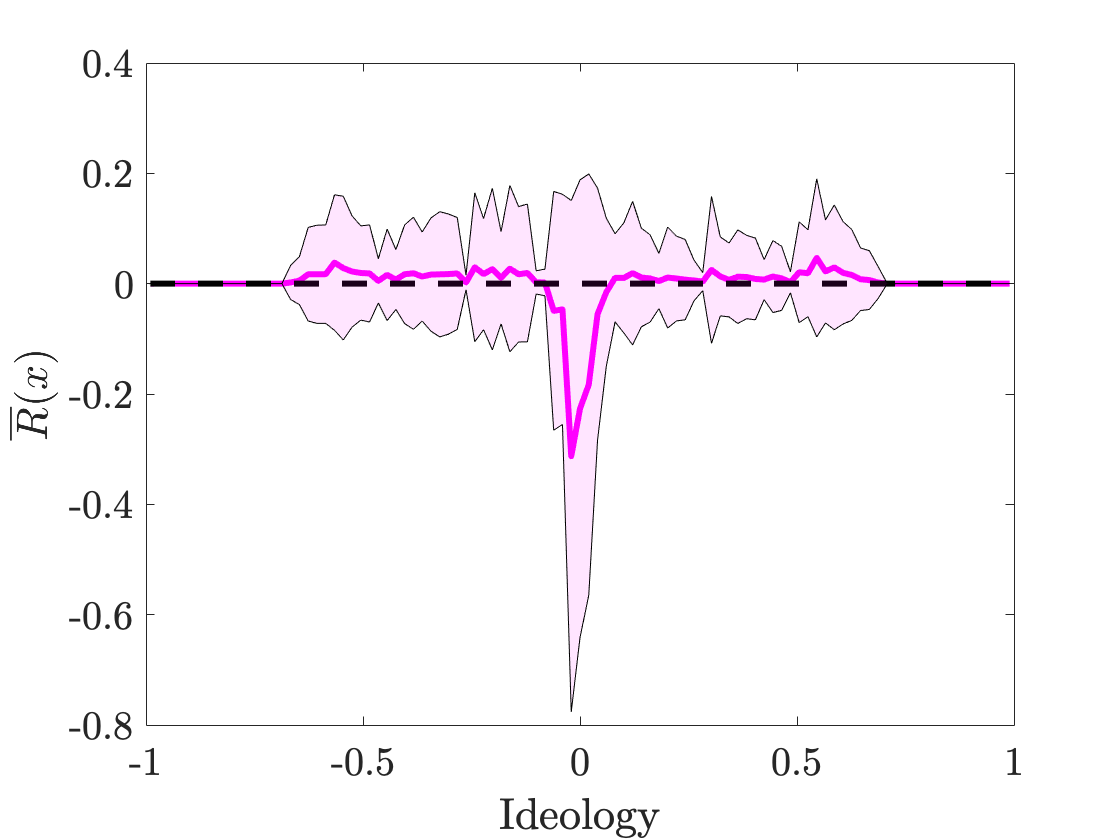}
	\caption{$n_M = 100$}
	\end{subfigure}
	\begin{subfigure}[t]{0.24\textwidth}
		\includegraphics[height=1.3in]{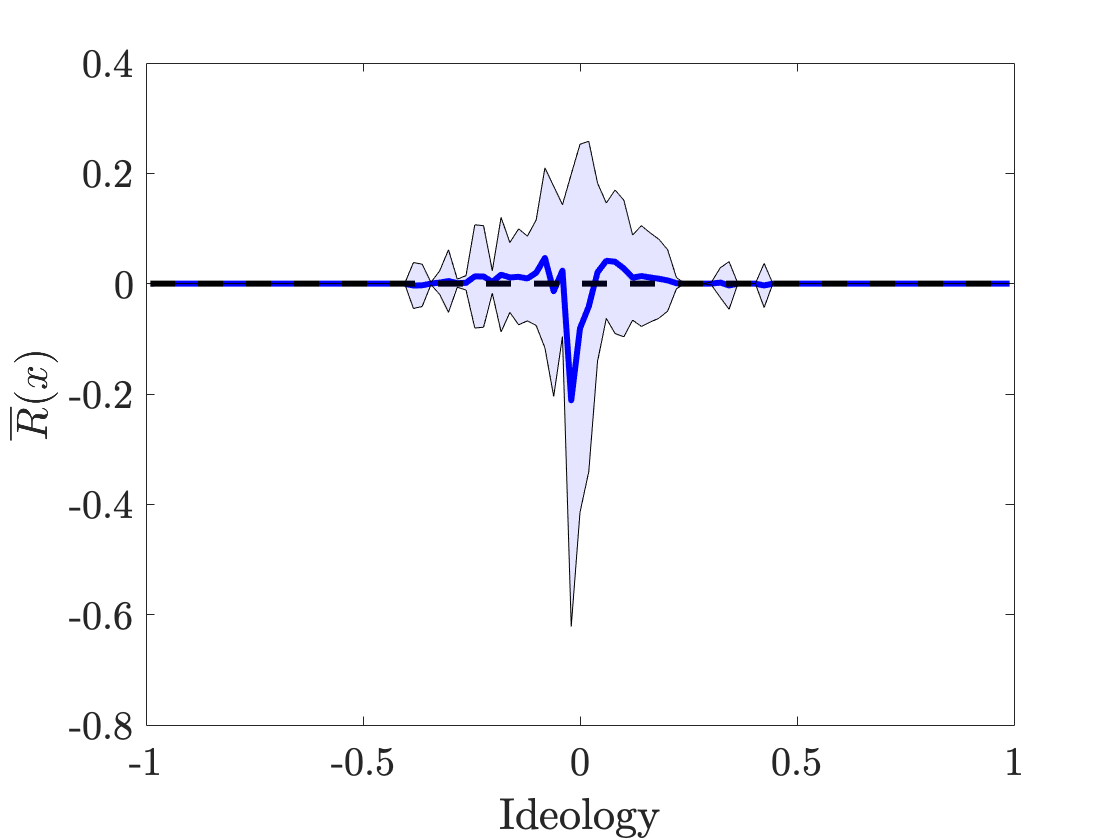}
	\caption{$n_M = 1$}
	\end{subfigure}
	\begin{subfigure}[t]{0.24\textwidth}
		\includegraphics[height=1.3in]{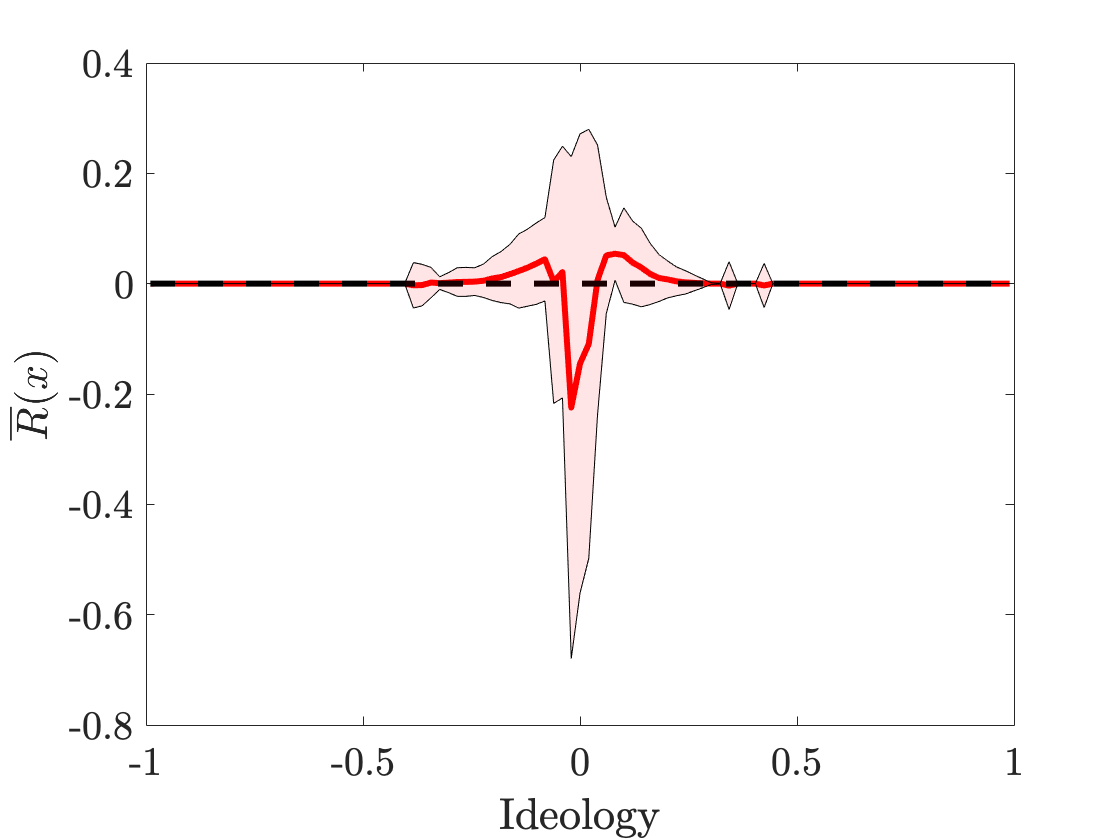}
	\caption{$n_M = 25$}
	\end{subfigure}
	\begin{subfigure}[t]{0.24\textwidth}
		\includegraphics[height=1.3in]{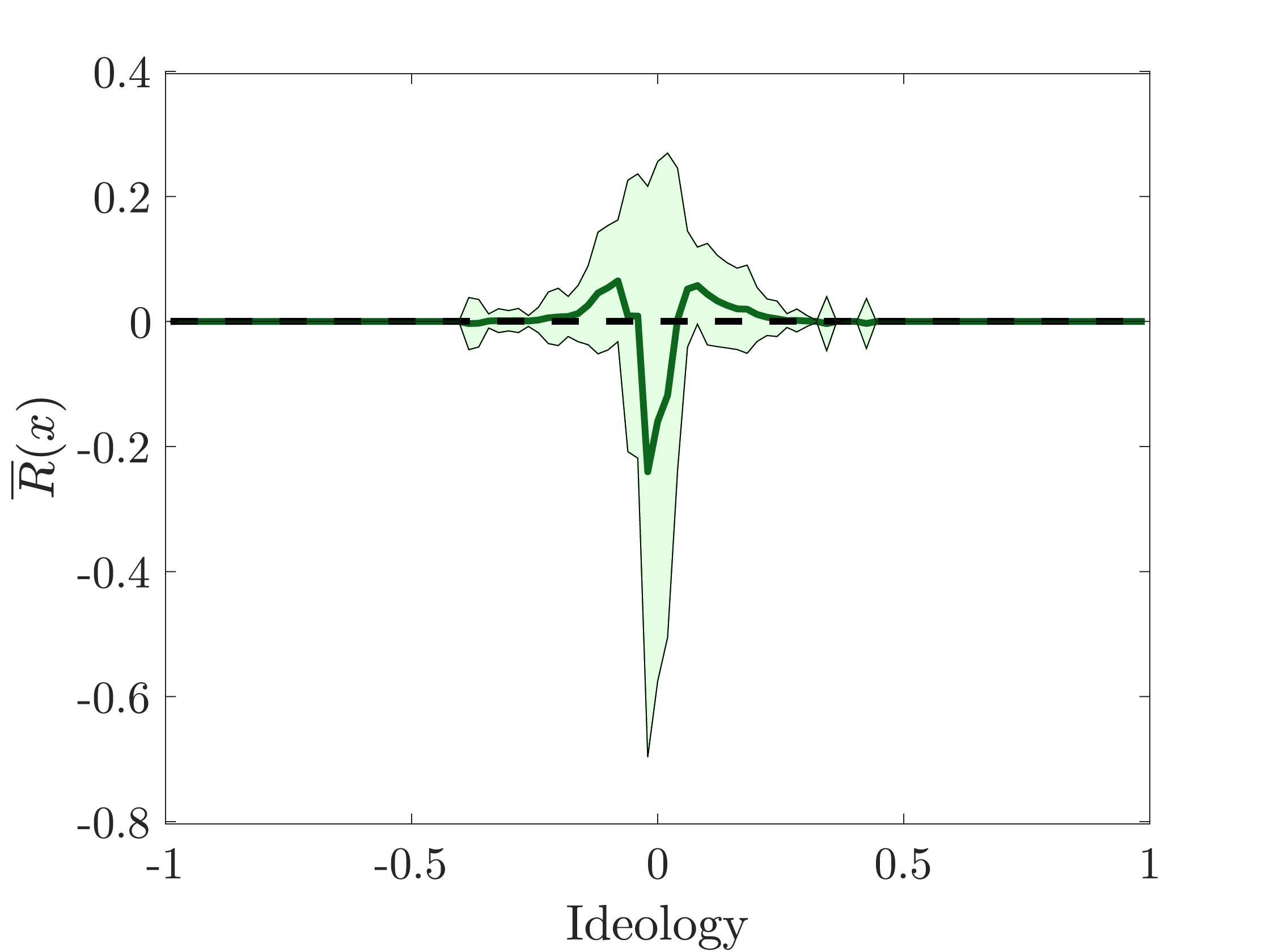}
	\caption{$n_M = 50$}
	\end{subfigure}
	\begin{subfigure}[t]{0.24\textwidth}
		\includegraphics[height=1.3in]{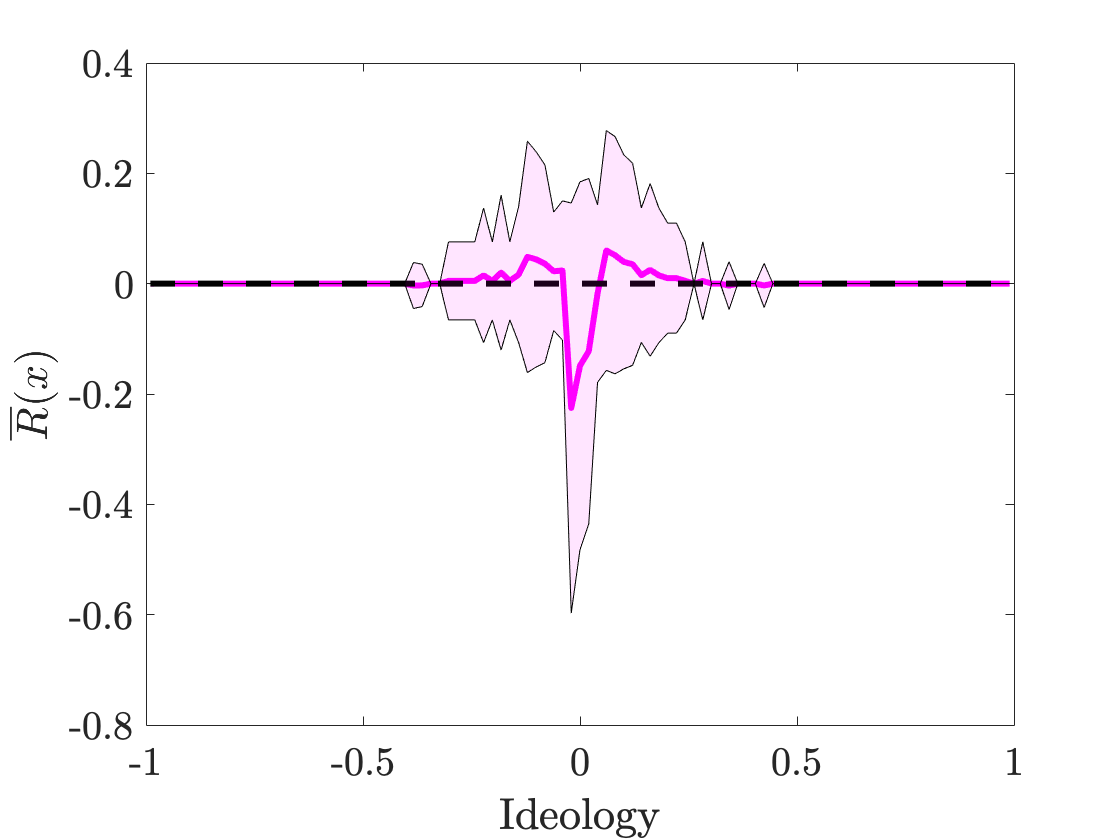}
	\caption{$n_M = 100$}
	\end{subfigure}
\caption{Impact on the partisan bias of the non-media accounts with media ideologies drawn from a probability distribution. In the top row [i.e., panels (a)--(d)], we show the media impact for $M=100$ media accounts, where we draw the ideological positions of each media account from the uniform distribution on $[-1,1]$. We vary the number $n_M$ of followers per media account. In the bottom row [i.e., panels (e)--(h)], we show the media impact for $M=100$ media accounts with ideological positions that we draw from a truncated Gaussian distribution on $[-1,1]$ with mean $0$ and standard deviation $0.5$. Both sets of simulations are on directed ER networks with $N=100$ non-media accounts. We average each media impact function $\overline{R}(x)$ over $200$ trials. We depict the standard deviations with the shaded regions. The dashed line at $\overline{R}(x)=0$ represents no media impact. Values of $\overline{R}(x)$ that lie below this line indicate that the media has decreased the prevalence of non-media accounts with ideology $x$ in comparison to the situation
in the absence of media; values of $\overline{R}$ above this line indicate that the media accounts have increased the prevalence of non-media accounts with ideological position $x$.
 }
\label{fig:impactvaryn}
\end{figure*}

As suggested by the individual trials in Fig.~\ref{fig:Reedexamples}, there is a relationship between media impact and time to convergence. As we illustrate in Fig.~ \ref{fig:ERconvergence}, our simulations indicate that larger media impact is positively correlated with longer time to convergence.


\subsection{Metastability and long-time dynamics}

In Sec.~\ref{sec:1dimpact}, we defined a numerical simulation to have ``converged" if $\lvert x^t_i - x^{t-1}_i\rvert < \text{TOL}=10^{-4}$ for all $i$. Given this definition, an important question to ask is whether (1) these ``converged" values have truly reached a stationary state or (2) our model will
subsequently reach a consensus state on extremely long time scales. To examine this question, we run numerical simulations of our model with a convergence tolerance of $\text{TOL}=0$ and impose a bailout time of $T=250,000$ time steps. If our simulations reach the bailout time, we record the ``converged" state as the state at the bailout time. We perform these simulations on the Reed College Facebook network with the same parameter values ($N=962$, $c=0.5$, and $x_M=0.9$) as in Fig.~\ref{fig:ReedArrows} and on directed ER networks with the same parameter values ($N=100$, $k=25$, $c=0.5$, and $x_M=0.9$) as in Fig.~\ref{fig:Constructedentrain}.

\begin{figure}[h]
	\includegraphics[width=0.5\textwidth]{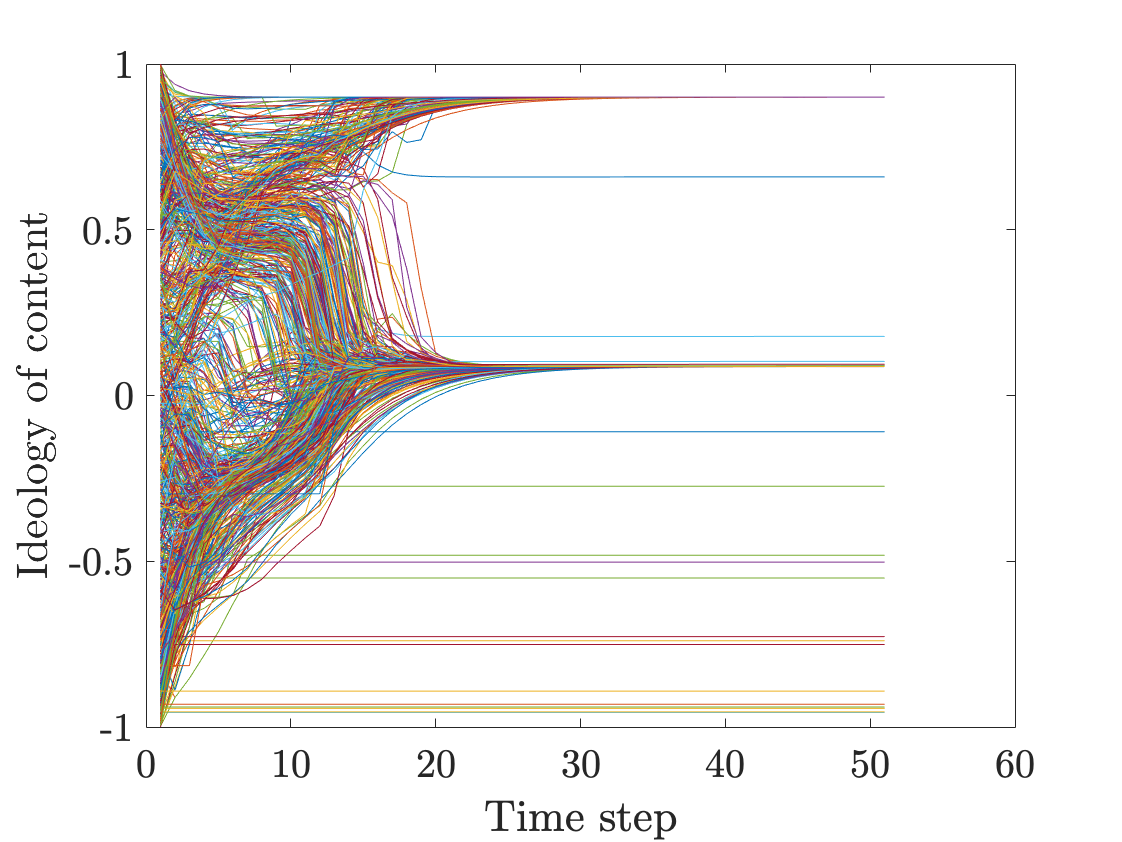}
	\includegraphics[width=0.5\textwidth]{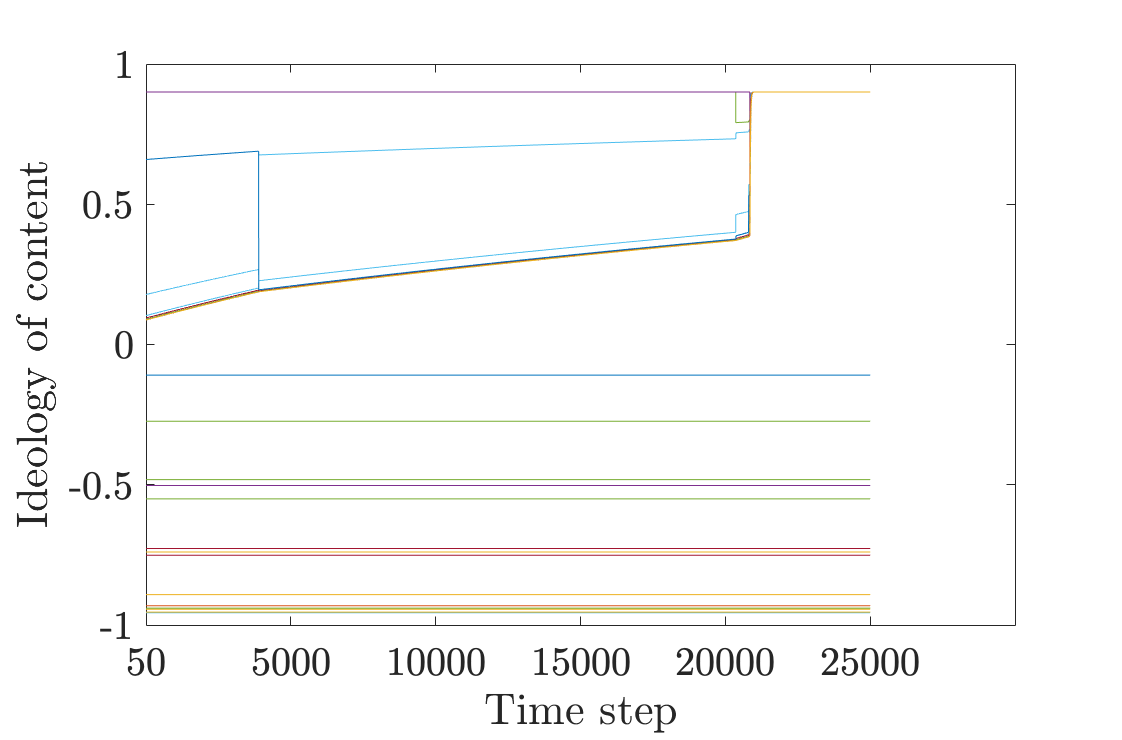}
	\caption{Example in which our model that exhibits multiple time scales in a single simulation of it on Reed College Facebook network with the same parameter values ($N=962$, $c=0.5$, and $x_M=0.9$) as in Fig.~\ref{fig:ReedArrows}. In this simulation, $M=9$ and $n_M=225$. In the top panel, we
	 set the convergence tolerance to $\text{TOL}=10^{-4}$; our simulation converges with this tolerance after $51$ time steps. In the bottom panel, we show the same network with the same initial data with the convergence tolerance set to $\text{TOL}=0.$ For this convergence tolerance, the two large ideology groups where ideology is positive eventually ``collapse" on a long time scale to a single ideology group, which is entrained to the media nodes. This ``collapse" takes over $2\times10^{4}$ time steps to occur; this is several orders of magnitude longer than the dynamics in the top panel.}
\label{fig:timescale-example}
\end{figure}

\begin{figure*}[h]
	\includegraphics[width=0.4\textwidth]{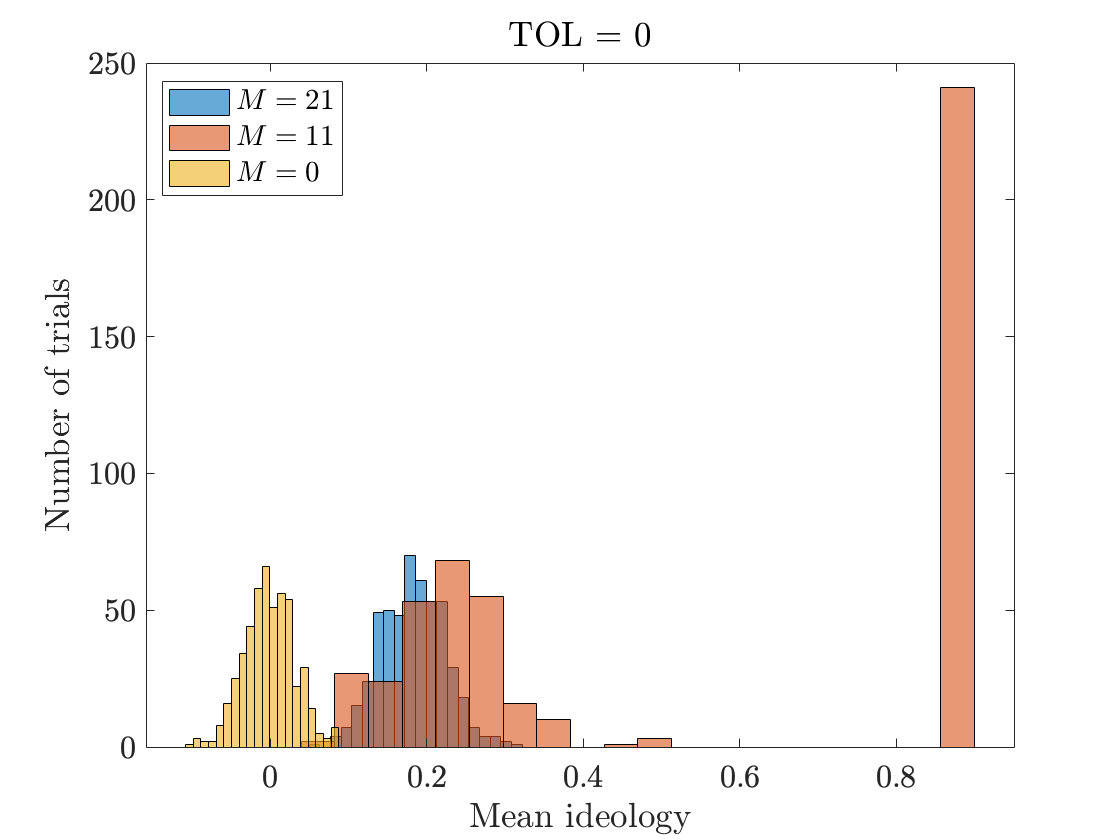}
	\includegraphics[width=0.4\textwidth]{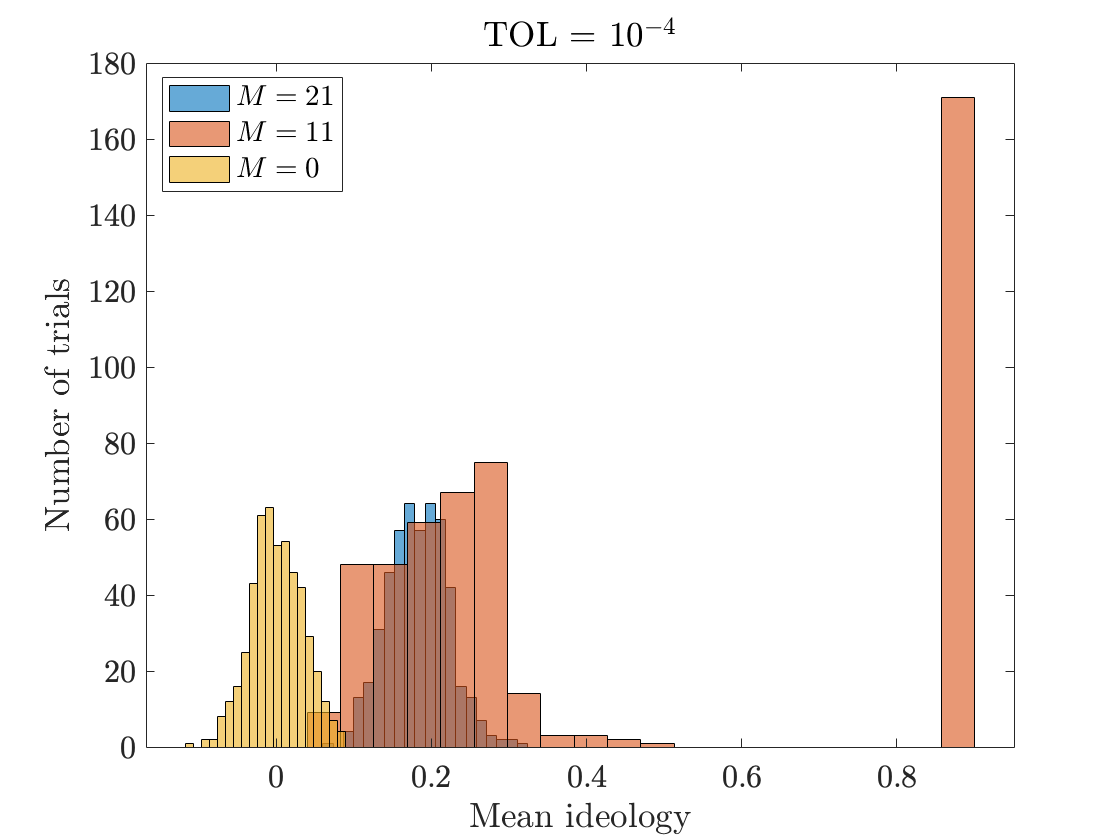}
	\includegraphics[width=0.4\textwidth]{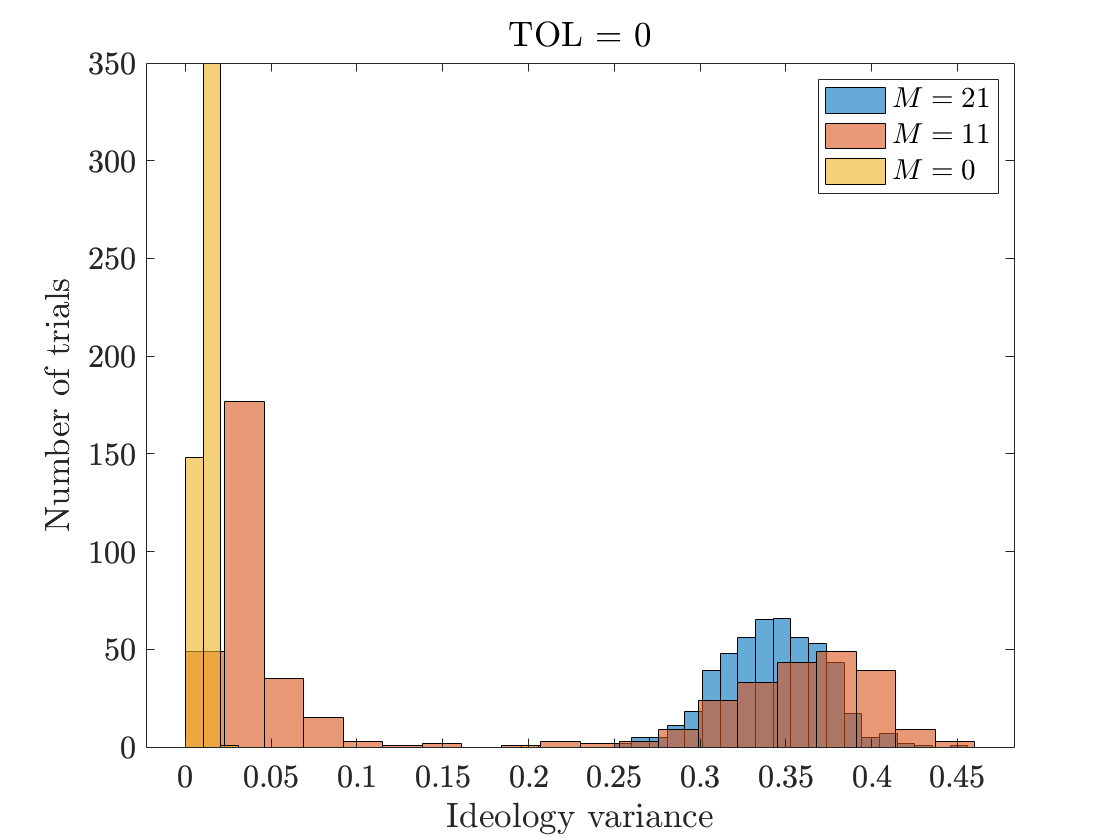}
	\includegraphics[width=0.4\textwidth]{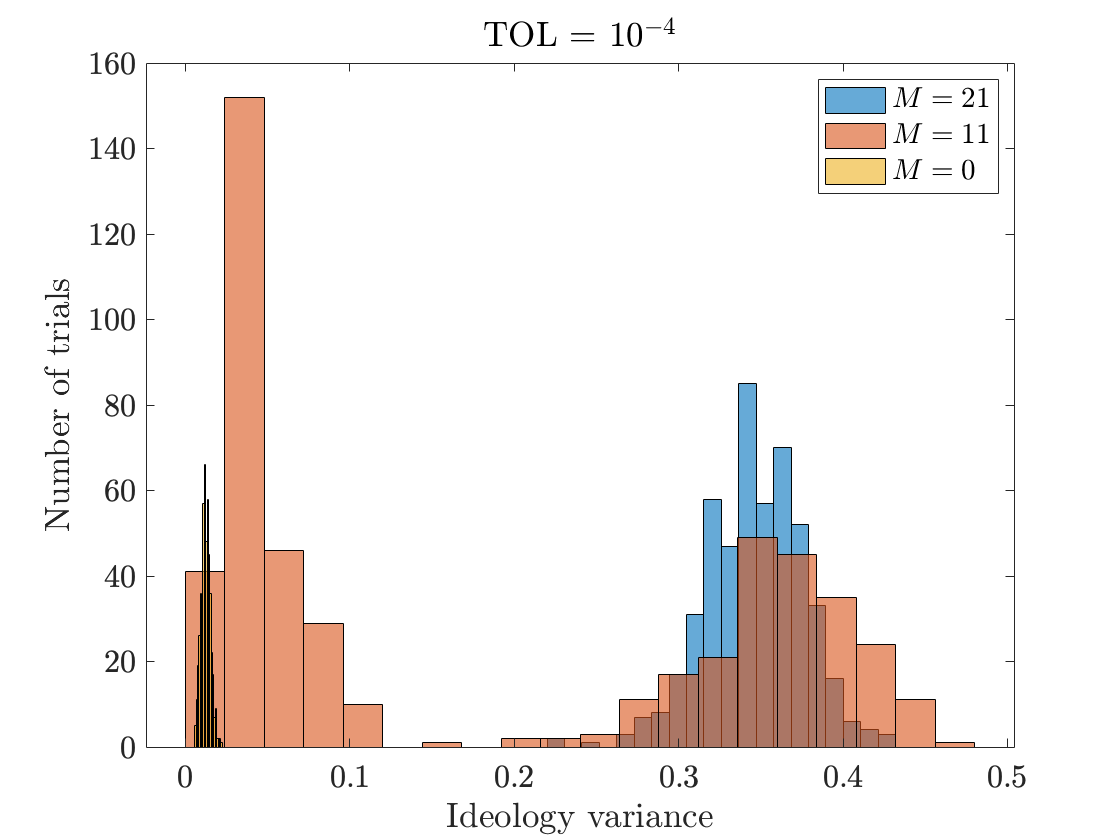}
	\caption{{
	We simulate our model on the Reed College Facebook network with the same parameter values ($N=962$, $c=0.5$, and $x_M=0.9$) as in Fig.~\ref{fig:ReedArrows} for $500$ trials for three levels of media impact: $M=0$ and $n_M=0$ (yellow), $M=11$ and $n_M=225$ (orange), and $M=21$ and $n_M=675$ (blue). In the top row, we show the distributions of the mean of ideology at convergence for convergence tolerances of (left) $0$ and (right) $10^{-4}$. In the bottom row, we show the distributions of the variance of the ideology at convergence for a convergence tolerances of (left) $0$ and (right) $10^{-4}$. Our histograms have $20$ bins. We observe that the qualitative behavior of the mean and variance of ideologies is the same for different tolerances for convergence.}}
\label{fig:timescale-dists-reed}
\end{figure*}

\begin{figure*}[h]
	\includegraphics[width=0.4\textwidth]{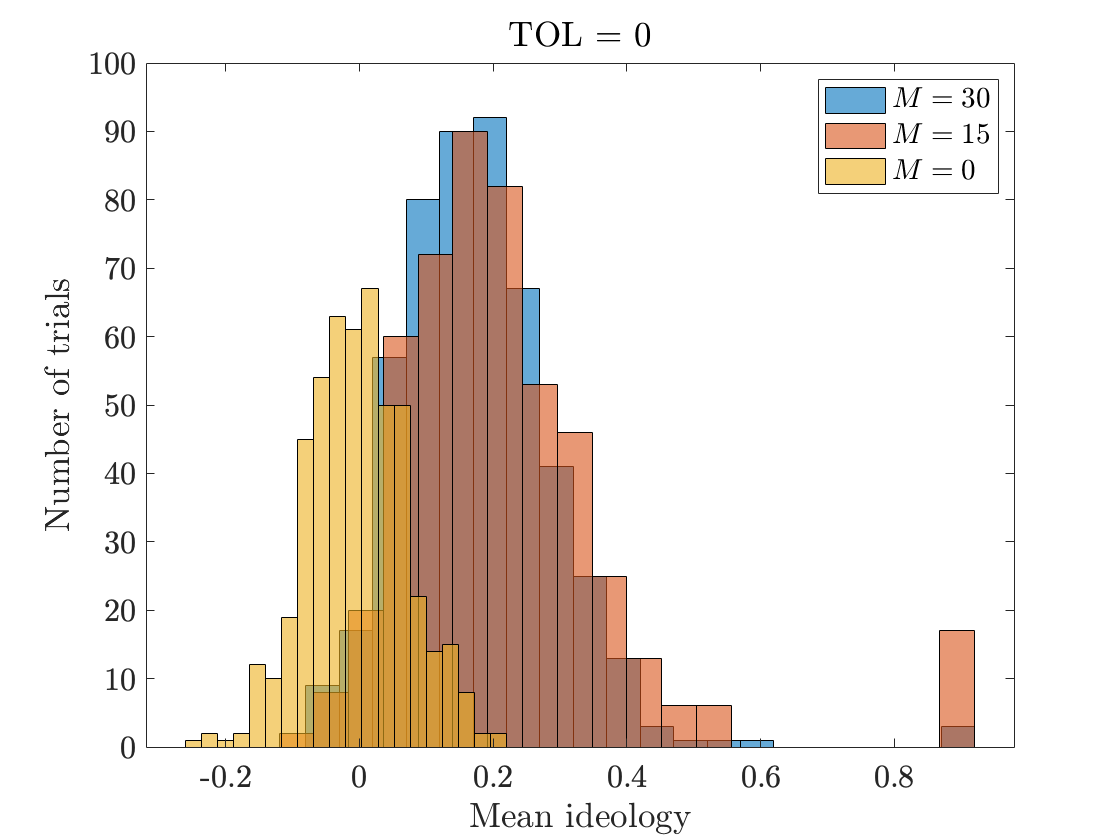}
	\includegraphics[width=0.4\textwidth]{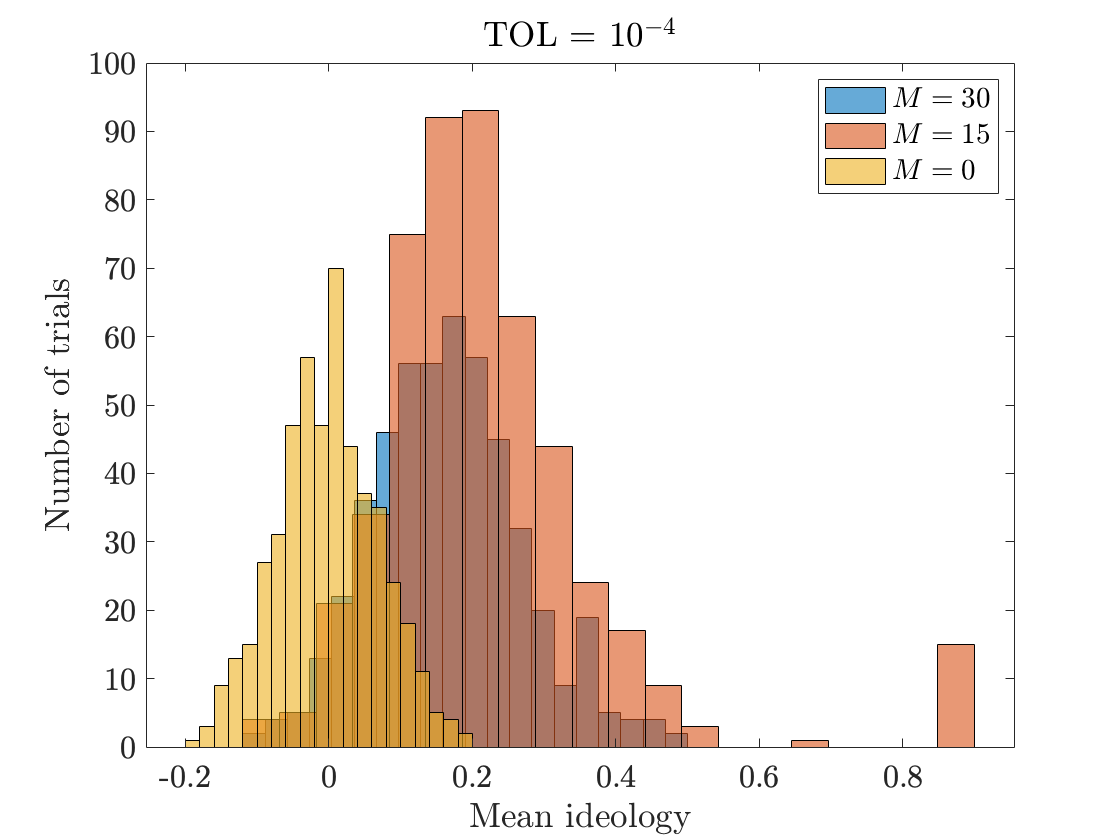}
	\includegraphics[width=0.4\textwidth]{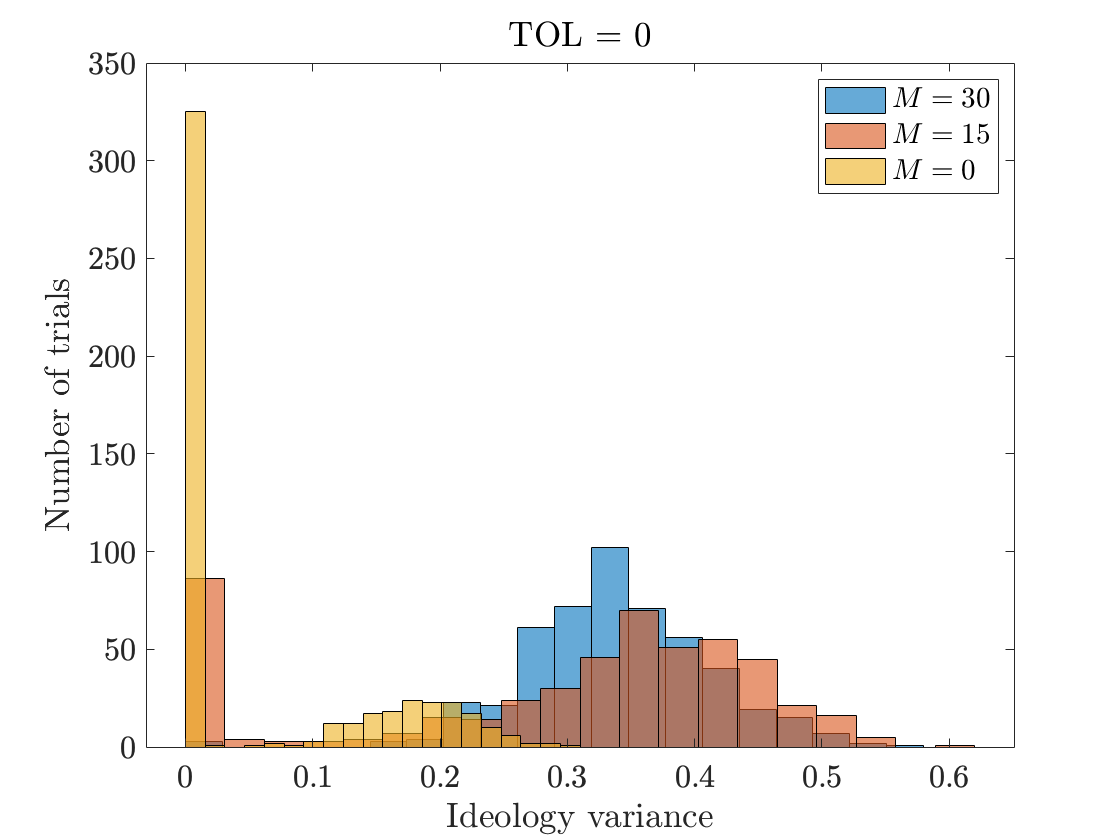}
	\includegraphics[width=0.4\textwidth]{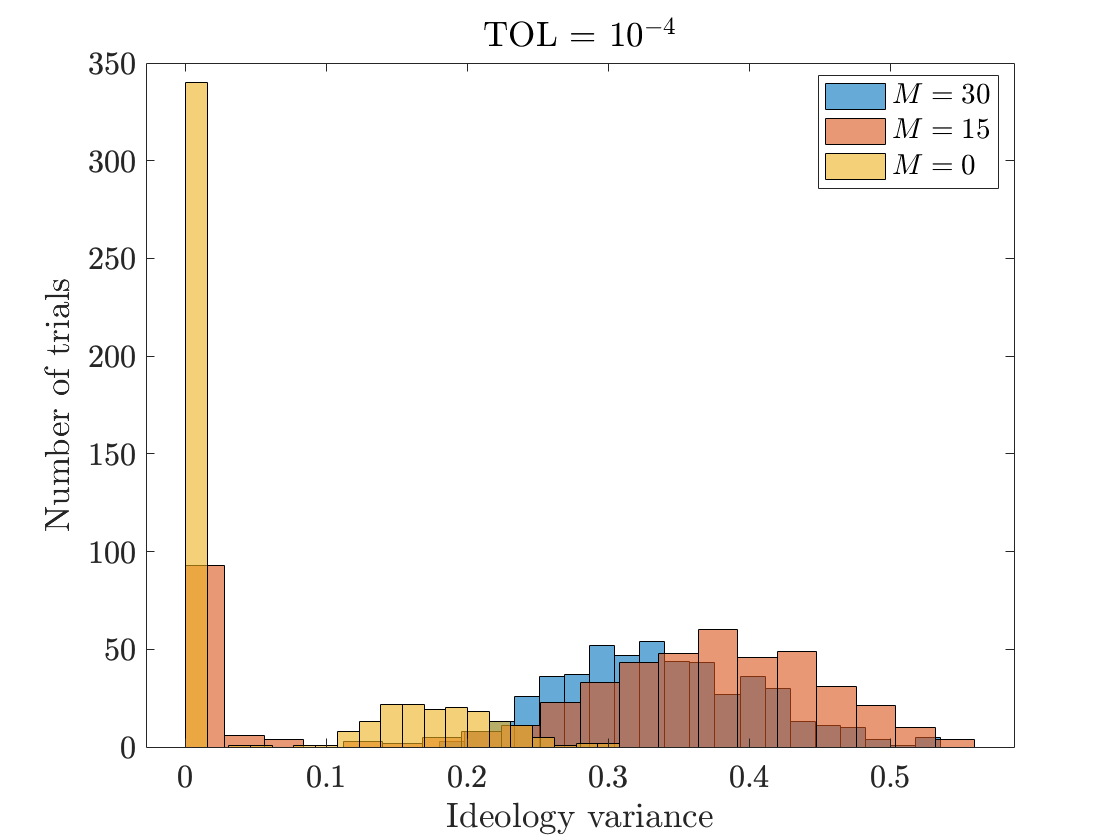}
	\caption{
	We simulate our model for $500$ trials, where we draw the network for each trial from a directed ER random-graph ensemble with $N=100$ nodes and an expected mean out-degree of $k=25$ (as in Fig.~\ref{fig:Constructedentrain}). We take $c=0.5$ and $x_M=0.9$, and we examine three levels of media impact: $M=0$ and $n_M=0$ (yellow), $M=15$ and $n_M=15$ (orange), and $M=30$ and $n_M=30$ (blue). 
	In the top row, we show the distributions of the mean ideology at convergence for convergence tolerances of $0$ (left) and $10^{-4}$ (right). In the bottom row, we show the distributions of the variance of the ideology at convergence for convergence tolerances of $0$ (left) and $10^{-4}$ (right). Our histograms have $20$ bins. For both tolerances, we observe that the mean ideology has the same qualitative behavior; the same is true of the variance of the ideology.
	}
\label{fig:timescale-dists}
\end{figure*}

Although the apparent ``convergence'' of our simulations does exhibit interesting
dynamics at multiple time scales in some trials (see Fig.~\ref{fig:timescale-example}), these cases are rare. The distributions for the mean have the same form regardless of whether we choose $\text{TOL}=10^{-4}$ or $\text{TOL}=0$. The same is true for the distributions of the variance.
In Figs.~\ref{fig:timescale-dists-reed} and \ref{fig:timescale-dists}, we show histograms of the mean ideology and variance of the ideology over $500$ trials in the Reed College Facebook network (see Fig.~\ref{fig:timescale-dists-reed}) and in directed ER networks (see Fig.~\ref{fig:timescale-dists}). Both figures illustrate that the qualitative dynamics are same for both tolerance levels. In yellow, we show the distributions without media impact ($M=0$ and $n_M=0$); in orange, we show the distributions when there are a moderate number of media accounts with a moderate number of followers; in blue, we show the distributions when there are large number of media accounts that each have a large number of followers. 
As in our simulations in Sec.~\ref{sec:1dimpact},
 the mean ideologies are higher in the presence of media nodes; and perfect entrainment (for which the mean ideology is $0.9$) occurs primarily when there are a moderate number of media accounts (e.g., $M=15$ for the ER networks), rather than when there are a larger number of media accounts (e.g., $M=30$ for the ER networks). 
 We supplement these observations by examining the distributions of the variance in ideology. Small variance in ideology corresponds to consensus states, and we observe these states only when media influence is absent (e.g., $M=0$) or low or moderate (e.g., $M=15$ for the ER networks). In other words, high levels of media influence (e.g., $M=30$ for the ER networks) do not result in ideological consensus (including as a result of entrainment to the media ideology), even at $0$ tolerance.

In the present paper, we focus on dynamics on short and medium time scales, as we believe that it is particularly relevant to simulate the dynamics of content spread over the course of a news cycle. However, the observed separation of time scales in some trials of our numerical experiments may have interesting implications for future studies on long-term effects of media impact in online social networks. 


\section{Extending our model to two ideological dimensions} \label{sec:2d}

We now move beyond the discussion of one ideological dimension and examine our content-spreading model with two ideological dimensions. Suppose that, at time $t$, each account has an ideological position of ${\bf x}_{i}\in[-1,1]\times[-1,1]$. We use the $\ell_2$ norm to calculate the distance $\mathrm{dist}({\bf x}_j,{\bf x}_i)=\| {\bf x}_j - {\bf x}_i \|_2$ in ideology between accounts $i$ and $j$. Depending on context, this two-dimensional ideology may represent two different aspects of a political-bias spectrum (e.g., the first dimension may represent socially liberal versus socially conservative, whereas the second may represent preferred economic policies that very between socialism and capitalism); or it may represent political bias on multiple issues (e.g., immigration reform and gun control). Given our ideological space and choice of metric on this space, we then write our model as in Eq.~\eqref{eqn:generalmodel}. 
 
 
\subsection{Simulations}

We simulate our content-spreading model with two ideological dimensions using a straightforward generalization of the simulations of our model with one ideological dimension. We add media accounts with a given ideology ${\bf x}_M$ and assign their followers uniformly at random from the non-media accounts. In our examples, we initialize non-media accounts with ideological positions that we draw uniformly at random for each trial. (Therefore, it is again true that there is no spatial ordering of ideology in the network structure.)

In Fig.~\ref{fig:2dimpact}, we show a heat map of simulations of media impact in our model with two ideological dimensions. We observe qualitatively similar results as in our simulations of our content-spreading model with one-dimensional ideologies. We again see that the largest media impact (i.e., entrainment) does not occur for the largest numbers of media accounts and followers per media account. Instead, the largest amount of entrainment occurs when there are a moderate number of media accounts, each of which has a moderate number of followers.


\section{Combining Media Bias and Quality}
\label{sec:quality}

Thus far, we have used our content-spreading model to study the effects of media ideology on content spreading in social networks. We now introduce the primary novel component of our work: incorporation of media content quality into our spreading model. Bessi et al.~\cite{bessi2014social} observed that individuals are more likely to spread low-quality content if it confirms or supports their existing biases, and we seek to incorporate such behavior into our content-spreading model.  

We again denote the state of account $i$ by ${\bf x}_i$. In our prior discussions, this state was a vector in $d$-dimensional ideological space (with $d = 1$ and $d = 2$ in our simulations). Now, however, we introduce an additional dimension into an account's current state to indicate the quality of shared content; therefore, we now take ${\bf x}_i \in [-1,1]^d\times[0,1]$. We allow content quality to lie on a continuous spectrum, with values near $x_{d+1,i}=0$ representing propaganda and disinformation (e.g., clickbait or so-called ``fake news'') and values closer to $x_{d+1,i}=1$ representing substantive, thorough, fact-based material. 

To focus our discussion, suppose that there is one ideological dimension, so $d=1$ and ${\bf x}_i = (x_{1,i},x_{2,i}) \in [-1,1] \times [0,1]$. As before, suppose that accounts adjust their ideological views only when they are exposed to content that is within a distance $c$ of their current ideology (i.e., only when $\mathrm{dist}(x_{1,i},x_{1,j})\leq c$). 
Additionally, we now 
suppose that accounts also consider content quality when we determine their receptiveness. Account $i$ decides whether the content of account $j$ is acceptable for spreading based on the distance $\mathrm{dist}(x_{1,i},x_{1,j})$ between their ideological positions. If this distance is very small (e.g., close to $0$), 
this content supports account $i$'s ideology, and $i$ is more likely to spread it even if it is not of high quality. However, if the distance in ideology is larger (e.g., close to $c$), 
account $i$ is more discerning and tends to spread the content only when it is of sufficiently high quality. 

We quantify content discernment in the following way. We calculate the minimum acceptable quality $q_{i,j}$ as a linear function of distance in ideology between account $i$ and account $j$. That is,
\begin{align}
	q_{i,j}=\frac{1}{c}\mathrm{dist}(x_{1,i},x_{1,j})\,.
\label{eqn:qualitydiscern}
\end{align}
With this functional form of $q_{i,j}$, account $i$ spreads content that confirms its bias exactly regardless of quality, because the minimum acceptable quality is $0$. If the content has an ideological position that equals the maximum receptiveness distance $c$ away from account $i$'s ideological position, then account $i$ spreads the content only if it is of quality $1$, the highest possible quality.

The ideological-position updating rule with quality discernment is
\begin{align}
	{x}_{1,i}^{t+1} = \frac{1}{\vert I_i \vert + 1} \left( {x}_{1,i}^{t} + \sum_{j=1}^{N+M}A_{ij}{x}_{1,j}^{t}g({\bf x}_i^{t},{\bf x}_{j}^{t})\right)\,,
\label{eqn:qualitymodel}
\end{align} 
where
\begin{align}
	g({\bf x}_i^{t},{\bf x}_j^{t}) = \begin{cases} 1\,, & \text{if } x_{2,j}\geq q_{i,j} \\
		0\,, & \text{otherwise}
		\end{cases}
\label{eqn:gfun}
\end{align}
and
\begin{align}
	I_i = \{ j \in \{1, \dots, N+M\} | A_{ij}=1; g({\bf x}_i,{\bf x}_j)=1\}\,.
\end{align}
We take the metric to be the $\ell_2$ norm $\mathrm{dist}(x_i,x_j)=\|x_{1,i}-x_{1,j}\|_2$, so we are using the same ideology metric as in our previous discussions. When a non-media node elects to spread content, we also adjust the quality of the content that it spreads with the update
\begin{align}
	{x}_{2,i}^{t+1} = \frac{1}{\vert I_i \vert + 1} \left( {x}_{2,i}^{t} + \sum_{j=1}^{N+M}A_{ij}{x}_{2,j}^{t}g({\bf x}_i^{t},{\bf x}_{j}^{t})\right)\,,
\label{quality-update}
\end{align} 
which, along with Eq.~\eqref{eqn:qualitymodel}, implies that 
\begin{align}
	{\bf x}_{i}^{t+1} = \frac{1}{\vert I_i \vert + 1} \left( {\bf x}_{i}^{t} + \sum_{j=1}^{N+M}A_{ij}{\bf x}_{j}^{t}g({\bf x}_i^{t},{\bf x}_{j}^{t})\right)\,.
\label{update-overall}
\end{align} 


\section{Measuring Impact from Multiple Sources: Media Bias and Quality} \label{sec:multi}

Earlier in our paper (see Sec.~\ref{sec:1dimpact}), we studied the effect of the numbers of media accounts and media followers per media account when media content has one fixed ideology. Our incorporation of content quality allows us to examine an important question of societal interest in a simple but plausible way: How do competing media accounts influence the outcome of opinions or ideological positions in a social network? This situation models a more realistic scenario of the influence of disparate news media (with heterogeneous political biases and content quality) in a social network. We extend our media entrainment summary diagnostic from our model with one ideological dimension to measure the impact of media accounts on the mean ideological position at convergence when there are multiple media sources. In this scenario, the quantity of interest is a function $\overline{R}({\bf x})$ that encodes
the impact of
each possible ideological position ${\bf x}$. 

To construct the function $\overline{R}({\bf x})$, we first establish a baseline function $r_0({\bf x})$ that encodes
the prevalence of each ideology for a given network in the absence of media accounts. Ideally, it make be desirable to
 take $r_0({\bf x})$ to be the probability that an account prefers content with ideological position ${\bf x}$. In practice, we construct a function $r_0({\bf x})$ by binning the ideology of content for each trial into bins of width $\delta {\bf x}$; we then count the number of times that there is content 
 in the ideology interval $[{\bf x}, {\bf x}+\delta {\bf x}]$, and we calculate a normalized histogram. Once we have constructed the baseline function $r_0({\bf x})$, we use our previous strategy (see Sec.~\ref{sec:1dimpact}) to construct a function $r_i({\bf x})$ that measures the prevalence of each ideology for a network in the presence of media accounts. That is, $r_i({\bf x})$ is the distribution that describes the probability that a non-media account has ideological position ${\bf x}$ when there are media accounts in the network. We then use these functions to construct the media impact function $\overline{R}({\bf x}) = r_i({\bf x})-r_b({\bf x})$. Positive values of $\overline{R}({\bf x})$ for ideology ${\bf x}$ indicate that the media has enhanced the prevalence of content with this ideology in the network, negative values indicate that the media has decreased the prevalence of content with this ideology in the network, and $0$ indicates no change.  

In our model, it is interesting to consider a variety of distributions of media ideologies. For example, we can draw these opinions from a convenient synthetic probability distribution, or we can determine them from empirical data. First, using one ideological dimension, we consider two examples in which we draw media ideologies from synthetic probability distributions. In Fig.~\ref{fig:impactvaryn}, we show the media impact functions when there are $M=100$ media accounts with ideologies that we draw (1) from a uniform distribution on $[-1,1]$ and (2) from a truncated Gaussian distribution on $[-1,1]$ with mean $0$ (before truncation) and standard deviation $0.5$. 
For each of these two examples, we examine the dependence of the media impact functions on the number $n_M$ of followers per media account. 

\begin{figure}[h!]
\includegraphics[height=2.5in]{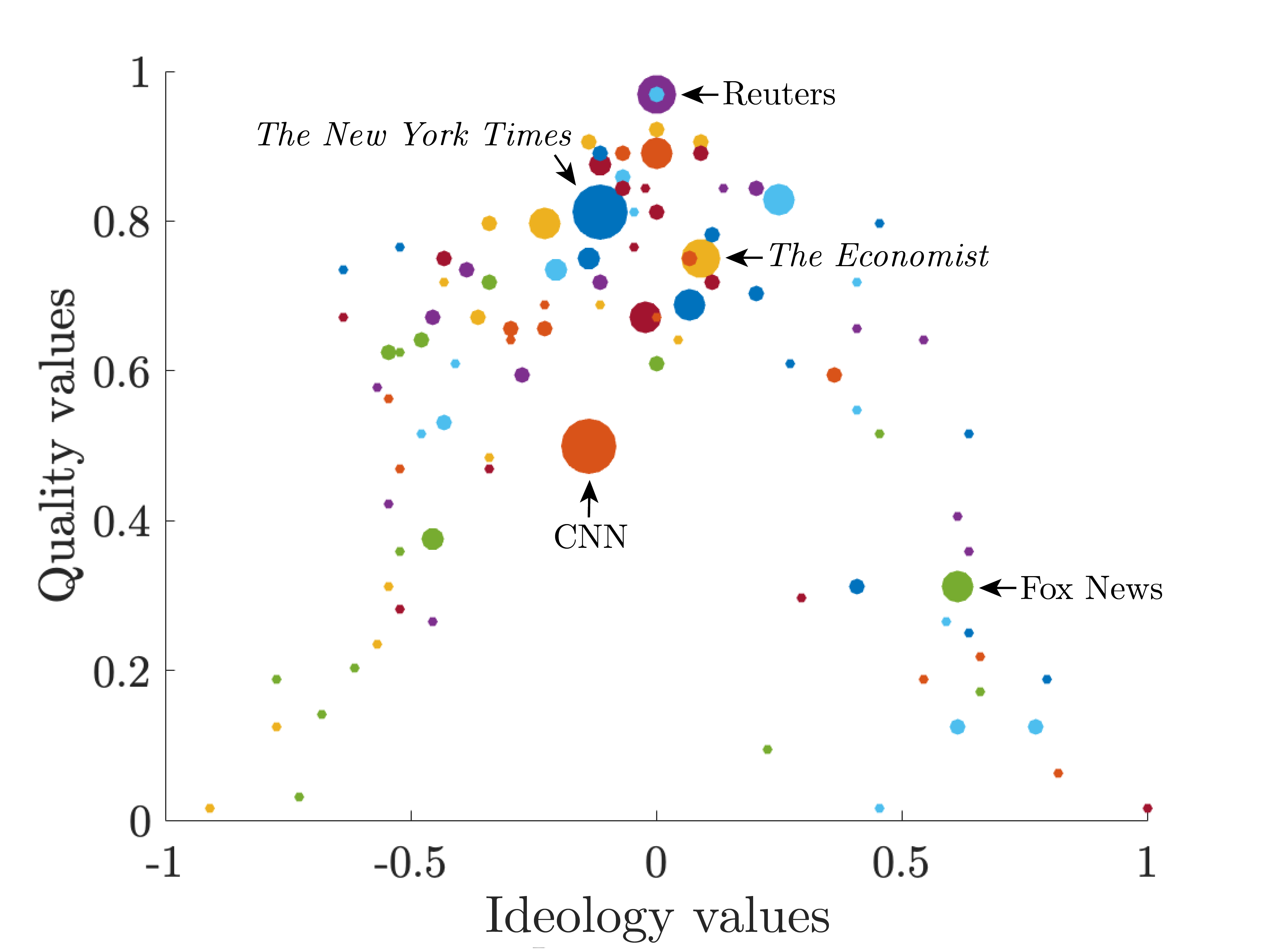}
\caption{Distribution of media accounts in (ideology, quality) space. We include $M = 103$ media accounts, which are hand-curated and available from the Ad Fontes Media Bias Chart \cite{mediabias}. Each colored dot represents one media account. The horizontal coordinate represents the ideological position of an account's political content, and the vertical coordinate represents the quality of its political content. The size of a dot represents the number of followers of the associated Twitter account. The five labeled accounts are the ones that are followed by at least $25$\% of the non-media accounts in our numerical experiments.
}
\label{fig:mediabiaschart}
\end{figure}

We then showcase our model using a media distribution that we generate from hand-curated empirical observations of real-world media sources that incorporate both political ideology (in one dimension) and content quality (in the other dimension). This example includes $M=103$ media accounts with ideological biases and qualities from Version 4.0 of the hand-curated Ad Fontes Media Bias Chart \cite{mediabias}. We rescale the ideology and quality chart coordinates so that they lie within the intervals $[-1,1]$ and $[0,1]$, respectively. In Fig.~\ref{fig:mediabiaschart}, we show these media content coordinates. These coordinates provide an illustrative example of a possible input to our model; one should not necessarily construe them as quantitatively representing the ``true" ideologies of the depicted sources. An important extension of our work is developing quantitative techniques (e.g., using sentiment analysis) to analyze bias and quality of content sources from real data. This is a difficult and interesting problem, and we leave it as future work.

\begin{figure*}
	\begin{subfigure}[t]{0.3\textwidth}
		\includegraphics[height=1.5in]{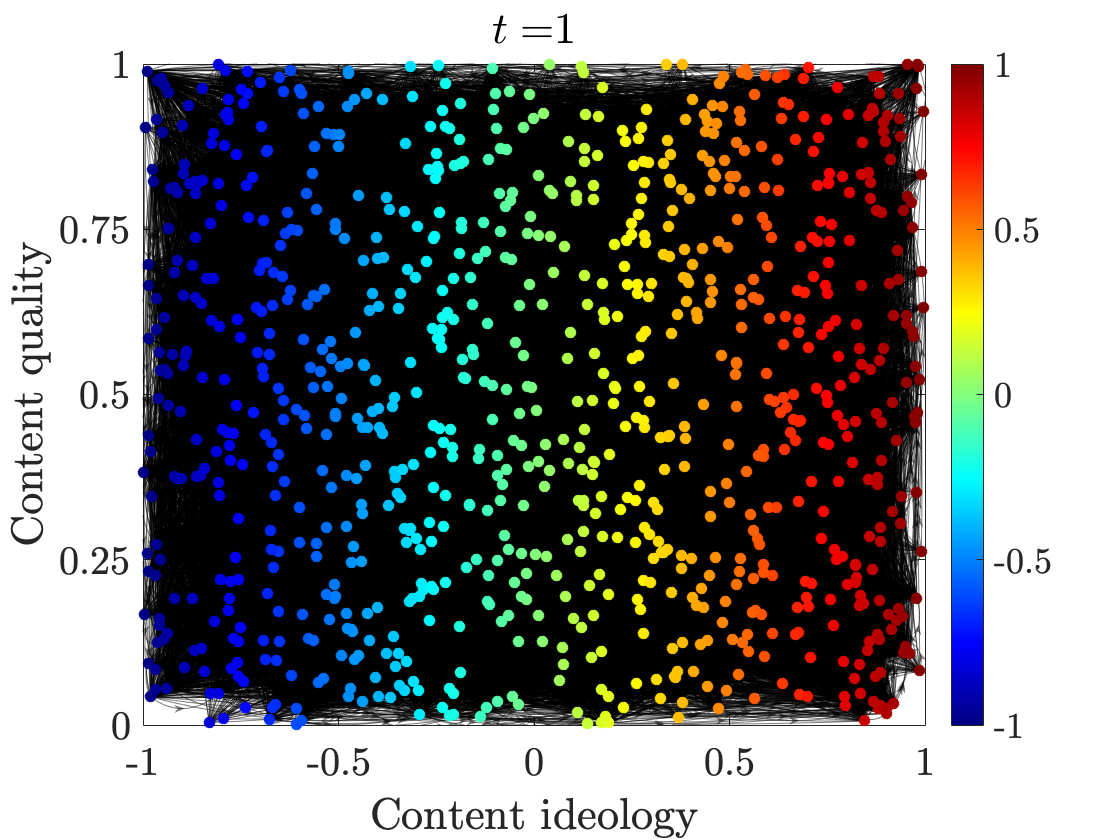}
	\end{subfigure}
	\begin{subfigure}[t]{0.3\textwidth}
		\includegraphics[height=1.5in]{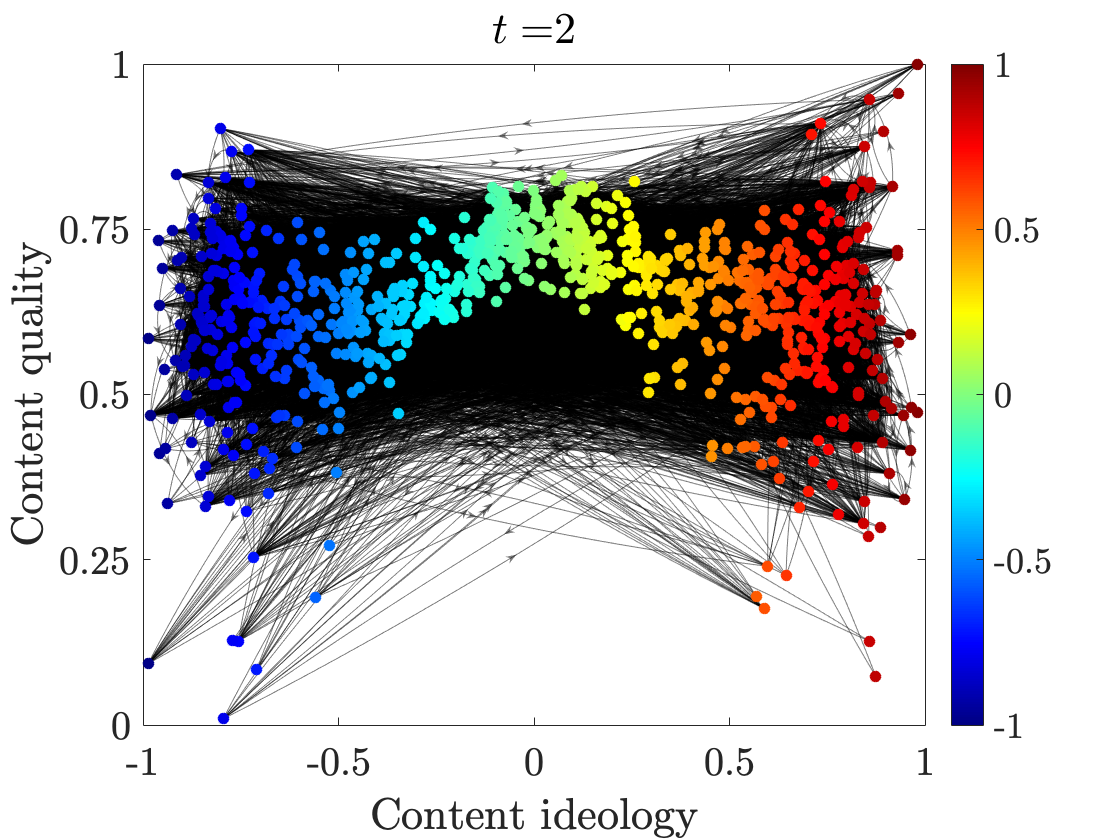}
	\end{subfigure}
	\begin{subfigure}[t]{0.3\textwidth}
		\includegraphics[height=1.5in]{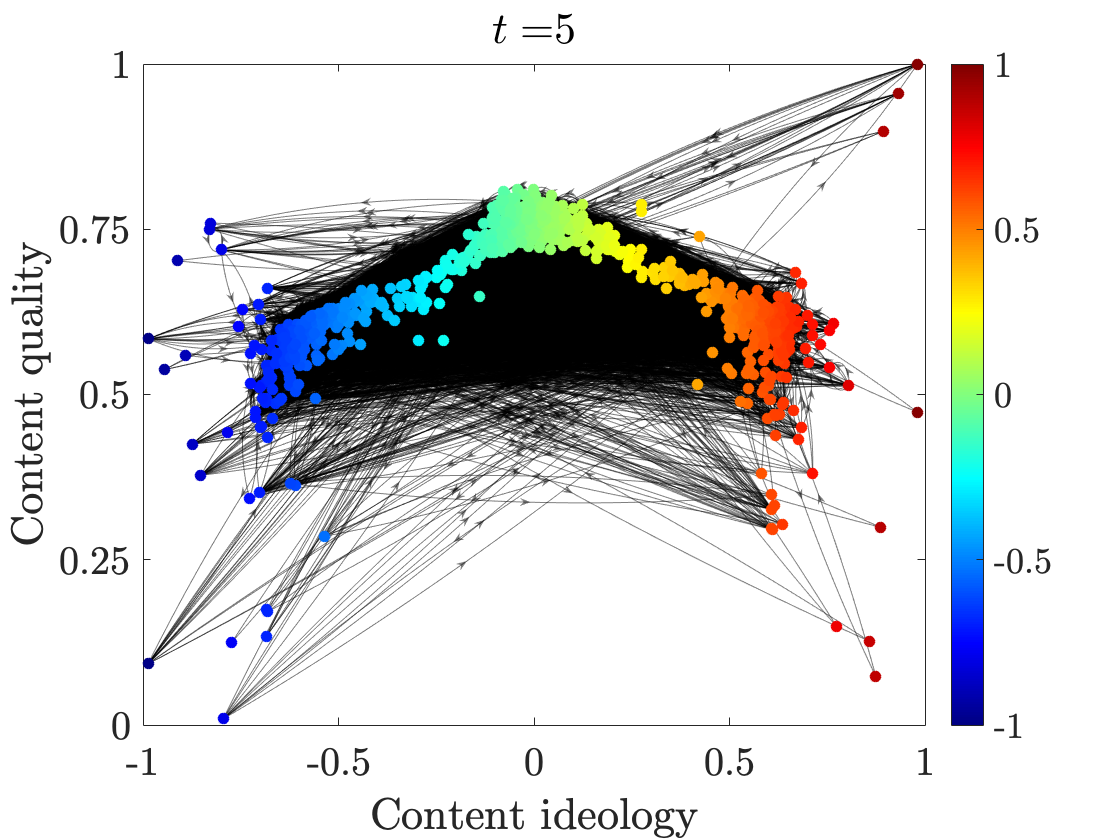}
	\end{subfigure}
	\begin{subfigure}[t]{0.3\textwidth}
		\includegraphics[height=1.5in]{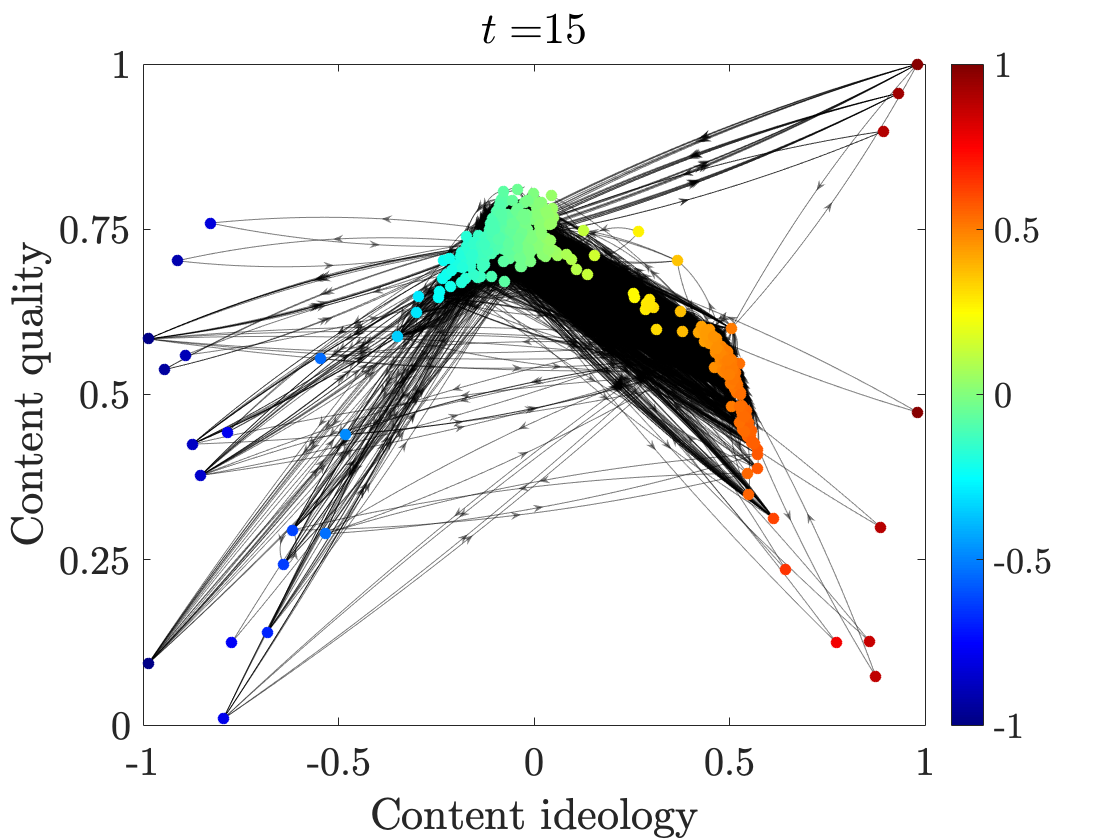}
	\end{subfigure}
	\begin{subfigure}[t]{0.3\textwidth}
		\includegraphics[height=1.5in]{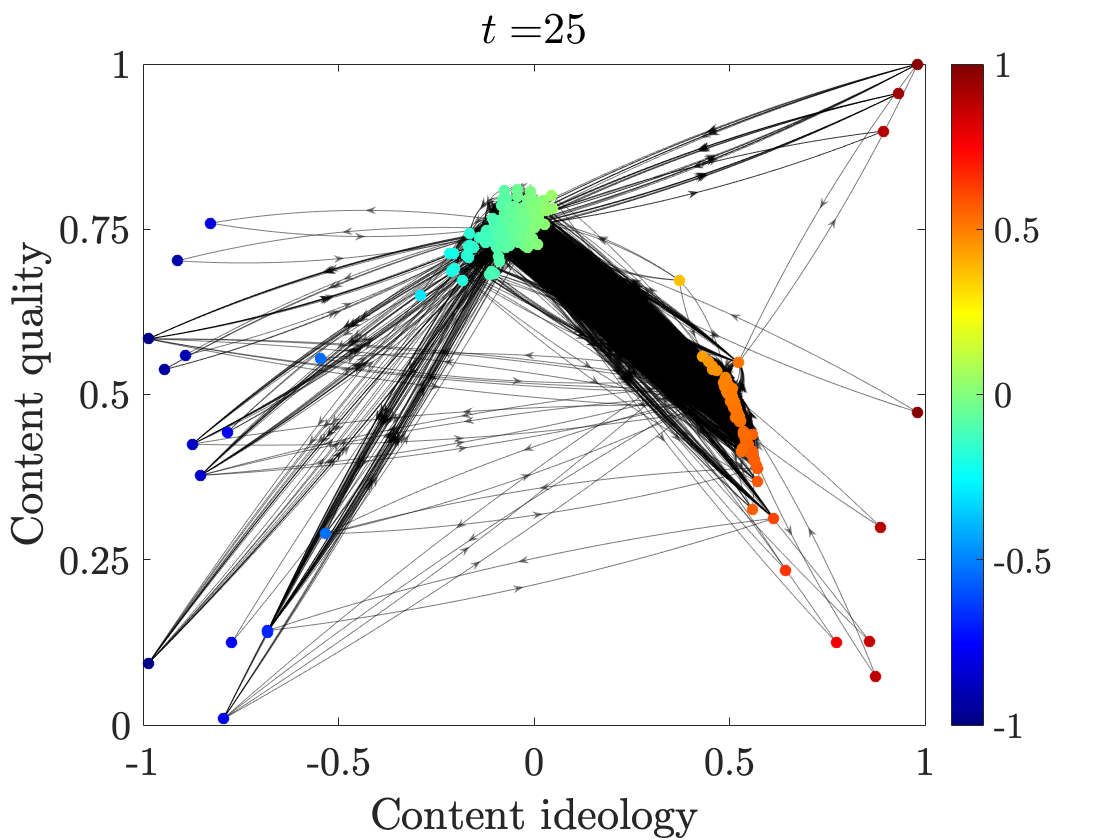}
	\end{subfigure}
	\begin{subfigure}[t]{0.3\textwidth}
		\includegraphics[height=1.5in]{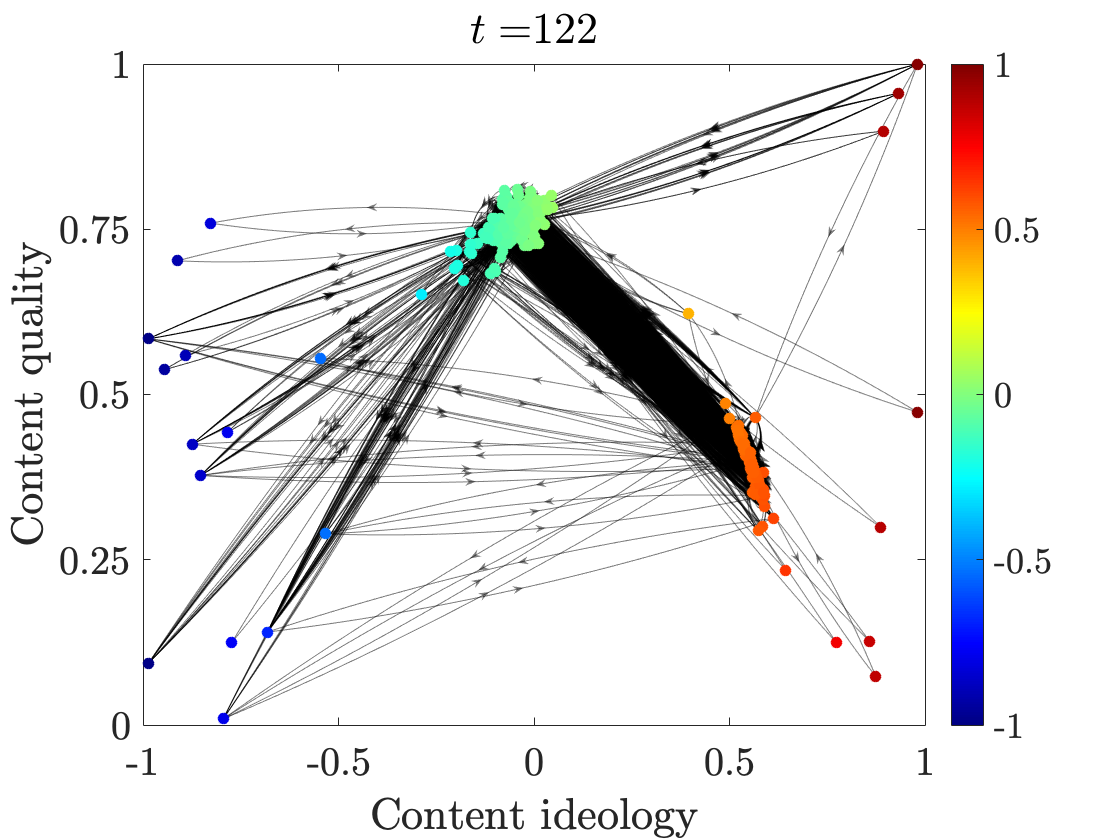}
	\end{subfigure}
\caption{Snapshots of temporal evolution of content ideology and quality in the Reed College network with $M = 103$ media accounts over $122$ time steps. We draw media ideologies and qualities from Version 4.0 of the Ad Fontes Media Bias Chart \cite{mediabias}. We choose the number of followers per media account to be proportional to the approximate number of followers that each media source had on Twitter on 15 February 2019 at 17:36 Pacific Standard Time (under the constraint that each media source in the model has at least one follower, so we round up to one follower for some accounts). Each dot in the figure represents a non-media account. The horizontal coordinate represents the ideological bias of each account's content at time $t$, and the vertical coordinate represents its content quality at time $t$. To aid visualization, the colors also signify  
ideological position. (The bluest color is the most liberal, and the reddest color is the most conservative.) We represent follower relationships as lines between the dots. This network architecture does not change over time, although the content that is spread by the non-media accounts does change over time. Because the ideology and quality of the content from the media accounts does not change, we do not show these nodes.
}
\label{fig:mediaexample}
\end{figure*}

In Fig.~\ref{fig:mediaexample}, we show the temporal evolution of one trial of our content-quality spreading model on the Reed College network with media account ideologies and qualities from the Media Bias Chart. We set each media account $j$ to have a number $n_{M,j}$ of followers that is proportional to its number of followers on Twitter on 15 February 2019 at 17:36 Pacific Standard Time (under the constraint that each media source in the model has at least one follower, so we round up to one follower for some accounts). We select followers for each media account by selecting the non-media accounts whose initial conditions are closest in ideological position to that of the media account. That is, for each media account $j$ (with $j \in\{1,\dots,M\}$) and each non-media account $i$ (with $i \in \{1,\dots,N\}$), we calculate the distance $\textrm{dist}(x_{1,i}^0,x_{1,M_j})$ in ideological position for all $i$ at $t=0$. From this set of all distances in ideological position from media account $j$, we select the $n_{M_j}$ smallest distances; these non-media accounts are the followers of media account $j$. In the simulations for which we use this media distribution as inputs, we observe the emergence of two primary communities (``echo chambers") of content: one in which the content is moderate ideologically and of fairly high quality (specifically, it is about
$0$ in ideology and about $0.75$ in quality) and one in which the opinion is more conservative but of lower quality (specifically, about $0.5$ in ideology and about $0.4$ in quality). The polarization that results from simulations of our model is a reflection of the polarization in the ideology and quality of the media account distribution. In our simulations, the account for Fox News (which has an ideological position of $0.613636$ and a quality of $0.3125$ in the Media Bias Chart and is followed by 28\% of the non-media accounts) is very popular and near the final location of the conservative community in (ideology, quality) space.

\begin{figure}[h!]
	\includegraphics[height=2.5in]{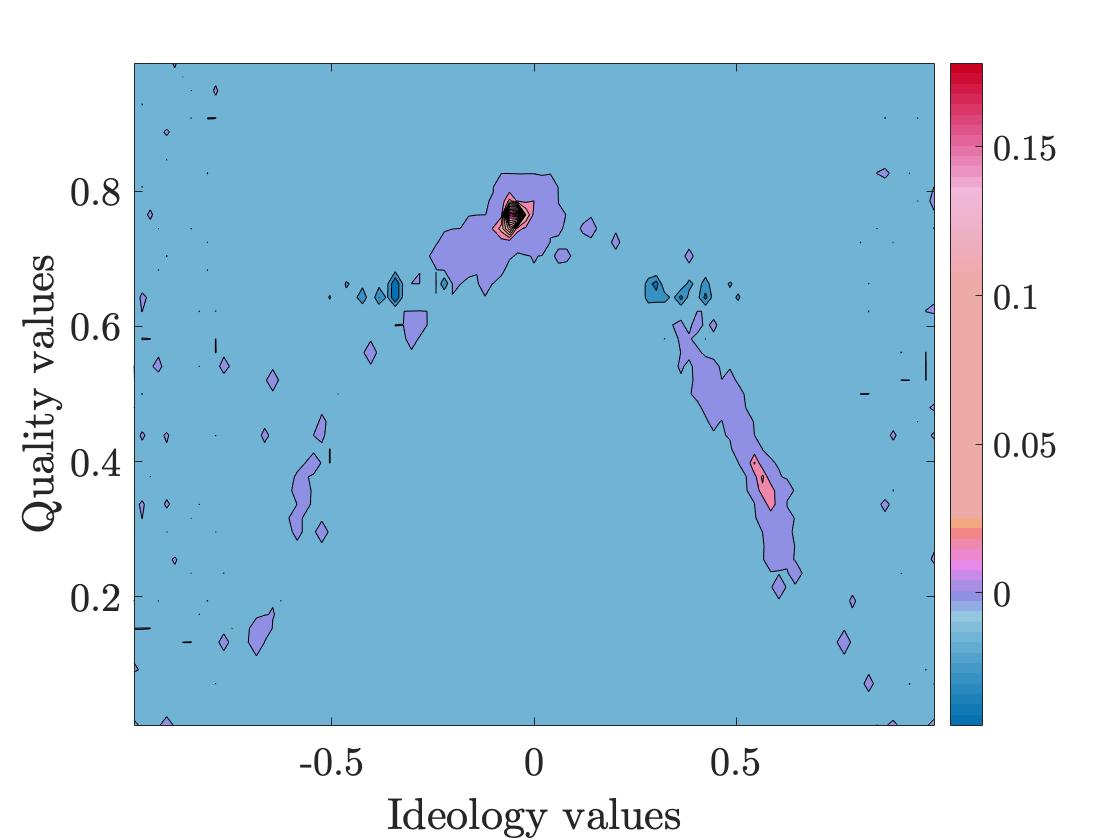}
\caption{Contour plot of the media impact function $\overline{R}(x)$ for ideology and quality in the Reed College Facebook network with $M = 103$ media accounts, where we draw media biases and qualities from Version 4.0 of the Ad Fontes Media Bias Chart \cite{mediabias} as in Fig.~\ref{fig:mediaexample}. The impact function $\overline{R}(x)$ illustrates the relative impact of the media accounts on the mean ideology and quality of content in the network. 
When $\overline{R}<0$ (in blue), the media has decreased the prevalence of content with the (ideology, quality) pair $(x_{1,i},x_{2,i})$ in comparison to the content in the absence of media; when $\overline{R}>0$ (in pink), the media has increased the prevalence of content with the (ideology, quality) pair $(x_{1,i},x_{2,i})$.
}
\label{fig:mediabiasimpact}
\end{figure}


\section{Conclusions and Discussion} \label{sec:discussion}

We introduced a model for the spread of content in an online social network, with both media and non-media nodes, that accounts for the effects of both ideology and content quality on the spreading dynamics. We based the spreading dynamics of our model on a bounded-confidence mechanism, such that accounts spread content that is sufficiently close to their current ideological position, as reflected by the content that they shared in the immediately previous time step. In the most sophisticated version of our model, the quality of content determines how close an account's ideological position should be to the content's ideological position for the account to share it; therefore, low-quality content is shared only when it supports an account's existing ideological biases. As far as we are aware, our model is the first mathematical model to explicitly incorporate the effects of media quality on spreading dynamics. This is a key novelty of our work.

We conducted simulations for our content-spreading model for media content with both one and two ideological dimensions. Using results from our simulations, we quantified the level of media ``entrainment'' (i.e., how much the media affects the ideological positions of non-media accounts) for a variety of network architectures. We examined how the amount of media entrainment of the non-media accounts in a 
network depends on the numbers of media accounts and followers per media account for that network. We found that media impact increases when one increases the number of non-media accounts, the mean number of non-media connections, or the receptiveness of non-media accounts. We also observed a relationship between media entrainment and convergence time in our model, with higher levels of media entrainment correlating positively with longer convergence times. Finally, we simulated a version of our model that accounts for content quality in the spreading dynamics. Using a hand-curated set of media inputs from real media outlets and their online followerships, we demonstrated that this version of our model produces polarization in both ideology and quality of content. Specifically, it yields a community of high-quality content in the center of the political spectrum and a conservative community of low-quality content. Our model provides a useful step towards increasing understanding of how media content quality affects the spread of online content. 

Our model is a simplistic model for the spread of media content in online social networks. This is a complex system, and naturally it is not appropriate to view our model as a perfectly accurate mathematical description of such phenomena.
Instead, our model provides a starting point for exploring the mechanisms that contribute to content spreading dynamics and echo chambers. There are many worthwhile ways to generalize our model. For example, our assumption that accounts choose whether to spread content based on a universal confidence parameter $c$ is a naive simplification, as is the homogeneity of non-media accounts in general. Spreading behavior surely depends on individual characteristics, as has been explored in models that include zealots \cite{waagen2015effect} and in other models of social dynamics \cite{yy-book,porter2016,castellano2009statistical}. Augmenting our model of content spreading by incorporating account heterogeneity is important future work. In such efforts, we expect that it will be insightful to explore the effects of structural homophily ({for example, as was explored using the DeGroot model in \cite{dandekar2013biased}}). Individuals in social networks are free to choose which accounts to follow (and which accounts to stop following or never follow), and such choices are sometimes driven by the desire to follow accounts that have similar ideologies \cite{bakshy2015exposure, bessi2014social}. This can increase structural homophily and exaggerate echo chambers in networks \cite{bessi2016homophily,del2016echo}, because edges are more likely to arise (and persist) between nodes with similar ideologies.
 
There are also other interesting avenues for extending our work. For example, one can generalize the dynamics from multiple ideological dimensions by allowing different weightings of the ideological dimensions. One can also incorporate spreading through multiple types of social media by generalizing our model to multilayer networks \cite{kivela2014multilayer} or develop models of media influence that go beyond pairwise interactions by generalizing it to hypergraphs or simplicial complexes \cite{renaud2019}. It will also be interesting to generalize recent work on bounded-confidence opinion dynamics that coevolve with network structure \cite{brede2019} to include the effects of media. It is also important to explore the influence of time-dependence in network architecture \cite{holme2012temporal} on the dynamics of our content updating rule, as accounts can follow new accounts and unfollow accounts (e.g., using a simplified setting such as network rewiring, as in adaptive voter models \cite{durrett12}).

A key strength of our model is its flexibility, as one can formulate generalizations of it (such as ones that we just discussed) in a straightforward way. Furthermore, one can already readily use our model to study many interesting phenomena.
For example, although we only examined a few case studies of media probability distributions in the present paper,
one can draw the number of media accounts, number of followers per account, and ideologies and quality of media content from any probability distribution --- either synthetic ones or ones that are inferred from empirical data --- and it is important to explore these ideas in future work. It is particularly desirable for media nodes in a content-spreading model to produce content that follows some distribution of ideologies and qualities, rather than always having the same parameter values for such features. It would also be very interesting to include media nodes with ideologies $x_M$ that are inferred from data, such as through sentiment analysis on news stories on a given topic (as described in \cite{batrinca2015social} and references therein). Using procedures such as topic modeling and sentiment analysis can generate a probability distribution of media accounts in ideological space as an output, which can then be fed into our model. There has been some recent progress in this direction. For example, Ye and Skiena \cite{ye2019mediarank} 
measured media bias and quality from the web pages and tweets of media sources, and Albanese et al.~\cite{albanese2019data} created a 
two-dimensional 
model based on semantic analysis to quantify media influence on voting.
Another important issue is to examine transient dynamics of content spreading, as it is necessary to go beyond our focus on the properties of ideological positions in long times or at convergence.

Developing mathematical approaches for analyzing and quantifying the dynamics of content spreading
has ramifications for how to mitigate the spread of undesired content and promote the spread of desired content online. One potential impact is the development of control strategies and ``fake-news filters'' (e.g., by flagging content that is below some threshold value on the quality axis in a model like ours) that are reminiscent of spam filters. Our work provides a step toward the development of novel algorithms to encourage selective spreading of high-quality or desirable content. Another area for which further development of such models may also be fruitful is in bot detection, where most existing algorithms rely on network measures, followership data, activity rates, or linguistic cues \cite{ferrara2016rise} (all of which are straightforward to manipulate by malicious actors), rather than directly exploiting spreading dynamics. Advances in these modeling efforts will also yield insights into the theory of online content spreading 
 and help bridge the gap between simple spreading models and realistic investigations of spreading on social media.


\section*{Acknowledgements}

MAP was supported by the National Science Foundation (grant \#1922952) through the Algorithms for Threat Detection (ATD) program. We thank Franca Hoffmann, Alex Pan, and two anonymous referees for their helpful comments. 






%


\end{document}